\documentclass[aps,superscriptaddress, showpacs,preprintnumbers, superscriptaddress, nofootinbibt,twocolumn]{revtex4}
\usepackage{eurosym}
\usepackage{amsfonts}
\usepackage{amssymb,amsmath,enumitem}
\usepackage{graphicx,color,epstopdf}
\usepackage{subfigure}
\usepackage{ulem}
\usepackage{tabularx}
\usepackage{booktabs}

\def\cL{\mathcal{L}}
\def\be{\begin{equation}}
\def\ee{\end{equation}}
\def\beg{\begin{align}}
\def\eeg{\end{align}}
\def\bea{\begin{eqnarray}}
\def\eea{\end{eqnarray}}

\def\nn{\nonumber \\}

\def\l{\left}
\def\r{\right}
\newcommand{\f}[2]{\frac{#1}{#2}}

\begin{document}

\title{Black hole solutions in modified gravity induced by quantum metric fluctuations}
\author{Jin-Zhao Yang}
\email{yangjch6@mail3.sysu.edu.cn}
\affiliation{School of Physics, Sun Yat-Sen University, Xingang Road, Guangzhou 510275,
P. R. China,}
\author{Shahab Shahidi}
\email{s.shahidi@du.ac.ir}
\affiliation{School of Physics, Damghan University, Damghan, Iran,}
\author{Tiberiu Harko}
\email{tiberiu.harko@aira.astro.ro}
\affiliation{Astronomical Observatory, 19 Ciresilor Street, 400487 Cluj-Napoca, Romania,}
\affiliation{Department of Physics, Babes-Bolyai University, Kogalniceanu Street,
Cluj-Napoca 400084, Romania,}
\affiliation{School of Physics, Sun Yat-Sen University, Xingang Road, Guangzhou 510275,
P. R. China,}

\author{Shi-Dong Liang}
\email{stslsd@mail.sysu.edu.cn}
\affiliation{School of Physics, Sun Yat-Sen University, Guangzhou 510275, People's
Republic of China,}
\affiliation{State Key Laboratory of Optoelectronic Material and Technology, and
Guangdong Province Key Laboratory of Display Material and Technology, \\
Sun Yat-Sen University, Guangzhou 510275, People's Republic of China,}

\date{\today }

\begin{abstract}
The inclusion of the quantum fluctuations of the  metric in the geometric action is a promising avenue for the understanding of the quantum properties of gravity. In this approach the metric is decomposed in the sum of a classical and of a fluctuating part, of quantum origin, which can be generally expressed in terms of an arbitrary second order tensor, constructed from the metric, and from the thermodynamic quantities describing matter fields. In the present paper we investigate the effects of  the quantum fluctuations on the spherically symmetric static black hole solutions of the modified field equations, obtained from a variational principle, by assuming that the quantum correction tensor is given by the coupling of a scalar field to the metric tensor.  After reformulating the field equations in a dimensionless form, and by introducing a suitable independent radial coordinate, we obtain their solutions numerically. We detect the formation of a black hole from the presence of a singularity in the metric tensor components. Several models, corresponding to different functional forms of the scalar field potential, are considered. The thermodynamic properties of the black hole solutions (horizon temperature, specific heat, entropy and evaporation time due to Hawking luminosity) are also investigated.
\end{abstract}

\maketitle
\tableofcontents

\section{Introduction}

The problem of the quantization of the gravitational field is still unsolved, representing a major subject of research in present day physics. Several approaches have been proposed to tackle this problem, involving different theoretical assumptions. One of the possible approaches to the problem of the quantization of gravity is the semiclassical approach, in  which matter is described quantum mechanically, while gravity is treated classically, via the Hilbert-Einstein action, $S=\int{\left(-R/2\kappa ^2\right)\sqrt{-g}d^4x}$, where $R$ is the Ricci scalar, and $\kappa $ is the gravitational coupling constant, even that fundamentally it is of quantum nature. Hence in such a theory gravity is coupled to matter by means of the semiclassical Einstein equations
\be\label{1}
R_{\mu \nu}-\frac{1}{2}g_{\mu \nu}R=\frac{8\pi G}{c^4}\left\langle\Psi\left|\hat{T}_{\mu\nu}\right|\Psi \right\rangle,
\ee
obtained by replacing  in Einstein's gravitational field equations the classical matter energy-momentum tensor $T_{\mu\nu}$ by the expectation value of the corresponding quantum operator in a given quantum state $\Psi$. The idea of semiclassical gravity was introduced in \cite{Moller} and \cite{Ros}, and its different aspects and implications have been further developed and discussed in \cite{sg1,sg2, sg3,sg3a, sg4,sg5,sg5a, sg6,sg7,sg8,sg9,sg10,sg11}.

It follows from Eq.~(\ref{1}) that the energy-momentum tensor $T_{\mu \nu}$  of the gravitating system is obtained in the classical limit from its quantum analogue by means of the $\left<\Psi \right |\hat{T}_{\mu \nu}\left |\Psi \right>=T_{\mu \nu}$.
One can derive the semiclassical Einstein equations ~(\ref{1}) from the variational principle
$\delta \left(S_g+S_{\psi}\right)=0$ \cite{sg3a},
where $S_g=\left(1/16\pi G\right)\int{R\sqrt{-g}d^4x}$ is the general relativistic action of the gravitational field, while the quantum part of the action is given by
\be\label{9}
S_{\Psi}=\int{\left[{\rm Im}\left \langle \dot{\Psi}|\Psi\right \rangle-\left \langle \Psi |\hat{H}|\Psi \right \rangle +\alpha \left(\left \langle \Psi |\Psi \right \rangle -1\right) \right]dt},
\ee
where $\hat{H}$ is the Hamiltonian operator, while $\alpha $ is a Lagrange multiplier. Another approach to semiclassical gravity has been proposed in \cite{sg3a},  and can  be obtained by introducing a coupling between the classical curvature scalar of the space-time and the quantum fields. More specifically, one can consider an action containing a geometry-quantum matter coupling of the form
$\int{RF\left(\left<f(\phi)\right>\right)_{\Psi}\sqrt{-g}d^4x}$, where $F$ and $f$ are arbitrary functions, and $\left(\left<f(\phi)\right>\right)_{\Psi}=\left<\Psi (t)\right|f[\phi (x)]\left|\Psi (t)\right>$. Then, by taking into account the effect of the geometry-quantum matter coupling, the effective semiclassical Einstein equations become \cite{sg3a}
\bea\label{2nn}
R_{\mu \nu}-\frac{1}{2}Rg_{\mu \nu}&=&16\pi G\Big[\left< \hat{T}_{\mu \nu}\right> _{\Psi}+G_{\mu \nu}F-\nonumber\\
&& \nabla _{\mu}\nabla _{\nu} F+
g_{\mu \nu}\Box F\Big].
\eea

One of the important consequences of the semi-classical gravitational model described by  Eq.~(\ref{2nn}) is that the energy-momentum tensor of the matter is not conserved anymore, since $\nabla _{\mu}\left< \hat{T}^{\mu \nu}\right> _{\Psi}\neq 0$. Thus, the model described by the field equations (\ref{2nn}) is related to a particle creation effect, in which there is an effective energy transfer from space-time to matter.

Another approach to quantization of physical systems is the stochastic quantization method, in which quantum fluctuations are described with the use of the Langevin equation. The Schr\"{o}dinger  equation was obtained by using stochastic methods in \cite{Nelson},  but the rigorous formulation of the stochastic quantization procedure  was introduced in \cite{Parisi}. In stochastic quantization  the quantum mechanical description of physical processes is
obtained as the limit,  with respect to a fictitious time variable $t$, of a hypothetical stochastic process described by a Langevin equation towards the thermal equilibrium. For reviews of the early developments in stochastic quantization see \cite{Dam} and \cite{Nam}, respectively. The study of the stochastic quantization of the gravitational field was initiated in \cite{R1,R2}, under the assumption that the metric tensor of general relativity satisfies the covariant Langevin equation given by \cite{R1}
\be
\dot{g}_{\alpha \beta}=-2i\left[R_{\alpha \beta}-\frac{\lambda +1}{2(2\lambda +1)}g_{\alpha \beta}R\right]+\xi _{\alpha \beta},
\ee
where $\lambda $ is a parameter, and $\xi _{\alpha \beta}$ is a stochastic source term. For recent developments in stochastic quantization of gravity see \cite{sg3a}.

An alternative semi-classical approach to quantum gravity was introduced in \cite{re8}, where it was shown that if in quantum gravity a metric  can be decomposed as the sum of the classical and quantum parts, then Einstein quantum gravity takes approximately the form of a modified gravity theory with a nonminimal interaction between geometry and matter. By making some assumptions for  the expectation value of a product of the quantum fluctuating metric at two points, in  \cite{re9} the effective semiclassical gravitational and scalar field Lagrangians were evaluated. For a vanishing expectation value of the first-order terms of the metric, the second order corrections were obtained. This type of quantum corrections also give rise to modified gravity.
The modified gravitational field equations and the modified conservation laws for the nonperturbative quantization of a metric were obtained in \cite{re11}, where it was shown that due to the quantum fluctuations a bounce universe can be obtained, and a decelerated expansion is also possibly in a dark energy dominated epoch. For other investigations of gravitational models with fluctuating metric see \cite{re10,re12,re13}. In \cite{re14}, after assuming that the expectation value of the quantum correction can be generally expressed in terms of an arbitrary second order tensor, the (classical) gravitational field equations in their general form have been derived. Cosmological models obtained by assuming that the quantum correction tensor is given by the coupling of a scalar field and of a scalar function to the metric tensor, and by a term proportional to the matter energy-momentum tensor, were investigated in detail.

Gravitational models with fluctuating metric lead naturally to modified gravity models with geometry-matter coupling. Such models have been previously proposed in the framework of classical physics as possible explanations of the cosmological observations that have led to a dramatic change of our understanding of the Universe.  High precision astronomical observations have indicated that the Universe underwent recently a transition to an accelerating, de Sitter type phase \cite{1n,2n,3n,4n,acc, Aad}. This observation is usually explained by postulating the existence of a major component of the Universe, the dark energy, which  can explain all the recent  cosmological observations \cite{PeRa03,Pa03}. However, to explain observations, a second component, called Dark Matter, is also necessary \cite{dm1,dm2}.

 On the other hand the possibility that dark energy as well as dark matter can be described as a generic property of an extended gravity theory,  going beyond general relativity, and its Hilbert-Einstein action formulation, cannot be rejected a priori. A large number of modified gravity theories generalizing Einstein's theory of general relativity have been proposed. One of the first investigations in this direction was represented by the $f(R)$ gravity theory, with action of the form \cite{Bu701,re4,Bu702,Bu703,re1,Bu704,Fel}
 \begin{equation*}
 S = \frac{1}{2\kappa ^2}\int {f(R) \sqrt{-g}d^4 x} +\int {L_m \sqrt{-g}d^4 x}.
 \end{equation*}
 This type of theories generalize  only the geometric part of the gravitational action, ignoring the deeper role the matter Lagrangian may play \cite{Mat}, and they involve only a  minimal coupling between matter and  geometry.

Modified gravity theories with arbitrary geometry-matter coupling were also proposed, the first one being the $f\left( R,L_{m}\right) $  modified gravity theory \cite{fL1,fL2,fL3,fL4}, with the gravitational action given by $S = \frac{1}{2\kappa ^2}\int {f\left(R,L_m\right) \sqrt{-g}d^4 x} $. Hence  matter becomes equivalent with geometry, and plays an effective role in modeling the geometry of space-time.  Another type of geometry-matter coupling is introduced in the
$f(R,T)$  gravity theory, with action given by $S=\int{\left[f(R,T)/2\kappa ^2+L_m\right]\sqrt{-g}d^4x}$ \cite{fT1,fT2}, where geometry and matter are coupled via  the trace $T$ of the energy-momentum tensor.  Many other gravitational theories with geometry-matter couplings have also been proposed, and studied extensively, like, for example, $f\left(R,T,R_{\mu \nu }T^{\mu \nu }\right)$ gravity theory, where $T_{\mu \nu }$ is the matter energy-momentum
tensor and $R_{\mu \nu} $ is the Ricci tensor, \cite{Har4,Odin}, hybrid metric-Palatini gravity theory $f(R,\mathcal{R})$ gravity,  where $\mathcal{R}$ is the Ricci scalar formed from a connection independent of the metric \cite{HM1,HM2,Revn1}, the
Weyl-Cartan-Weitzenb\"{o}ck (WCW) gravity theory \cite{WCW}, or the $f(Q,T)$ gravity \cite{fQ1,fQ2}, respectively.  Theories in which the torsion scalar $\tilde{T}$, essentially a geometric quantity, couples to the trace $T$ of the matter
energy-momentum tensor, called  $f(\tilde{T},\mathcal{T})$ gravity theories, have also been proposed \cite{HT}. Gravitational theories with higher derivative matter fields were considered in \cite{HDM}. For recent reviews of the generalized $f\left(R,L_{m}\right)$, $f(R,T)$, and Hybrid Metric-Palatini  type gravity theories see \cite{Revn} and \cite{book}, respectively.

The gravitational theories with geometry-matter coupling have the intriguing property that the four-divergence of the matter energy-momentum tensor is usually different of zero, $\nabla _{\mu}T^{\mu \nu}\neq 0$. From a thermodynamic  point of view the non-conservation of $T_{\mu \nu}$ can be interpreted physically within the formalism of the thermodynamics of open systems \cite{fT2, creat, Pavon}. Therefore one can assume that the energy and momentum balance equations in these gravitational theories describe irreversible matter production processes. Thus the non-conservation of $T_{\mu \nu}$  can be related to the irreversible energy transfer from the gravitational field to the newly created particle constituents.

One of the important predictions of the quantum field theory in curved space-times is the creation of particles from the cosmological vacuum   \cite{P1,Z1,P2,Full,P3}. In the quantum field theoretical approaches to gravity particles creation processes, where they naturally appear, are assumed to play a crucial role. According to quantum field theory in curved spacetimes in the expanding Friedmann-Robertson-Walker Universe quanta of the minimally-coupled scalar field are created from the cosmological vacuum \cite{P3, Lee, Park}.

Hence, the existence of particle production processes in both modified gravity theories with geometry-matter coupling and quantum theories of gravity in curved space-times suggests that a profound  relation between these two, allegedly distinct physical theories, may exist. In fact such a relation was found in \cite{re11}, where it was pointed out that in the framework of the  nonperturbative approach for the quantization of the metric, introduced in \cite{re8,re9,re10},  as a result of the quantum fluctuations of the metric, a particular type of $f(R,T)$ gravity naturally comes out. The Lagrangian of the theory, first introduced in \cite{re11} is given by
\begin{equation*}
L=\frac{1}{2\kappa ^2}\left[(1-\alpha )R+\left(L_m-\frac{\alpha}{2} T\right)\right]\sqrt{-g},
\end{equation*}
where $\alpha $ is a constant. This result suggests  the possibility of the phenomenological description of quantum  mechanical particle creation processes via the $f(R,T)$ or $f\left(R,L_m\right)$ gravities.  Such an interpretation may lead to a better understanding of the physical processes related to particle generation through the coupling between matter and geometry.

 Obtaining black hole type solutions of the Einstein field equations is of fundamental importance for the observational testing and theoretical understanding of gravitational theories. The first vacuum solution of general relativity was found by Karl Schwarzschild \cite{Sch1}, for the case of a static spherically symmetric central object. For a review of the exact solutions of the Einstein
field equations see \cite{0}. The properties of the gravitational force can be tested  by using the electromagnetic emissivity properties of
thin disks that exist around compact objects \cite{A1,
A2,A3,A4,A5,A6,A7,A8,A9,A10,A11,A12,A13,A14,A15,A16}. For the observational possibilities of testing black hole geometries by using electromagnetic
radiation see \cite{RB}.

Black hole type solutions have been extensively considered in  modified
theories of gravity. Since the existing literature is extremely extensive we refer the reader to the papers \cite{Br0, Br1, Br2, Br3, Bh1, Bh2, Bh3, Bh4, Bh5, Bh8, Bh10, Bh11, Bh12, Bh6, Bh9, Bh7, Bh13, Bh14, Bh14a, Bh14b, Bh14c, Bh15, Bh16, Bh17, Bh18, Bh19a, Bh19b, Bh20}. Thus,  it was shown in \cite{Bh15} that black-hole solutions are a generic property of the Einstein-scalar-Gauss-Bonnet theory with a coupling function $f(\phi)$.
 In \cite{Bh16} black hole solutions were studied in  $U(1)$ gauge-invariant scalar-vector-tensor theories with
second-order equations of motion. In \cite{Bh17}  four-dimensional conformal gravity models coupled with a
self-interacting conformally invariant scalar field were studied, and exact asymptotically anti-de Sitter black hole solutions and
asymptotically Lifshitz black hole solutions with dynamical exponents $z = 0$
and $z = 4$ of were obtained. In the Starobinsky modified gravity model the vacuum solutions around a spherically symmetric and static massive object  were studied with the use of a perturbative approach in \cite{Bh18}. Black holes generated by (non)linear electrodynamics in the presence of dilatonic fields were studied in \cite{Bh19a,Bh19b}, respectively.  Vacuum static spherically symmetric solutions in the Hybrid Metric-Palatini gravity theory in its scalar tensor representation were considered in \cite{Bh20}, by using a numerical approach to study the behavior of the metric functions and of the scalar field. The thermodynamic properties of the black hole solutions (horizon temperature, specific heat, entropy and evaporation time due to Hawking luminosity) were also investigated in detail. Static and spherically symmetric solutions in a gravity theory that extends the standard Hilbert-Einstein action with a Lagrangian constructed from a three-form field $A_{\alpha \beta \gamma}$, which is related to the field strength and a potential term, were obtained numerically in \cite{Bh21}.

 It is the goal of the present paper to further investigate the physical and astrophysical implications of modified gravitational theories that appear due to the quantum fluctuations of the  metric, as introduced in \cite{re11,re8,re9,re10}, and further developed in \cite{re14}. We assume, as a starting point of our investigations, that the quantized gravitational field can be described, within a semiclassical approximation,  by a general quantum  metric that can be decomposed as the sum of the classical and of a stochastic fluctuating part, of quantum origin.   The Einstein quantum gravity corresponding to this decomposition of the metric leads, at the classical level, to an effective theory similar to the  modified gravity models with a nonminimal coupling between matter and geometry, which have been previously investigated in \cite{fL1, fT1,Har4}. To obtain explicit predictions from the model we assume that the expectation value of the quantum correction can be represented as a second order tensor $K_{\mu \nu}$, which can be obtained from the metric and from the thermodynamic quantities characterizing the matter content of the Universe. Then, as first step in our study,  from the first order quantum gravitational action we derive the classical gravitational field equations in their general form. These equations represent an effective formulation of the quantum gravity model with stochastically fluctuating metric.

  In our study we will consider a specific choice for the fluctuation tensor $K_{\mu \nu}$, which is obtained by the coupling of a scalar field with the classical metric. This choice gives a particular version of the $f(R,T)$ gravity model \cite{fT1}. After obtaining, from a variational principle, the gravitational field equations, we will consider spherically symmetric static black hole solutions of the modified field equations. Due to their mathematical complexity the field equations can be investigated only numerically.  We rewrite the field equations in a dimensionless form, and after introducing a suitable independent radial coordinate, we investigate their solutions numerically. We consider two functional forms of the scalar field potential, assuming first that it is negligibly small, while in the second case it is of the Higgs type. The formation of a black hole is indicated by the presence of a singularity in the metric tensor components.  The thermodynamic properties of the black hole solutions (horizon temperature, specific heat, entropy and evaporation time due to Hawking luminosity) are also investigated.

  Our work is organized as follows.  The field equations induced by the quantum fluctuations of the metric are derived, in their general form, in Section~\ref{sect2}. The field equations resulting from the coupling of the metric fluctuation to the metric via a scalar field are also obtained. The static spherically symmetric gravitational equations in vacuum are presented in Section~\ref{sect3}. Numerical black hole solutions are obtained, for two choices of the scalar field potential, in Section~\ref{sect4}. The thermodynamic properties of the solutions are investigated in Section~\ref{sect5}. We discuss and conclude our results in Section~\ref{sect6}. In the present paper we use a system of units with $c=1$.

\section{Gravitational field equations in the presence of quantum metric fluctuations}\label{sect2}

In this Section we present first the general semiclassical effective gravitational field equations obtained in the presence of a fluctuating geometry. The field equations corresponding to a fluctuation tensor $K_{\mu \nu}$ given by the coupling between the classical metric and a scalar field are also obtained.

\subsection{Gravitational field equations from a fluctuating metric}

The quantum mechanical description of the natural phenomena is very successful at the level of molecules, atoms, elementary particles and classical fields, excluding gravitation. In the quantum approach all physical quantities must be represented by operators. In a quantum theory of gravity, it it exists, we should assume, as a starting point, that we can quantize the geometrical quantities associated to the gravitational field by associating to them some appropriately chosen field operators. Therefore, in a rigorous quantum theory of gravity,  the Einstein gravitational field equations must become some operator equations, represented generally as \cite{re8,re9,re10}
\begin{eqnarray}
\hat{R}_{\mu\nu}-\frac{1}{2}R\hat{g}_{\mu\nu}=\frac{8\pi G}{c^4}\hat{T}_{\mu\nu}.
\end{eqnarray}

This formulation of the quantized geometric gravity  corresponds to a full non-perturbative quantum gravity theory.  From the Einstein operator equations, after averaging them over all possible products of the metric operators $\hat {g}\left(x_1\right)...\hat{g}\left(x_n\right)$ we may be able to obtain some useful physical information \cite{re8,re9,re10}. If we introduce the Green functions $\hat{G}_{\mu\nu}$ of the quantized gravitational field, an exact quantum approach requires to solve the infinite set of operator equations,
\begin{eqnarray*}
\left<Q|\hat{g}(x_1)\hat{G}_{\mu\nu}|Q\right>&=&\left<Q|\hat{g}(x_1) \hat{T}_{\mu\nu}|Q\right>,\\
\left<Q|\hat{g}(x_1)\hat{g}(x_2) \hat{G}_{\mu\nu}|Q\right>&=&\left<Q|\hat{g}(x_1)\hat{g}(x_2) \hat{T}_{\mu\nu}|Q\right>,\\
\dots&=&\dots.
\end{eqnarray*}
In these equations $\left|Q\right>$ denotes the quantum state of the gravitational field. Moreover, $\left|Q\right>$ may not correspond to the ordinary vacuum state of standard quantum field theory. Up to now no exact solutions of the operator equations  have been found, and actually it seems that they cannot be solved analytically. Therefore, in order to investigate the physical implications of quantized gravitational field models it is necessary to resort to approximate or numerical methods  \cite{re8}-\cite{re10}.

An interesting approach for the study of quantum gravitational phenomena was suggested in \cite{re8}. Its basic idea consists in the decomposition of the metric tensor  operator $\hat{g}_{\mu\nu}$ into the sum of two terms. The first is the average of the classical metric tensor $g_{\mu \nu}$, while the second is given by a fluctuating tensorial part $\delta\hat{g}_{\mu\nu}$.  Hence we can write
\begin{eqnarray}\label{c1}
\hat{g}_{\mu\nu}=g_{\mu\nu}+\delta\hat{g}_{\mu\nu}.
\end{eqnarray}

 In order to obtain a theory that can be handled either analytically or numerically we introduce another approximation. More exactly, we assume that the average of the fluctuating  part of the metric, which describes the quantum effects, is described with the help of a classical tensorial quantity  $K_{\mu\nu}$, so that the average of the quantum fluctuations of the metric become
\be\label{c2}
\left<\delta \hat{g}_{\mu\nu}\right>=K_{\mu\nu}\ne 0.
\ee

Therefore, after ignoring higher order fluctuations of the metric field, we can obtain the effective semiclassical Lagrangian of the gravitational field  that includes the effects of the quantum fluctuations in the form \cite{re8}
\begin{align}\label{2}
L&=\sqrt{-\hat{g}}\mathcal{L}_g\left(\hat{g}_{\mu\nu}\right)+
\sqrt{-\hat{g}}\mathcal{L}_m\left(\hat{g}_{\mu\nu}\right)\nonumber\\
&\approx \sqrt{-g}(\mathcal{L}_g+\mathcal{L}_m)+\left[\frac{\delta(\sqrt{-g}\mathcal{L}_g)}{\delta g^{\mu\nu}}+
\frac{\delta(\sqrt{-g}\mathcal{L}_m)}{\delta g^{\mu\nu}}\right]\delta\hat{g}^{\mu\nu}\nonumber\\&=\frac{1}{2\kappa ^2}\sqrt{-g} \left(R+G_{\mu\nu}\delta\hat{g}^{\mu\nu}\right)+
\sqrt{-g}\Bigg(\mathcal{L}_m-\frac{1}{2}T_{\mu\nu}\delta\hat{g}^{\mu\nu}\Bigg),
\end{align}
where we have denoted \(\kappa ^2=8\pi G/c^4\), and $\mathcal{L}_g\left(\hat{g}_{\mu\nu}\right)$ is the quantized Lagrangian  of the gravitational field. Moreover,  by $\mathcal{L}_m\left(\hat{g}_{\mu\nu}\right)$ we have denoted the matter Lagrangian, while $T_{\mu\nu}$ is the matter energy-momentum tensor, defined as
\begin{eqnarray}
T_{\mu\nu}=-\frac{2}{\sqrt{-g}}\frac{\delta \left(\sqrt{-g}\mathcal{L}_m\right)}{\delta g^{\mu\nu}}.
\end{eqnarray}

 Hence, to summarize our approach to quantum gravity, we go from the full quantum operator version of the Einstein field equations to a classical formulation by performing two steps. The first step is the decomposition of the metric in two components, the classical part and the fluctuating part, while the second step is the replacement of the fluctuating part by its average value. Hence in this way we have arrived at an effective  semiclassical theory of the gravitational field, entirely expressed in terms of classical geometrical and physical quantities. On the other hand it is important to point out that in this formalism we the exact form of the effective quantum perturbation tensor \(K^{\mu\nu}\)  cannot be fixed from first principles. Hence the expression of $K_{\mu \nu}$ remains arbitrary, and it must be obtained from physical or geometrical considerations.

After performing the variation with respect to the metric the gravitational field equations corresponding to the first order corrected quantum Lagrangian (\ref{2}) are obtained as
\begin{eqnarray}\label{genfe}
G_{\mu\nu}&=&\kappa^2\bigg(T_{\mu\nu}+\gamma^{\alpha\beta}_{\mu\nu}T_{\alpha\beta}-\frac{1}{2}g_{\mu\nu}T_{\alpha\beta}K^{\alpha\beta}\nonumber\\
&&+2\f{\delta^2\cL_m}{\delta g^{\alpha\beta}\delta g^{\mu\nu}}K^{\alpha\beta}-\f{1}{2}\cL_mKg_{\mu\nu}+\f{1}{2}KT_{\mu\nu}\nonumber\\
&&+\cL_mK_{\mu\nu}\bigg)-
\frac{1}{2}\bigg(2\gamma^{\alpha\beta}_{\mu\nu}G_{\alpha\beta}-g_{\mu\nu}G_{\alpha\beta}K^{\alpha\beta}\nonumber\\
&&+g_{\mu\nu}\nabla_\alpha\nabla_\beta K^{\alpha\beta}-KR_{\mu\nu}+RK_{\mu\nu}\bigg)\nonumber\\
&&+\frac{1}{2}\bigg[\nabla_\alpha\nabla_{(\nu}K^\alpha_{\mu)}+\Box Kg_{\mu\nu}-\Box K_{\mu\nu}-\nabla_\nu\nabla_\mu K\bigg],\nonumber\\
\end{eqnarray}
where we have introduced the notations \(K=g_{\mu\nu}K^{\mu\nu}\)and \(A_{\alpha\beta}\delta K^{\alpha\beta}=\delta g^{\mu\nu}(\gamma^{\alpha\beta}_{\mu\nu}A_{\alpha\beta})\), respectively. Moreover, \(A_{\alpha\beta}\) is given by either $R_{\alpha \beta}$, or $T_{\alpha \beta }$. Depending on the adopted model, the quantity  \(\gamma^{\alpha\beta}_{\mu\nu}\)  could be an algebraic tensor, an operator,  or their combination.

\subsection{Scalar field-metric coupling}

  In order to investigate the astrophysical implications of the modified gravity models induced by the quantum metric fluctuations on the black hole structure, in the present paper we will consider the case in which the expectation value of the quantum fluctuation tensor $K^{\mu\nu}$ can be expressed in the form \cite{re14},
\be\label{K1}
K^{\mu\nu}=\alpha\phi (x)g^{\mu\nu},
\ee
where $\phi (x)$ is an arbitrary function, and $\alpha$ is a constant. More exactly, we assume that $K_{\mu \nu}$ is proportional to the classical metric tensor. The particular case $\phi ={\rm constant}$ was investigated in \cite{re8} and \cite{re11}, respectively. Therefore in our approach, due to the quantum fluctuations of the metric tensor, an extra component, proportional to the classical one, does appear in the total metric.

 As for the function $\phi (x)$, in the following we assume that it is a physical scalar field that couples to the metric, in the presence of a  self-interaction potential $V(\phi)$. Hence, to the general quantum  perturbed gravitational Lagrangian (\ref{2}) we must add a supplementary Lagrangian
\bea
\mathcal{L}_{\phi}=-\frac{1}{2}\partial _{\mu }\phi \partial ^{\mu}\phi -V(\phi),
\eea
 representing the source term of the scalar field. Hence, in the presence of quantum fluctuations that can be described by the coupling between  a scalar field and the classical metric,  for the effective Lagrangian of the semiclassical gravitational field we obtain the expression
\begin{align}\label{Lalpha}
 \mathcal{L}_{tot}=\frac{1}{2\kappa^2}(1-\alpha\phi)R+\mathcal{L}_m-\frac{\alpha}{2}\phi T-\frac{1}{2}\partial _{\mu }\phi \partial ^{\mu}\phi -V(\phi).
\end{align}

By varying the action  (\ref{Lalpha}) with respect to the metric tensor $g_{\mu \nu}$, and taking into account that
\bea
g^{\alpha\beta}\frac{\delta T_{\alpha\beta}}{\delta g^{\mu\nu}}
= g_{\mu\nu}\mathcal{L}_m-2T_{\mu\nu},
\eea
we obtain the semiclassical effective quantum Einstein gravitational field equations in the form
\begin{align}\label{fealpha}
G_{\mu\nu}&=\frac{\kappa ^2}{1-\alpha\phi}\Bigg[(1+3\alpha\phi)T_{\mu\nu}
-\frac{\alpha}{2}\phi (T+2\mathcal{L}_m) g_{\mu\nu}\nonumber\\&+
\nabla_\mu \phi \nabla_\nu \phi
-\frac{1}{2}g_{\mu\nu}(\nabla _{\beta }\phi \nabla ^{\beta}\phi +2V)\nonumber\\&+\frac{\alpha}{\kappa^2}(\Box\phi g_{\mu\nu}-\nabla_\mu\nabla_\nu\phi)\Bigg].
\end{align}

 By varying the Lagrangian  (\ref{Lalpha})  with respect to the scalar field $\phi $ we obtain the generalized Klein-Gordon equation of the model,  given by
\bea\label{KGa}
\Box\phi-\frac{\alpha}{2\kappa^2}R-\frac{\alpha}{2}T-\frac{\textmd{d} V}{\textmd{d} \phi}=0.
\eea

By contracting the Einstein field equations (\ref{fealpha}) we obtain
\begin{align}\label{Ralpha}
R=&\frac{\kappa ^2}{\alpha\phi-1}\bigg[(1+\alpha\phi)T-4\alpha\phi\mathcal{L}_m\nonumber\\&-\nabla _{\mu }\phi \nabla ^{\mu}\phi -4V+\frac{3\alpha}{\kappa^2}\Box\phi\bigg].
\end{align}
Therefore we can reformulate the gravitational field equations as
\begin{align}
R_{\mu\nu}&=\frac{\kappa^2}{1-\alpha\phi}\Bigg[(1+3\alpha\phi)T_{\mu\nu}-\frac{1}{2}(1+2\alpha\phi)Tg_{\mu\nu}\nonumber\\
&+\phi\mathcal{L}_m g_{\mu\nu}+\nabla_\mu\phi\nabla_\nu\phi+V(\phi)g_{\mu\nu}\nonumber\\&-\frac{\alpha}{2\kappa^2}(2\nabla_\mu\nabla_\nu\phi+\Box\phi g_{\mu\nu})\Bigg].
\end{align}

\subsubsection{The vacuum case}

In the following we will consider vacuum solutions of the semiclassical Einstein equations in the presence of quantum fluctuations, described by the coupling between a scalar field and the classical metric tensor. Therefore, in the absence of matter, the gravitational field equations (\ref{fealpha}) reduce to
\begin{align}\label{fev}
G_{\mu\nu}&=\frac{\kappa ^2}{1-\alpha\phi}\Bigg[\nabla_\mu \phi \nabla_\nu \phi
-\frac{1}{2}g_{\mu\nu}(\nabla _{\beta }\phi \nabla ^{\beta}\phi +2V)\nonumber\\&+\frac{\alpha}{\kappa^2}(\Box\phi g_{\mu\nu}-\nabla_\mu\nabla_\nu\phi)\Bigg].
\end{align}
The contraction of the above equations gives
\begin{align}\label{Rvac}
R=\frac{\kappa ^2}{1-\alpha\phi}\Bigg(\nabla _{\mu }\phi \nabla ^{\mu}\phi +4V-\frac{3\alpha}{\kappa^2}\Box\phi\Bigg).
\end{align}

In the vacuum the generalized Klein-Gordon equation takes the form
\be
\Box\phi - \frac{\alpha}{2\kappa^2} R - \frac{d V}{d \phi} = 0.
\ee

By eliminating with the use of Eq.~(\ref{Rvac}), the Ricci scalar from the generalized Klein-Gordon equation, we obtain
\begin{align}\label{KGv}
\Bigg[1&+\frac{3\alpha^2}{2\kappa^2(1-\alpha\phi)}\Bigg]\Box\phi\nonumber\\&=\frac{\alpha}{2\left(1-\alpha\phi\right)} \Big(\nabla^\beta \phi \nabla_\beta\phi + 4V\Big)+\frac{d V}{d \phi}.
\end{align}

 \section{The spherically symmetric static field equations in modified gravity induced by the quantum metric fluctuations}\label{sect3}

In the present Section we  present the static and spherically symmetric field equations obtained from  Eqs.~(\ref{fev}) and \eqref{KGv}, respectively, for an arbitrary scalar field potential.

\subsection{The static spherically symmetric gravitational field equations}

 In order to study the black hole type solutions in the modified gravity induced by the quantum metric fluctuations, we introduce  spherical coordinates defined as $(t,r,\theta,\varphi)$.  For the space--time metric we adopt the standard form
\be
ds^2 = -e^\nu  dt^2 +e^\lambda dr^2 + r^2 (d\theta^2 + \sin^2 \theta d\varphi^2).
\ee
Due to the static character of the metric $\nu$ and $\lambda$ are functions of the radial coordinate $r$ only.

The vacuum Einstein field equations and the generalized Klein-Gordon equation in static spherically symmetry are obtained as
\begin{align}\label{f1}
-e^{-\lambda}&\left(\frac{1}{r^2}-\frac{\lambda '}{r}\right)+\frac{1}{r^2}=\frac{1}{1-\alpha\phi}\Bigg[\kappa ^2\left(\frac{e^{-\lambda}}{2}\phi '^2+V(\phi)\right)\nonumber\\&-\frac{1}{2}\alpha e^{-\lambda}\left(2\phi^{\prime\prime}-\lambda^\prime\phi^\prime+\frac{4\phi^\prime}{r}\right)\Bigg],
\end{align}

\begin{align}\label{f2}
e^{-\lambda}&\left(\frac{\nu '}{r}+\frac{1}{r^2}\right)-\frac{1}{r^2}=\frac{1}{1-\alpha\phi}\Bigg[\kappa ^2\left(\frac{e^{-\lambda}}{2}\phi '^2-V(\phi)\right)\nonumber\\&+\frac12\alpha e^{-\lambda}\left(\nu^\prime\phi^\prime+\frac{4\phi^\prime}{r}\right)\Bigg],
\end{align}
\begin{align}\label{f3}
-\frac12e^{-\lambda}&\left[\nu ''+\left(\nu '-\lambda '\right)\left(\frac{\nu^\prime }{2}+\frac{1}{r}\right)\right]\nonumber\\
&=\frac{1}{1-\alpha\phi}\Bigg\{\kappa ^2\left(\frac{e^{-\lambda}}{2}\phi '^2+V(\phi)\right)\nonumber\\&-\alpha e^{-\lambda}\left[\phi^{\prime\prime}+\left(\frac{\nu^\prime}{2}-\frac{\lambda^\prime}{2}+\frac1r\right)\phi^\prime\right]\Bigg\},
\end{align}
\begin{align}\label{f4}
e^{-\lambda}&\Bigg[1+\frac{3\alpha^2}{2\kappa^2(1-\alpha\phi)}\Bigg]\Bigg[\phi^{\prime\prime}+\Bigg(\frac{\nu^\prime}{2}-\frac{\lambda^\prime}{2}+\frac2r\Bigg)\phi^\prime\Bigg]\nonumber\\&=\frac{\alpha}{2\left(1-\alpha\phi\right)} \Big(e^{-\lambda}\phi^{\prime2}+ 4V\Big)+\frac{d V}{d \phi}.
\end{align}
In order to simplify the mathematical formalism we represent the metric tensor component $e^{-\lambda}$ as
\be
e^{-\lambda}=1-\frac{2Gm(r)}{r},
\ee
where we have introduced the effective mass $m(r)$ of the field.

\subsection{Dimensionless form of the field equations}

In the following  we also define the dimensionless scalar field $\psi$, given by
\be
\psi=1-\alpha\phi.
\ee

In order to solve the field equations numerically, it is useful to introduce a set of dimensionless variables $(\eta, \zeta, \alpha, M, U)$ defined as follows,
\begin{align}\label{d1}
&r=\frac{2GM_0}{c^2}\frac{1}{\eta}, \qquad \psi'=\frac{4\pi}{M_0c^2\kappa^2}\zeta,\qquad \alpha=\kappa\beta,\nonumber\\
&m(r)=M_0 M(\eta),\qquad V=\left(\frac{4\pi}{M_0c^2\kappa^3}\right)^2U.
\end{align}

Hence the static spherically symmetric gravitational field equations in the presence of the coupling between a scalar field and the metric can be written as
\begin{widetext}
\begin{eqnarray}\label{deq1}
\frac{d\psi}{d\eta}&=&-\frac{1}{\eta^2}\zeta,
\end{eqnarray}
\begin{eqnarray}\label{deq2}
\frac{d\zeta}{d\eta}&=&\frac{\zeta}{\eta}+\frac{\zeta}{\eta(1-M\eta)}+\frac{(2\psi\eta+\zeta)\zeta^2}{2\eta^3\psi(2\psi+3\beta^2)}-\frac{\zeta^2}{\eta^2\psi}-\nn
&&\frac{\zeta }{\eta^3(1-M\eta)\psi}U+\frac{2\beta^2(2\psi\eta+\zeta)}{\eta^3(1-M\eta)\psi(2\psi+3\beta^2)}U+\frac{\beta(2\psi\eta+\zeta)}{\kappa\eta^3(1-M\eta)(2\psi+3\beta^2)}\frac{dU}{d\phi},
\end{eqnarray}
\begin{eqnarray}\label{deq3}
\frac{d\nu}{d\eta}&=&\frac{1}{\eta}-\frac{1}{\eta(1-M\eta)}+\frac{\zeta}{\eta(1-M\eta)(2\psi\eta+\zeta)}+\frac{3\zeta}{\eta(2\psi\eta+\zeta)}-\nn
&&\frac{\zeta^2}{\beta^2\eta^2(2\psi\eta+\zeta)}+\frac{2U}{\eta^2(1-M\eta)(2\psi\eta+\zeta)},
\end{eqnarray}
\begin{eqnarray}\label{deq4}
\frac{dM}{d\eta}&=&-\frac{\zeta(1-M\eta)}{\eta^2(2\psi\eta+\zeta)}+\frac{\zeta}{\eta^2(2\psi\eta+\zeta)}+\Bigg[\frac{1-M\eta}{\eta^4\psi(2\psi+3\beta^2)}-\frac{2(1-M\eta)}{\eta^3\psi(2\psi\eta+\zeta)}-\frac{1-M\eta}{\beta^2\eta^3(2\psi\eta+\zeta)}\Bigg]\zeta^2+\nn
&&\Bigg[\frac{4\beta^2}{\eta^4\psi(2\psi+3\beta^2)}-\frac{2}{\eta^3(2\psi\eta+\zeta)}-\frac{2\zeta}{\eta^4\psi(2\psi\eta+\zeta)}\Bigg]U+\frac{2\beta}{\kappa\eta^4(2\psi+3\beta^2)}\frac{dU}{d\phi},
\end{eqnarray}
\end{widetext}

In these variables the the limit $r\rightarrow \infty$ is obtained for $\eta \rightarrow 0$, while the point $r=0$ corresponds to $\eta \rightarrow \infty$. The system of differential equations (\ref{deq1})-(\ref{deq4}) must be solved with the initial conditions $M(0)=1$, $\lambda (0)=0$, $\nu (0)=0$, $\psi (0)=\psi _0$, and $\zeta (0)=\zeta _0$, respectively, where we have assumed that the geometry around the black hole is asymptotically flat, taking the Minkowski form at infinity.

\section{Numerical black hole solutions in modified gravity induced by the quantum metric fluctuations}\label{sect4}

  Eqs.~(\ref{deq1})-(\ref{deq4}) represent a coupled strongly nonlinear system of ordinary differential equations. Generally they cannot be solved analytically. In order to obtain their solutions we must resort to numerical methods. In the following we will present the numerical solutions of the vacuum static spherically symmetric equations in the modified gravity induced by the quantum metric fluctuations, corresponding to two choices of the scalar field potential, corresponding to a vanishing scalar field potential, and to a Higgs type potential, respectively. The existence of a black hole is indicated by the presence of a singular point $\eta =\eta _s$ in the equations (the event horizon), for which $e^{-\lambda \left(\eta _s\right)}=0$, and $e^{\nu \left(\eta _s\right)}=0$, respectively.

\subsection{The case $V(\phi)$=0}

In the case of a vanishing scalar field potential the system of differential equations describing the geometric and physical properties in the vacuum outside of  massive objects in the presence of quantum fluctuations is given by
	\begin{eqnarray}\label{vzero1}
	\frac{d\psi}{d\eta}&=&-\frac{1}{\eta^2}\zeta,
	\end{eqnarray}
	\begin{eqnarray} \hspace{-0.6cm}\frac{d\zeta}{d\eta}&=&\frac{\zeta}{\eta}+\frac{\zeta}{\eta(1-M\eta)}+\frac{(2\psi\eta+\zeta)\zeta^2}{2\eta^3\psi(2\psi+3\beta^2)}-\frac{\zeta^2}{\eta^2\psi},
	\end{eqnarray}
	\begin{eqnarray} \frac{d\nu}{d\eta}&=&\frac{1}{\eta}-\frac{1}{\eta(1-M\eta)}+\frac{\zeta}{\eta(1-M\eta)(2\psi\eta+\zeta)}+\nonumber\\
&&\frac{3\zeta}{\eta(2\psi\eta+\zeta)}-\frac{\zeta^2}{\beta^2\eta^2(2\psi\eta+\zeta)},
	\end{eqnarray}
	\begin{eqnarray}\label{vzero4}
\hspace{-0.4cm}&&\frac{dM}{d\eta}=-\frac{\zeta(1-M\eta)}{\eta^2(2\psi\eta+\zeta)}+\frac{\zeta}{\eta^2(2\psi\eta+\zeta)}+\nonumber\\
\hspace{-0.4cm}&&\Bigg[\frac{1-M\eta}{\eta^4\psi(2\psi+3\beta^2)}-\frac{2(1-M\eta)}{\eta^3\psi(2\psi\eta+\zeta)}-\frac{1-M\eta}{\beta^2\eta^3(2\psi\eta+\zeta)}\Bigg]\zeta^2.\nonumber\\
	\end{eqnarray}

The solutions of the nonlinear system of equations (\ref{vzero1})-(\ref{vzero4}) are obtained numerically, for different initial conditions $\zeta _0$ and $\psi _0$, and for different values of the coupling parameter $\beta$, describing the interaction between the scalar field and the effective fluctuating component of the quantum metric. The formation of a black hole with radius of the event horizon $\eta _s$ is detected from the conditions $e^{\nu \left(\eta _s\right)}=0$, and  $e^{-\lambda\left(\eta_s\right)}=1-M\left(\eta_s\right)\eta_s$, respectively.

\begin{figure*}[htbp]
	\centering
	\includegraphics[width=8.9cm]{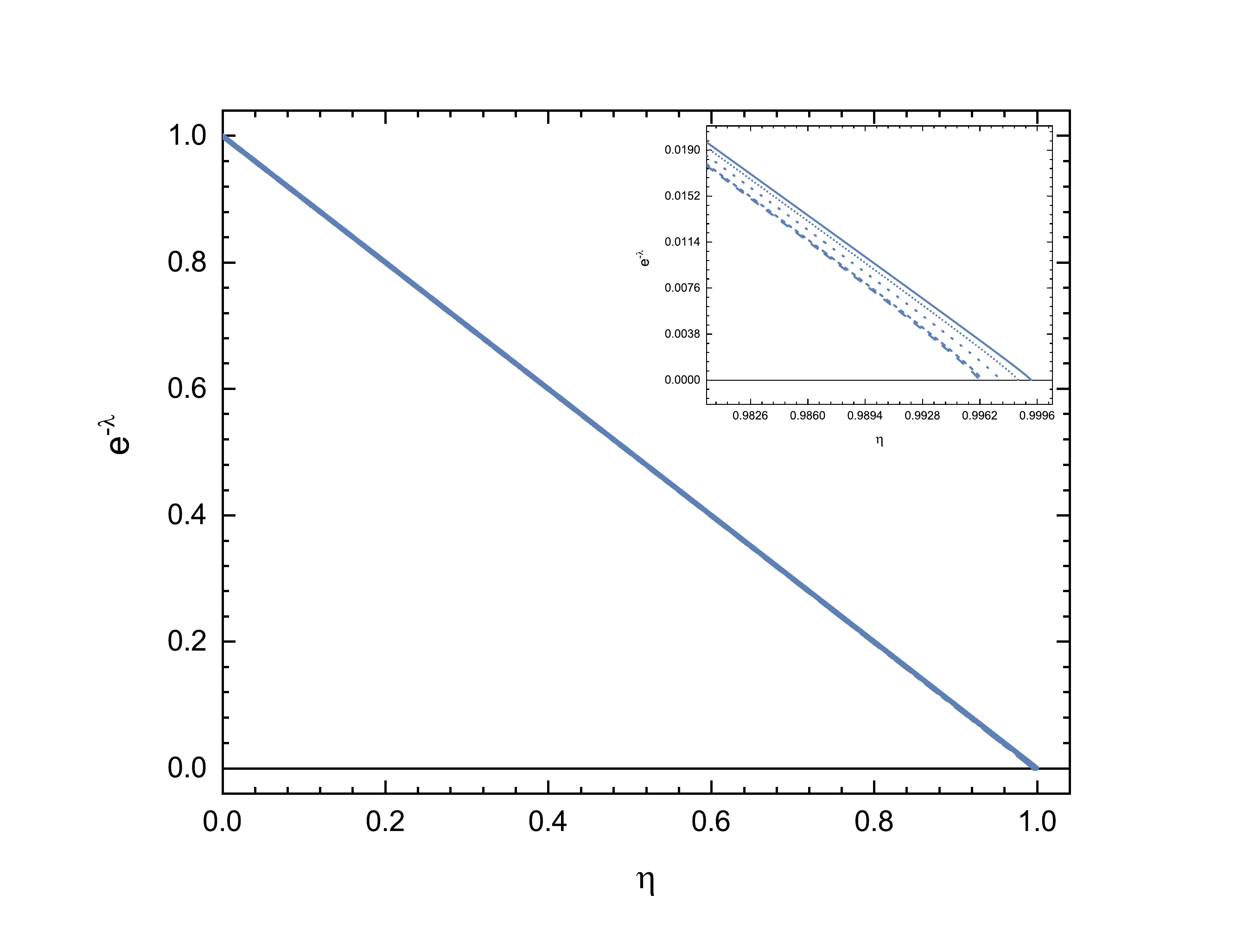}
\includegraphics[width=8.9cm]{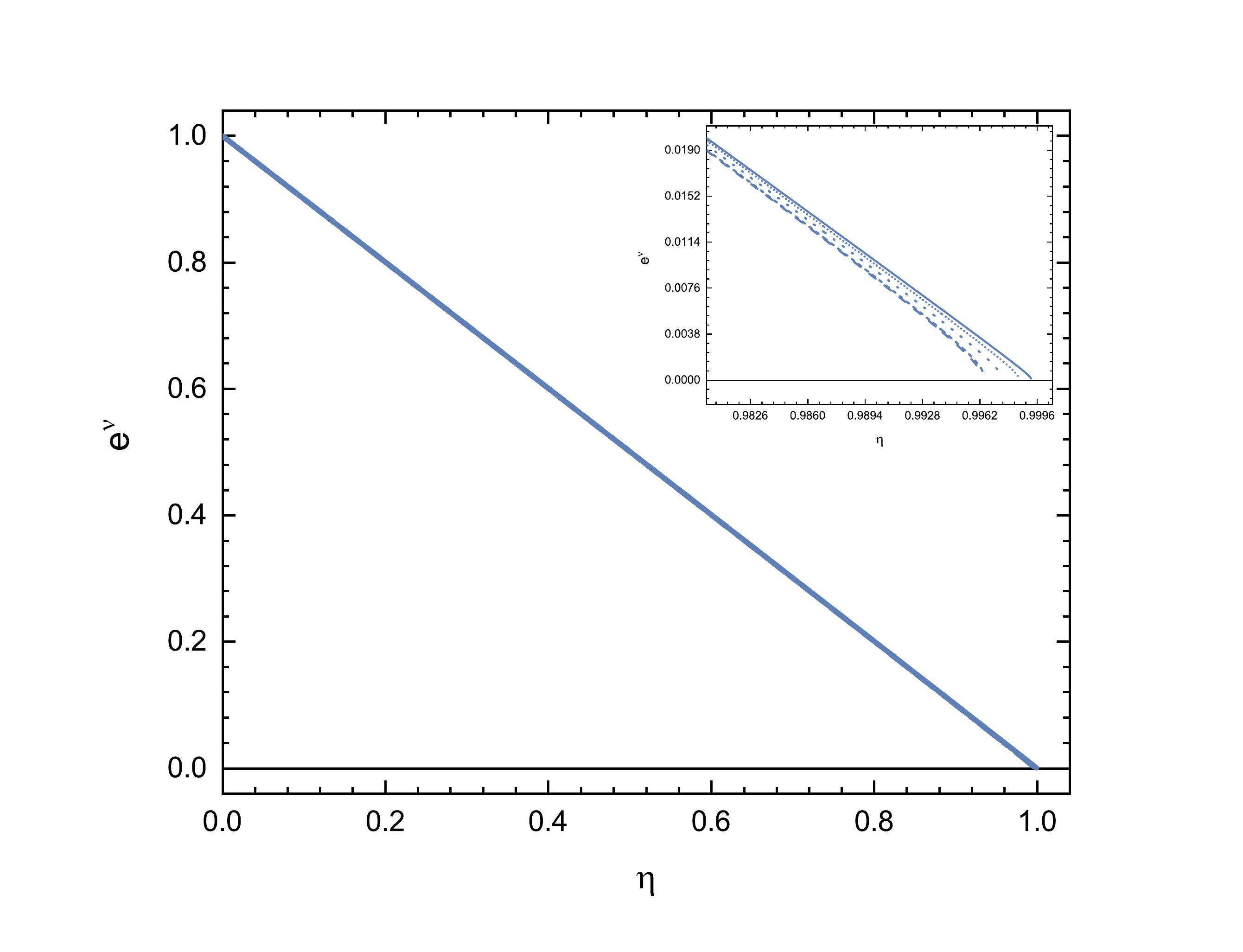}
			\caption{Variations of  the metric tensor component $e^{-\lambda}$ (left figure) and $e^{\nu}$ (right figure) for the case of a zero potential coupling between the scalar field and the effective quantum fluctuating metric for $\psi_0=0.5$, and $\zeta_0=1\times 10^{-15}$ respectively, and for different values of $\beta $: $\beta=0.0265$ (solid curve), $\beta=0.03$ (dotted curve), $\beta=0.04$ (short dashed curve), $\beta=0.1$ (dashed curve), and $\beta=100$ (long dashed curve). The additional Figures shows the variation for $\eta$ close to $1$.}\label{fb3}
\end{figure*}

The variations of the metric tensor components are presented in Fig.~\ref{fb3}. As one can see from the Figures, both metric tensor components reach simultaneously the zero value for a finite $\eta$, indicating the formation of a black hole, with the event horizon located at the singular point $\eta _s$. At large distances from the black hole center (small values of $\eta$, the variations of the metric tensor components are independent on the numerical values of $\beta$. However, near the singular point $\eta _s=1$ such a dependence does appear, leading to the modification of the position of the event horizon as compared to the Schwarzschild case.

\begin{figure}[htbp]
		\centering
		\includegraphics[width=8.9cm]{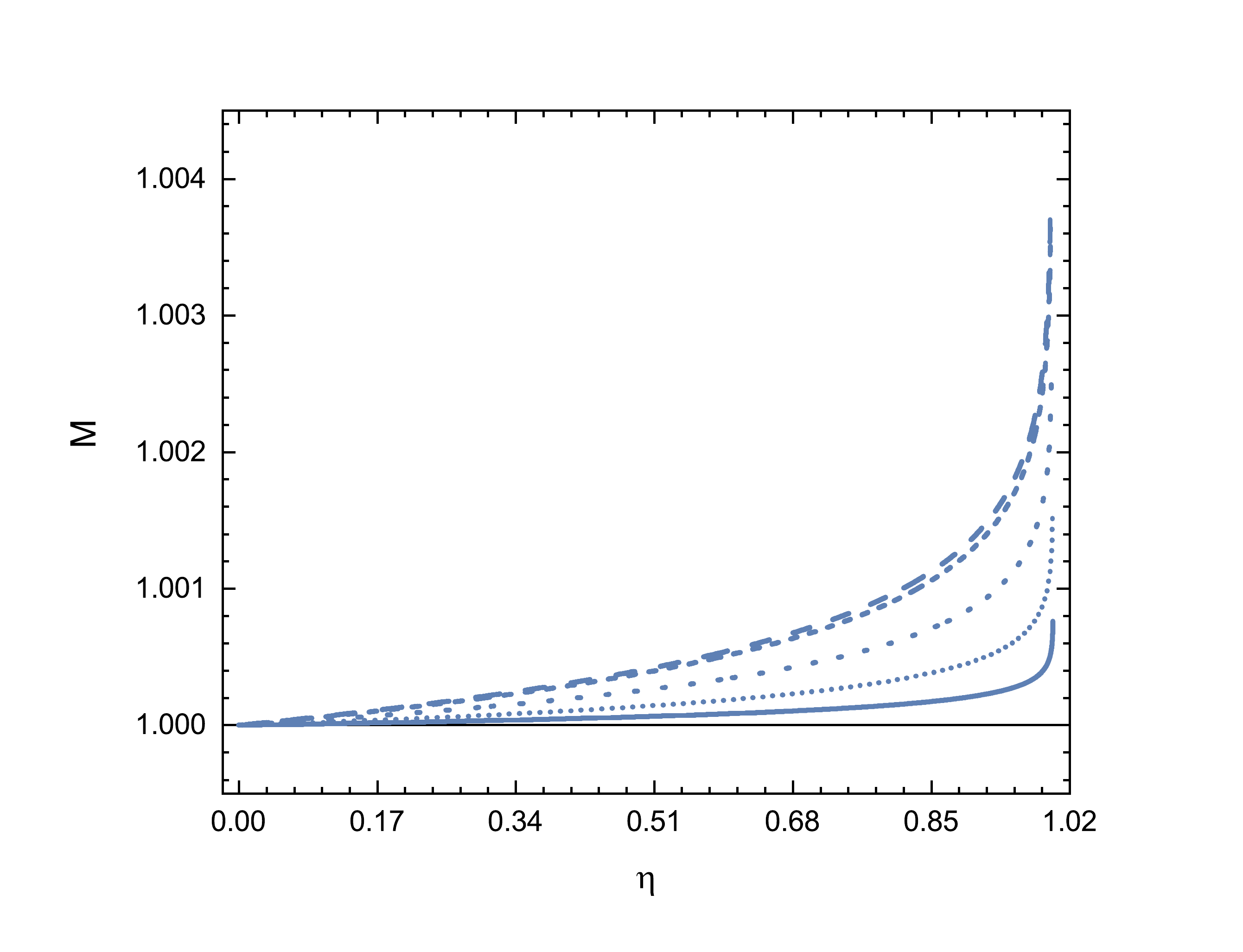}	
			\caption{Variation of effective mass function $M$ for the case of a zero potential coupling between the scalar field and the effective quantum fluctuating metric for $\psi_0=0.5$ and $\zeta_0=1\times 10^{-15}$, respectively, and for different values of $\beta$: $\beta=0.0265$ (solid curve), $\beta=0.03$ (dotted curve), $\beta=0.04$ (short dashed curve), $\beta=0.1$ (dashed curve), and $\beta=100$ (long dashed curve), respectively.}
			\label{fb2}
	\end{figure}

The variation of the effective mass $M$ of the black hole is represented in Fig.~\ref{fb2}. As compared to the mass at infinity, the mass of the black hole increases once we approach the event horizon. This increase is due to the presence of the quantum fluctuations, generating an effective contribution to the energy-momentum tensor of the gravitational field. The variation of the effective mass depends significantly on the coupling constant between the scalar field and the metric, and its numerical value increases rapidly when approaching the event horizon.

\begin{figure}[htbp]
	\centering
		\includegraphics[width=8.2cm]{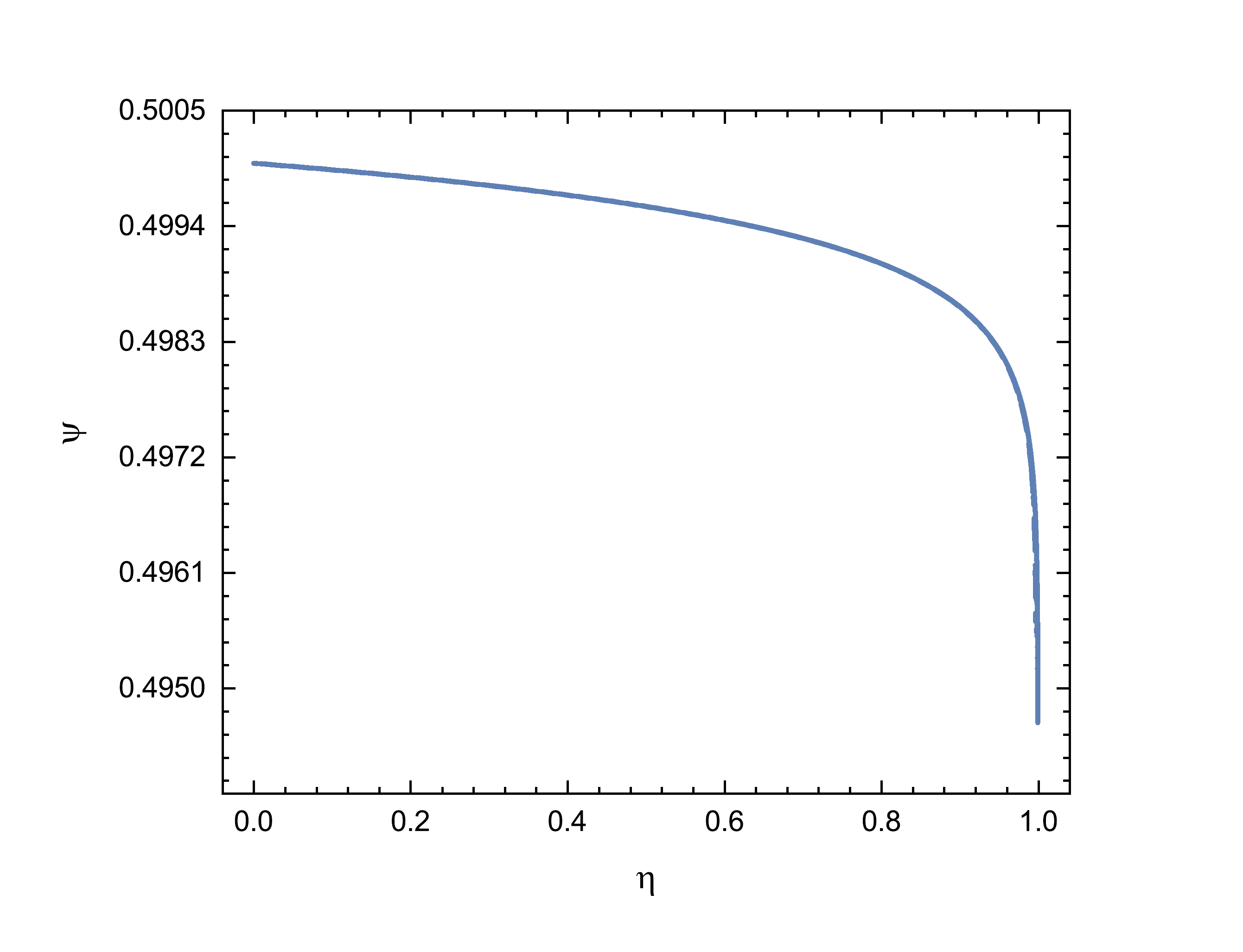}
			\caption{Variation of the scalar field $\psi$ for the case of a zero potential coupling between the scalar field and the effective quantum fluctuating metric for $\psi_0=0.5$, and $\zeta_0=1\times 10^{-15}$, respectively, and for different values of $\beta$: $\beta=0.0265$ (solid curve), $\beta=0.03$ (dotted curve), $\beta=0.04$ (short dashed curve), $\beta=0.1$ (dashed curve), $\beta=100$ (long dashed curve).}
		\label{fb1}
\end{figure}

The variation with respect to the inverse of radial coordinate of the modified scalar field $\psi$ is represented in Fig.~\ref{fb1}. The redefined scalar field $\psi$  takes lower values when approaching the event horizon, and its variation is basically independent on the initial conditions and the numerical values of $\beta$.

\begin{figure}[htbp]
	\centering
	\includegraphics[width=250pt]{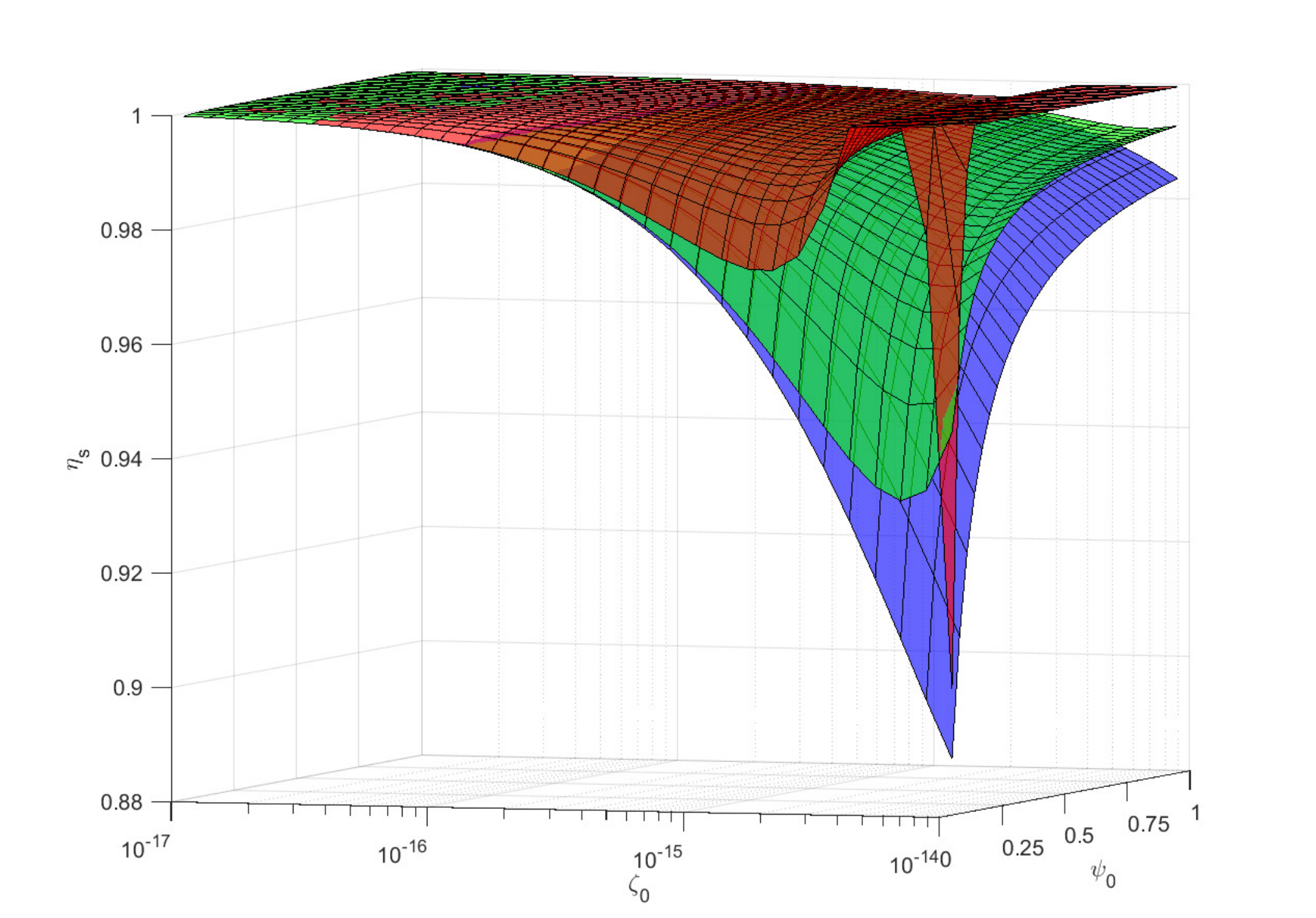}
	\caption{Variation of the horizon of a black hole in the presence of quantum metric fluctuations induced by the coupling between a zero potential scalar field and the metric tensor as a function of the initial conditions at infinity of the field  $\left(\psi_0,\zeta_0\right)$ and for some specific values of $\beta$. The surface in red is for $\beta=0.05$, the surface in green is for $\beta=0.1$,  and the surface in blue is for $\beta=1$, respectively.}
	\label{fb5}
\end{figure}

	The variation of the position of the event horizon of a static spherically symmetric black hole as a result of the variations of the initial conditions at infinity of the scalar field are presented, for a selected number of values of $\beta$, in Fig.~\ref{fb5}. As one can see from the Figure, the simultaneous variations of the initial conditions of the field, as well as of the coupling parameter $\beta$ can lead to an important modification in the location of the event horizon of the black hole.

A number of explicit numerical values of $\eta_s$ are presented in Table~\ref{tt1} for $\beta =0.1$, and for different values of the initial conditions at infinity of the scalar field, and its derivative.

\begin{table*}[htbp]
	\centering
	 \begin{tabular}{|>{\centering}p{70pt}|>{\centering}p{70pt}|>{\centering}p{70pt}|>{\centering}p{70pt}|>{\centering}p{70pt}|>{\centering\arraybackslash}p{70pt}|}
		\hline
		$\psi_0$ & $\zeta_0$ & $\eta_{s}$ & $\psi_0$ & $\zeta_0$ & $\eta_{s}$\\
		\hline
		$0.05$ & $1\times 10^{-14}$ & $0.94685$ & $0.05$ & $1\times 10^{-15}$ & $0.97444$\\
\hline
		$0.1$ & $1\times 10^{-14}$ &  $0.96303$ & $0.05$ & $3\times 10^{-15}$ & $0.94881$\\
\hline
		$0.2$ & $1\times 10^{-14}$ & $0.97715$ & $0.05$ & $5\times 10^{-15}$ & $0.93718$\\
\hline
		$0.4$ & $1\times 10^{-14}$ & $0.98647$ & $0.05$ & $7\times 10^{-15}$ & $0.93485$\\
\hline
		$0.8$ & $1\times 10^{-14}$ & $0.99214$ & $0.05$ & $9\times 10^{-15}$ & $0.94074$\\
		\hline
		\end{tabular}
	\caption{Location of the event horizon $\eta_s$ of the black holes in the presence of quantum metric fluctuations induced by the coupling between a zero potential scalar field and the metric tensor for different values of $\psi_0$ and $\zeta_0$, and for fixed $\beta=0.1$.}
	\label{tt1}	
\end{table*}

\subsubsection{Interpolating functions}

In order to facilitate the comparison of the predictions of the modified gravity model induced by the quantum metric fluctuations we have also obtained some analytical representations of the basic geometric and physical parameters. The interpolating functions are obtained for the range of the parameters used in the previous Section. The variation of the effective mass  $M(\eta)$ of the black hole can be obtained as
	\begin{eqnarray}
M(\eta)&=&a_1\left(\zeta _0,\psi _0\right)\eta^3+b_1\left(\zeta _0,\psi _0\right)\eta^2+\nn
&&c_1\left(\zeta _0,\psi _0\right)\eta+d_1\left(\zeta _0,\psi _0\right),
	\end{eqnarray}
where the coefficients $(a_1,b_1,c_1,d_1)$ are given by
\begin{eqnarray*}
a_1\left(\zeta _0,\psi _0\right) &=& -3.157\times 10^{-3}+1.988\times 10^{12}\frac{\zeta_0}{\psi_0}-\nn
&&4.484\times 10^{-3}\ln{\psi_0}+ 3.966\times 10^{12}\psi_0\zeta_0,\\
b_1\left(\zeta _0,\psi _0\right) &=& 3.888\times 10^{-3}-1.76\times 10^{12}\frac{\zeta_0}{\psi_0}+\nn
&&5.84\times 10^{-3}\ln{\psi_0}- 5.07\times 10^{12}\psi_0\zeta_0,\\
c_1\left(\zeta _0,\psi _0\right) &=&-1.457\times 10^{-3}+ 8.389\times 10^{11}\frac{\zeta_0}{\psi_0}-\nn
&&2.004\times 10^{-3}\ln{\psi_0}+ 1.993\times 10^{-3} \psi_0 ,\\
d_1\left(\zeta _0,\psi _0\right) &=& 1 -3.021\times 10^{10}\frac{\zeta_0}{\psi_0} +\nn
&&  1.298\times 10^{-4}\ln{\psi_0} -1.114\times 10^{11}\psi_0\zeta_0,
\end{eqnarray*}
with the Pearson correlation coefficient given by $R^2 = 0.9997$.
The metric tensor coefficient $e^{-\lambda}$ can be obtained from the definition
\bea
\hspace{-0.4cm}e^{-\lambda} &=& 1-M\eta = a_2\left(\zeta _0,\psi _0\right)\eta^4+b_2\left(\zeta _0,\psi _0\right)\eta^3+\nonumber\\
\hspace{-0.4cm}&&+c_2\left(\zeta _0,\psi _0\right)\eta^2+d_2\left(\zeta _0,\psi _0\right)\eta+e_2\left(\zeta _0,\psi _0\right),
\eea
where $(a_2,b_2,c_2,d_2,e_2)$ are coefficients depending on $\left(\zeta _0,\psi_0\right)$ so that $a_2=-a_1$, $b_2=-b_1$, $c_2=-c_1$, $d_2=-d_1$ and $e_2=1$, respectively.

We can fit $e^{\nu}$ with the help of the function
	\begin{eqnarray}
e^{\nu(\eta)}&=&a_3\left(\zeta _0,\psi _0\right)\eta^2+b_3\left(\zeta _0,\psi _0\right)\eta+\nn
&&c_3\left(\zeta _0,\psi _0\right),
\end{eqnarray}
where the coefficients $(a_3,b_3,c_3)$ are given by
\begin{eqnarray*}
	a_3\left(\zeta _0,\psi _0\right) &=& 1.279\times 10^{-3}-2.431\times 10^{12}\frac{\zeta_0}{\psi_0}+\nn
	&&1.061\times 10^{-3}\ln{\psi_0}- 1.161\times 10^{12}\psi_0\zeta_0,\\
b_3\left(\zeta _0,\psi _0\right) &=& -1.001+2.415\times 10^{12}\frac{\zeta_0}{\psi_0}-\nn
	&&6.142\times 10^{-4}\ln{\psi_0}+7.529\times 10^{11}\psi_0\zeta_0,\\
	c_3\left(\zeta _0,\psi _0\right) &=&1- 1.011\times 10^{11}\frac{\zeta_0}{\psi_0}+\nn
	&&1.311\times 10^{-4}\ln{\psi_0}-1.236\times 10^{11} \psi_0\zeta_0 ,
\end{eqnarray*}
with the Pearson correlation coefficient equal to $R^2 = 0.9999$.

\subsection{The Higgs type potential $V(\phi)=-\frac{\mu ^2}{2}\phi^2+\frac{\xi}{4} \phi^4$}

As a second example of vacuum solutions of the gravitational field equations induced by the quantum metric fluctuations, with the fluctuation tensor given by the coupling between a scalar field and the metric,   we consider that the self-interaction potential associated to the scalar field is of the  Higgs-type, having the form
\be
V (\phi) = -\frac{\mu ^2}{2}\phi ^2+\frac{\xi }{4}\phi ^4,
\ee
where $\mu ^2$ and $\xi$  are constants.  The Higgs potential plays an essential role in particle physics. From a physical point of view we may assume that $-\mu ^2$ represents the mass of the scalar field particle associated to the gravitational metric fluctuations. For the strong interaction case the Higgs self-coupling constant $\xi$ takes the value $\xi \approx 1/8$ \cite{Aad}, a numerical value which was obtained from accelerator experiments. But it is obvious that in the case of the gravitational models with fluctuating quantum metric the value of $\xi$ may be very different from the one inferred from elementary particle physics.

The Higgs type potential can be written into a dimensionless form as
\begin{equation}
U(\psi)=-\frac{\chi ^2}{\beta^2}(1-\psi)^2+\frac{\sigma}{\beta^4}(1-\psi)^4,
\end{equation}
where $\chi ^2=\mu ^2M_0^2\kappa ^4/32\pi ^2$, and $\sigma=\xi M_0^2\kappa ^2/64 \pi ^2$, respectively. For the Higgs potential the static spherically symmetric  gravitational field equations in modified gravity with metric fluctuations escribed by the coupling between a scalar field and the metric take the form
\begin{eqnarray}
	\frac{d\psi}{d\eta}&=&-\frac{1}{\eta^2}\zeta,
	\end{eqnarray}
	\begin{eqnarray} \hspace{-0.8cm}&&\frac{d\zeta}{d\eta}=\frac{\zeta}{\eta}+\frac{\zeta}{\eta(1-M\eta)}+\frac{(2\psi\eta+\zeta)\zeta^2}{2\eta^3\psi(2\psi+3\beta^2)}-\frac{\zeta^2}{\eta^2\psi}\nonumber\\
\hspace{-0.8cm}&&-\frac{\zeta}{\eta^3(1-M\eta)\psi} \Bigg[-\frac{\chi ^2}{\beta^2}(1-\psi)^2+\frac{\sigma}{\beta^4}(1-\psi)^4\Bigg]\nonumber\\
\hspace{-0.8cm}&&+\frac{2(2\psi\eta+\zeta)}{\eta^3(1-M\eta)\psi(2\psi+3\beta^2)}\Bigg[-\chi ^2(1-\psi)^2+\frac{\sigma}{\beta^2}(1-\psi)^4\Bigg]\nn
\hspace{-0.8cm}	&&+\frac{2\psi\eta+\zeta}{\eta^3(1-M\eta)(2\psi+3\beta^2)}\Bigg[-2\chi ^2(1-\psi)+\frac{4\sigma}{\beta^2}(1-\psi)^3\Bigg],\nn
	\end{eqnarray}
\begin{eqnarray}
	\frac{d\nu}{d\eta}&=&\frac{1}{\eta}-\frac{1}{\eta(1-M\eta)}+\frac{\zeta}{\eta(1-M\eta)(2\psi\eta+\zeta)}\nn
&&+\frac{3\zeta}{\eta(2\psi\eta+\zeta)}- \frac{\zeta^2}{\beta^2\eta^2(2\psi\eta+\zeta)}+\nn
&&\frac{2}{\eta^2(1-M\eta)(2\psi\eta+\zeta)}\times \nn
&&\Bigg[-\frac{\chi ^2}{\beta^2}(1-\psi)^2+\frac{\sigma}{\beta^4}(1-\psi)^4\Bigg],
	\end{eqnarray}
	\begin{eqnarray}
\hspace{-0.9cm}&&	\frac{dM}{d\eta}=-\frac{\zeta(1-M\eta)}{\eta^2(2\psi\eta+\zeta)}+\frac{\zeta}{\eta^2(2\psi\eta+\zeta)}+\nn
\hspace{-0.9cm}&&\Bigg[\frac{1-M\eta}{\eta^4\psi(2\psi+3\beta^2)}-\frac{2(1-M\eta)}{\eta^3\psi(2\psi\eta+\zeta)}-\frac{1-M\eta}{\beta^2\eta^3(2\psi\eta+\zeta)}\Bigg]\zeta^2+\nn
\hspace{-0.9cm}&&\Bigg[\frac{4\beta^2}{\eta^4\psi(2\psi+3\beta^2)}-\frac{2}{\eta^3(2\psi\eta+\zeta)}-\frac{2\zeta}{\eta^4\psi(2\psi\eta+\zeta)}\Bigg]\times\nn
\hspace{-0.5cm} &&\Bigg[-\frac{\chi ^2}{\beta^2}(1-\psi)^2+\frac{\sigma}{\beta^4}(1-\psi)^4\Bigg]+\nn
\hspace{-0.5cm}&& \frac{2}{\eta^4(2\psi+3\beta^2)}\Bigg[-2\chi ^2(1-\psi)+\frac{4\sigma}{\beta^2}(1-\psi)^3\Bigg].
	\end{eqnarray}

\begin{figure*}[htbp]
	\centering
	\includegraphics[width=8.5cm]{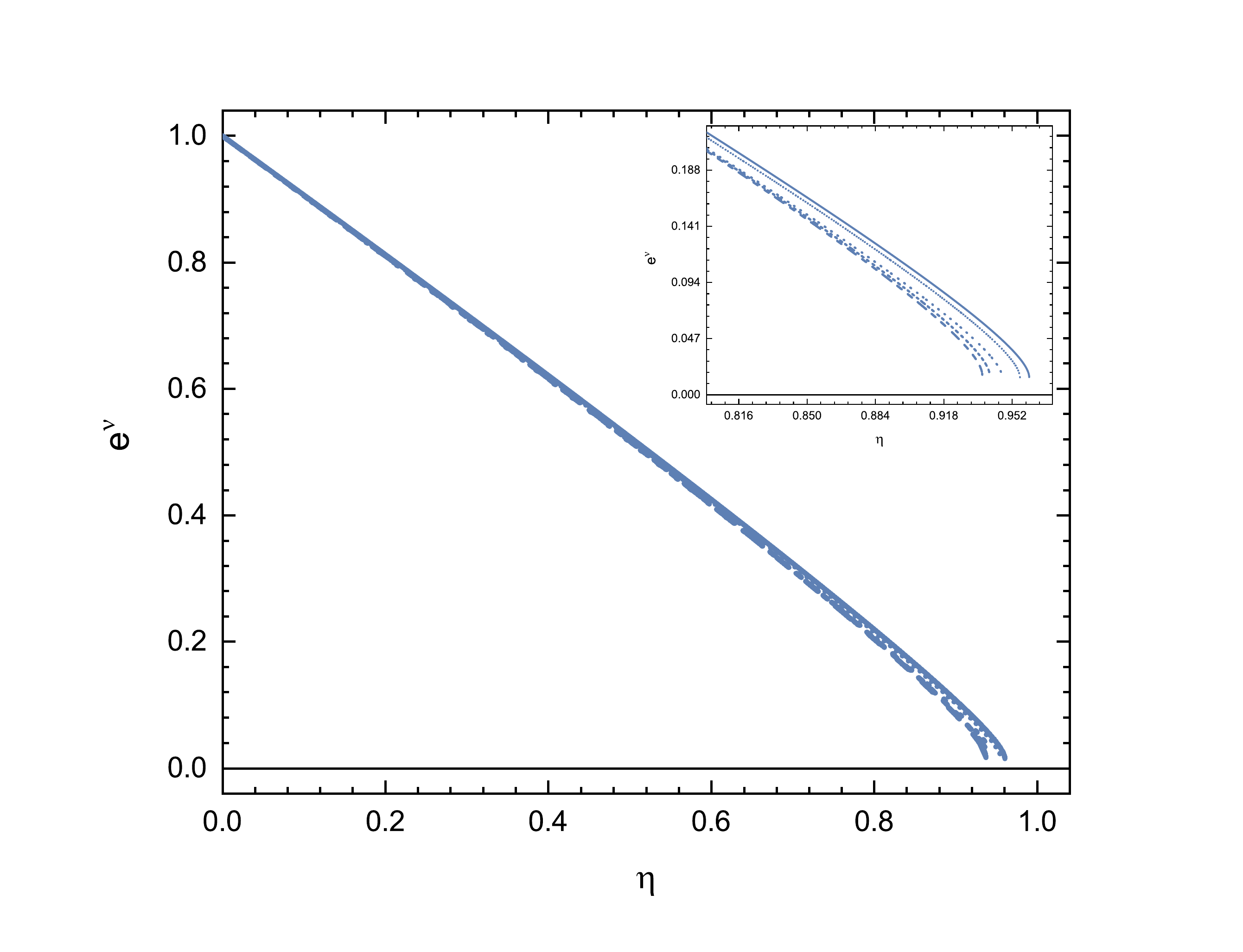}
	\includegraphics[width=8.5cm]{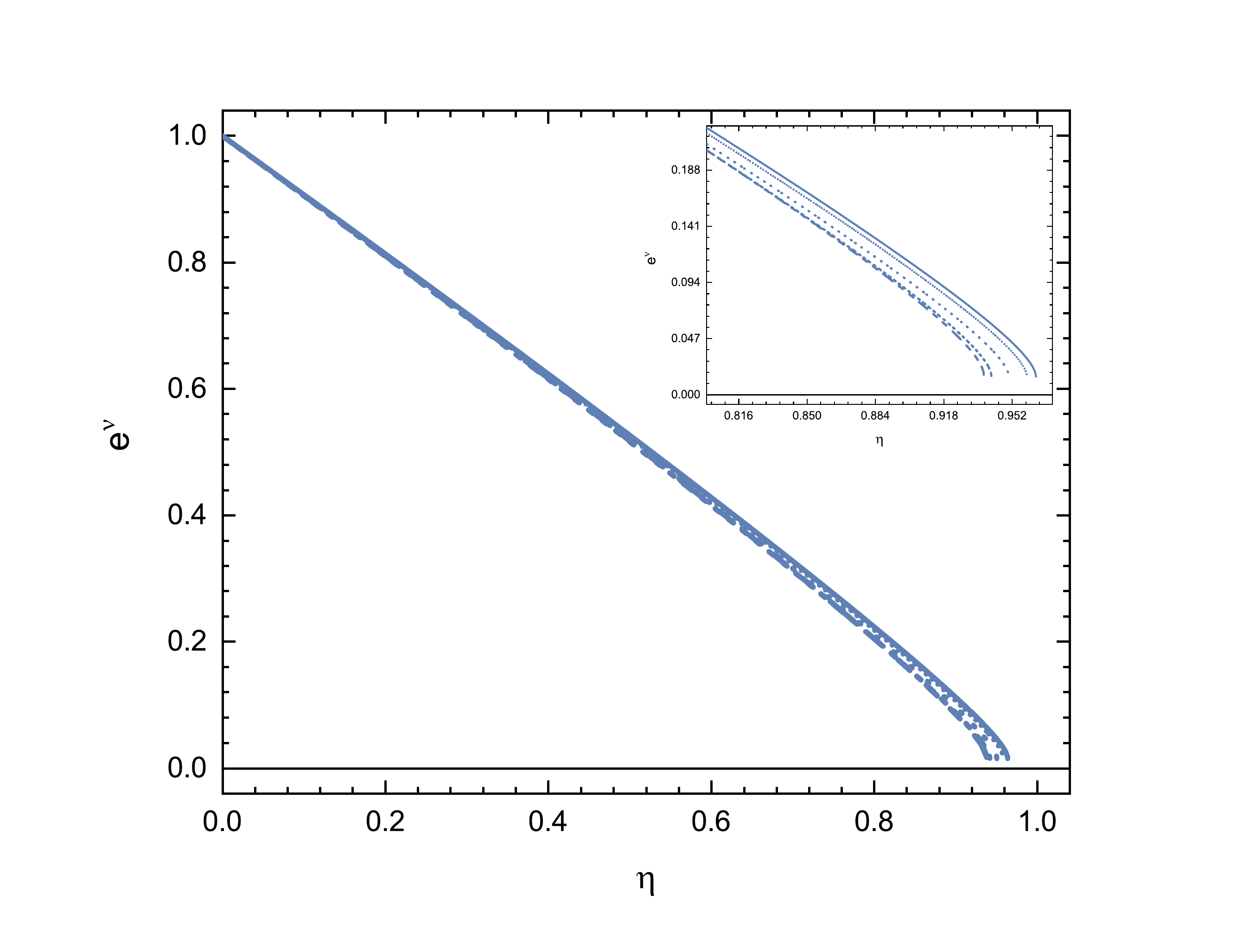}
		\label{ff18}
	\caption{Variation of the metric tensor component $e^{-\lambda}$ in the presence of quantum metric fluctuations induced by the coupling between a scalar field with a Higgs type potential and the metric tensor, for  $\beta=0.1$, $\psi_0=0.1$ and $\zeta_0=5\times 10^{-15}$. In the Figure on the left, $\sigma=1\times 10^{-26}$, while $\chi$ takes the values: $\chi=1\times 10^{-12}$ (solid curve), $\chi=3\times 10^{-12}$ (dotted curve), $\chi=5\times 10^{-12}$(short dashed curve), $\chi=7\times 10^{-12}$( dashed curve), $\chi=9\times 10^{-12}$ (long dashed curve). The additional Figure shows the variation of $e^{-\lambda}$ in a range of $\eta$ close to $1$. In the Figure on the right, $\sigma=9\times 10^{-26}$, while $\chi$ takes the values:: $\chi=1\times 10^{-12}$ (solid curve), $\chi=3\times 10^{-12}$ (dotted curve), $\chi=5\times 10^{-12}$ (short dashed curve), $\chi=7\times 10^{-12}$ (dashed curve), $\chi=9\times 10^{-12}$ (long dashed curve). The additional Figure shows the variation of $e^{-\lambda}$ in a range of $\eta$ close to $1$.}\label{fg18}
\end{figure*}

\begin{figure*}[htbp]
	\centering
	\includegraphics[width=8.5cm]{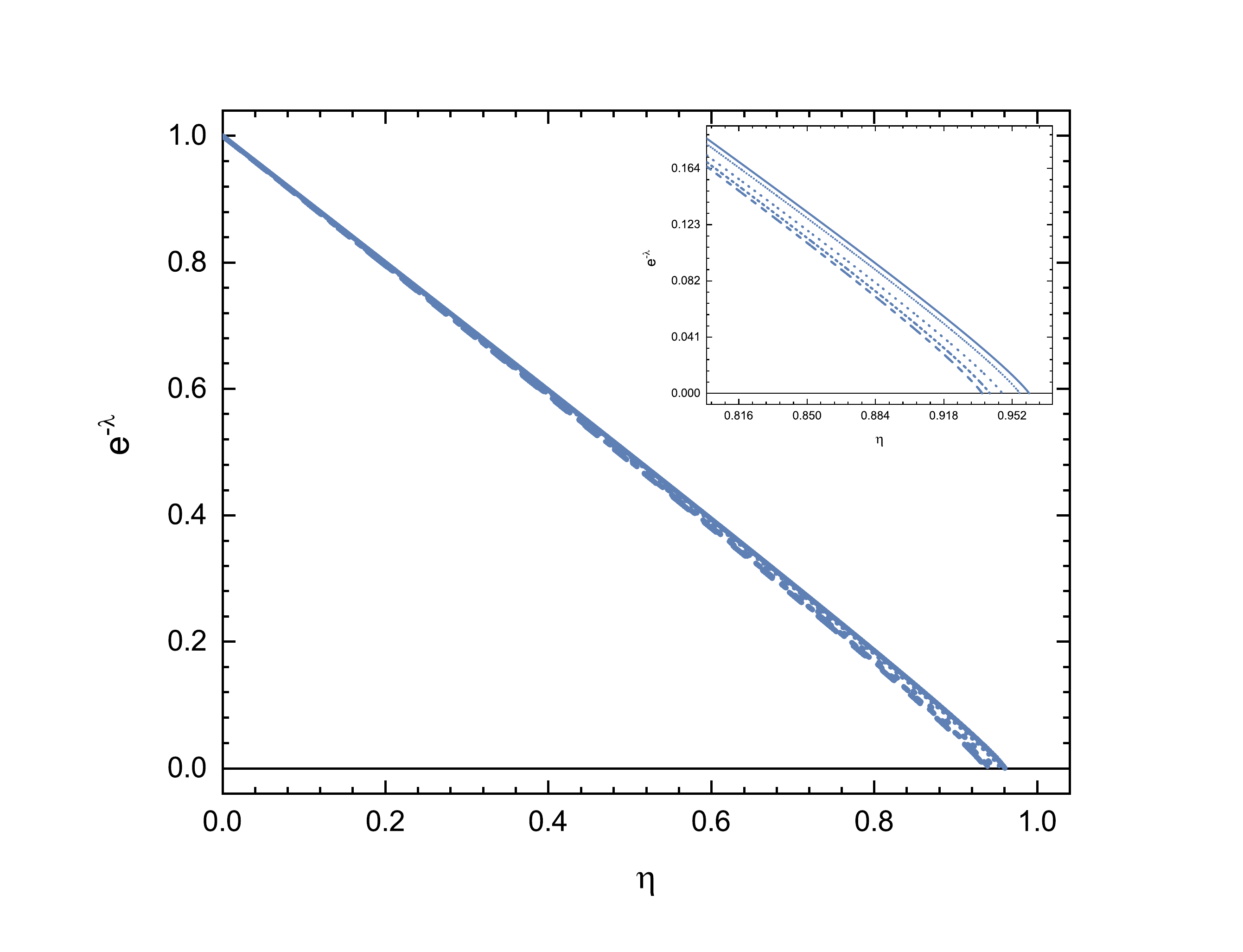}
	\includegraphics[width=8.5cm]{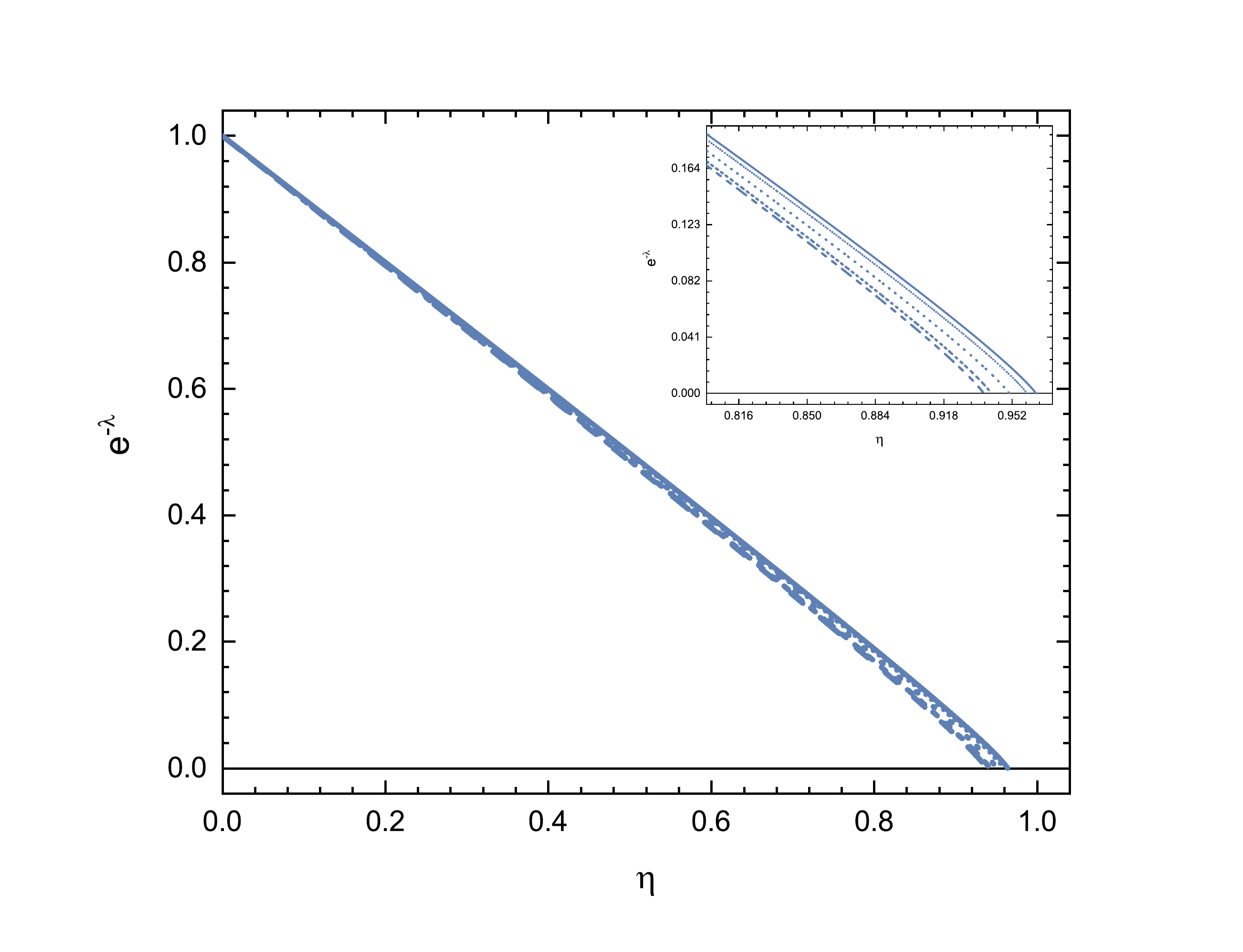}
\caption{Variation of the metric tensor component $e^\nu$ in the presence of quantum metric fluctuations induced by the coupling between a scalar field with a Higgs type potential and the metric tensor for $\beta=0.1$ , $\psi_0=0.1$ and $\zeta_0=5\times 10^{-15}$. Left panel: $\sigma=1\times 10^{-26}$, while $\chi$ takes the values: $\chi=1\times 10^{-12}$ (solid curve), $\chi=3\times 10^{-12}$ (dotted curve), $\chi=5\times 10^{-12}$(short dashed curve), $\chi=7\times 10^{-12}$( dashed curve), $\chi=9\times 10^{-12}$ (long dashed curve). The additional Figure shows the variation of $e^{-\lambda}$ in a range of $\eta$ close to $1$. Right panel: $\sigma=9\times 10^{-26}$, while $\chi$ takes the values:: $\chi=1\times 10^{-12}$ (solid curve), $\chi=3\times 10^{-12}$ (dotted curve), $\chi=5\times 10^{-12}$ (short dashed curve), $\chi=7\times 10^{-12}$ (dashed curve), $\chi=9\times 10^{-12}$ (long dashed curve). The additional Figure shows the variation of $e^{-\lambda}$ in a range of $\eta$ close to $1$.}\label{fg17}
\end{figure*}

 We detect the presence of a black hole through the existence of a singularity in the metric tensor components at a finite $\eta =\eta _s$, which corresponds to the event horizon of the black hole. The variations of the metric tensor components $e^{-\lambda}$ and $e^{nu}$ in the presence of quantum metric fluctuations induced by the coupling between a scalar field with a Higgs type potential and the metric tensor are presented in Figs.~\ref{fg18} and \ref{fg17}, respectively. To obtain the Figures, we have fixed the scalar field - metric coupling constant, the initial conditions at infinity, as well as the numerical value of the parameter $\sigma $ of the potential, and we have varied the numerical values of $\chi$.

As one can see from the Figures, both $e^{-\lambda}$ and $e^{\nu}$ vanish simultaneously for a given $\eta =\eta _s$, indicating the presence of a singularity in the metric, and the formation of a black hole. For small values of $\eta$ the behavior of the metric tensor components is basically independent on the parameters of the potential, but once the event horizon is approached, the position of the singularity depends on the initial conditions, as well as on the parameters of the Higgs potential.

\begin{figure*}[htbp]
	\centering
		\includegraphics[width=8.2cm]{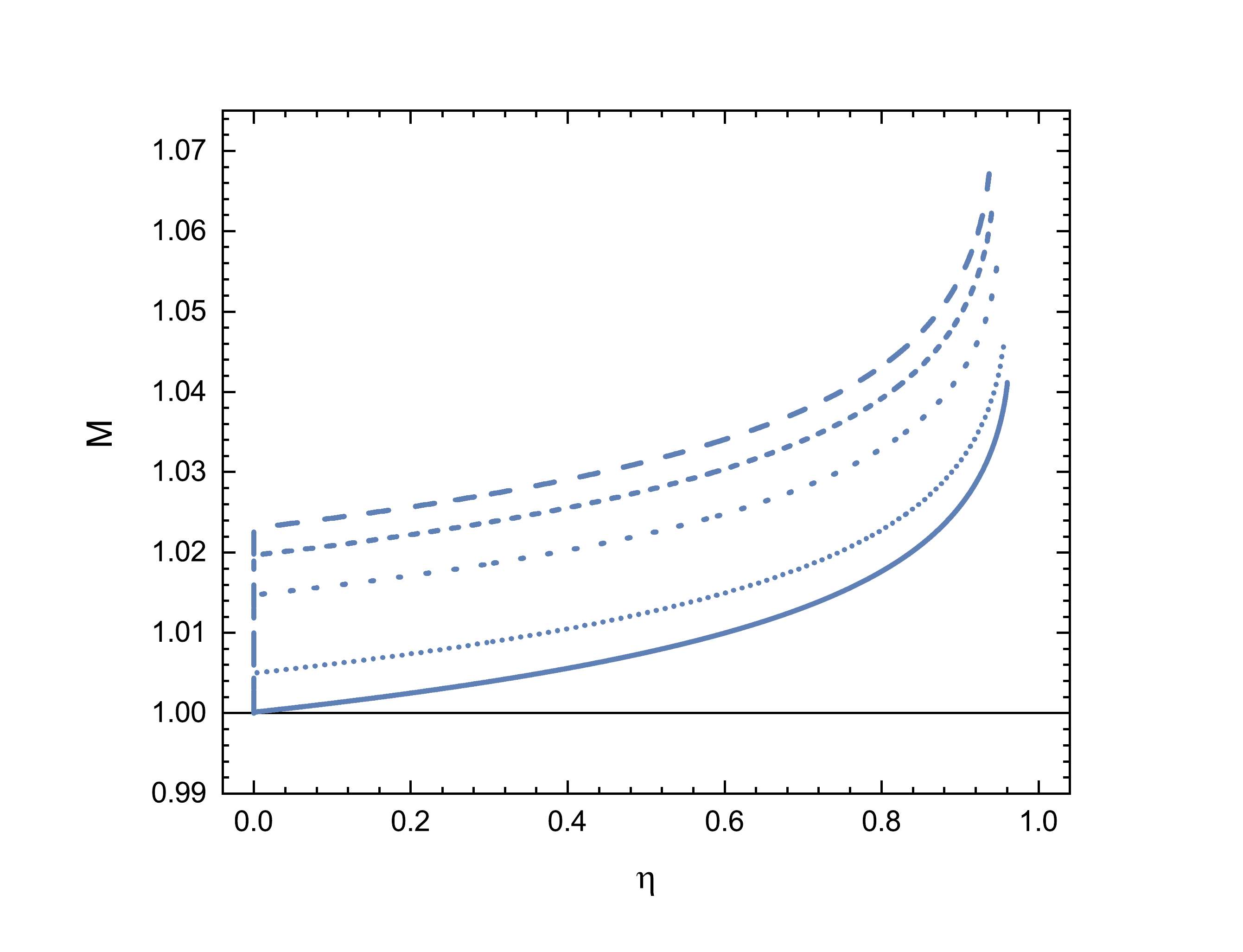}
		\includegraphics[width=8.2cm]{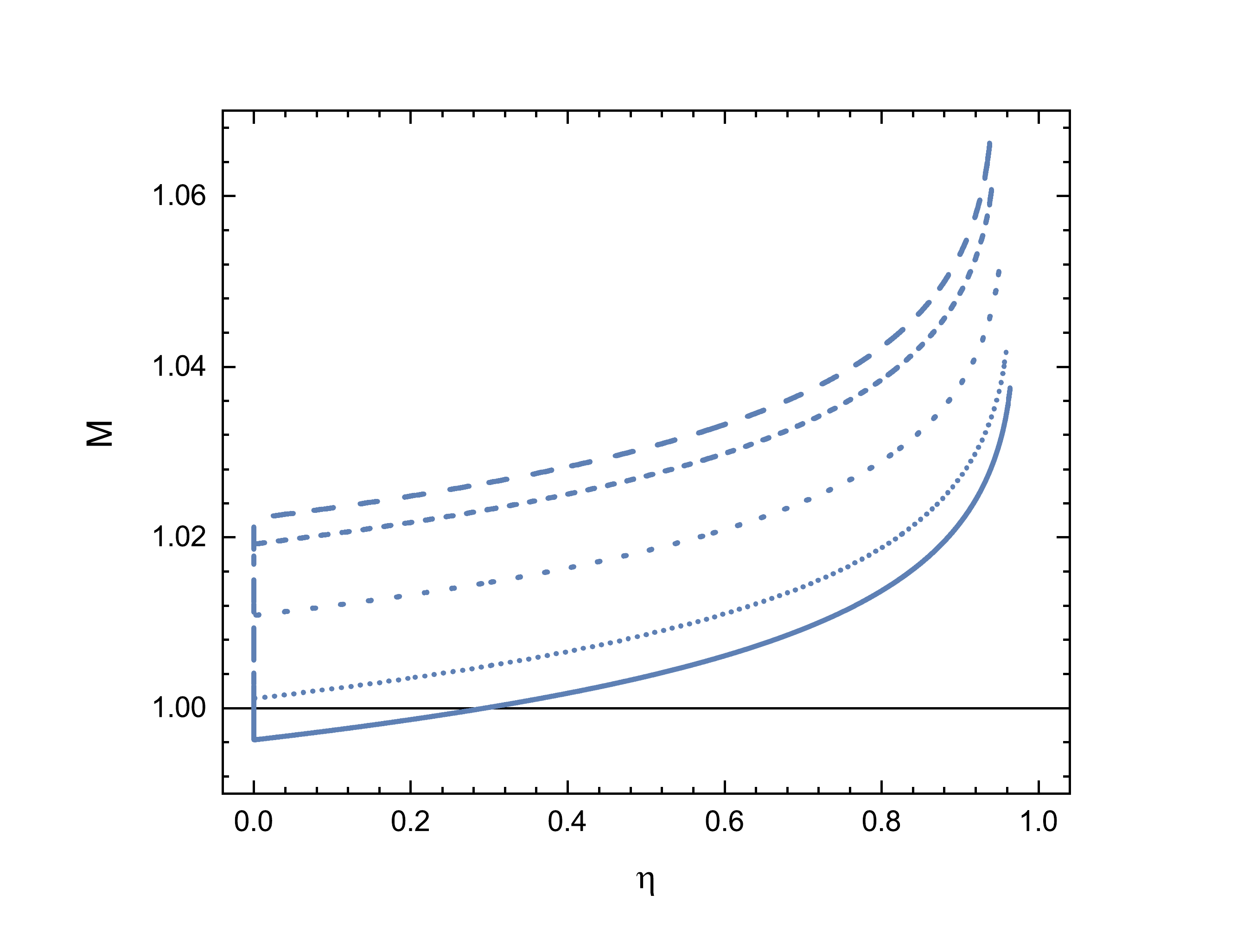}
		\caption{Variation of the effective mass $M$ in the presence of quantum metric fluctuations induced by the coupling between a scalar field with a Higgs type potential and the metric tensor for $\beta=0.1$ , $\psi_0=0.1$ and $\zeta_0=5\times 10^{-15}$. Left Figure:  $\sigma=1\times 10^{-26}$, while $\chi$ takes the values: $\chi=1\times 10^{-12}$ (solid curve), $\chi=3\times 10^{-12}$ (dotted curve), $\chi=5\times 10^{-12}$(short dashed curve), $\chi=7\times 10^{-12}$( dashed curve), $\chi=9\times 10^{-12}$ (long dashed curve). Right Figure: $\sigma=9\times 10^{-26}$, while $\chi$ takes the values:: $\chi=1\times 10^{-12}$ (solid curve), $\chi=3\times 10^{-12}$ (dotted curve), $\chi=5\times 10^{-12}$ (short dashed curve), $\chi=7\times 10^{-12}$ (dashed curve), $\chi=9\times 10^{-12}$ (long dashed curve).}\label{fg14}
\end{figure*}

The effective mass of the black hole in the presence of a Higgs potential is represented in Figs.~\ref{fg14}. As one can see from the Figures, the effective mass of the black hole is strongly dependent on the parameters of the Higgs potential. The energy related to the quantum fluctuations of the metric could add a significant contribution to the total mass of the black hole.

\begin{figure*}[htbp]
	\centering
	\includegraphics[width=250pt]{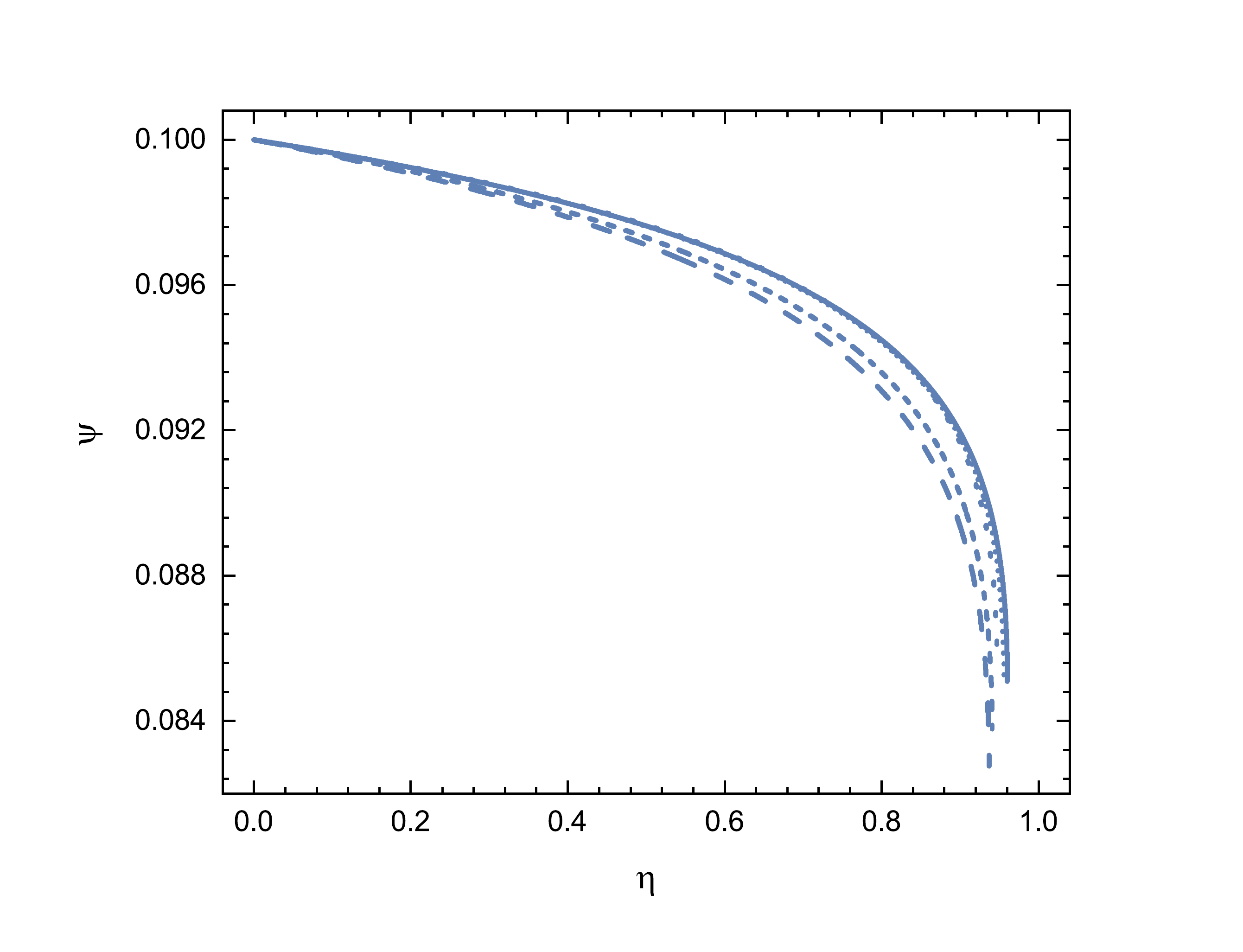}
		\includegraphics[width=250pt]{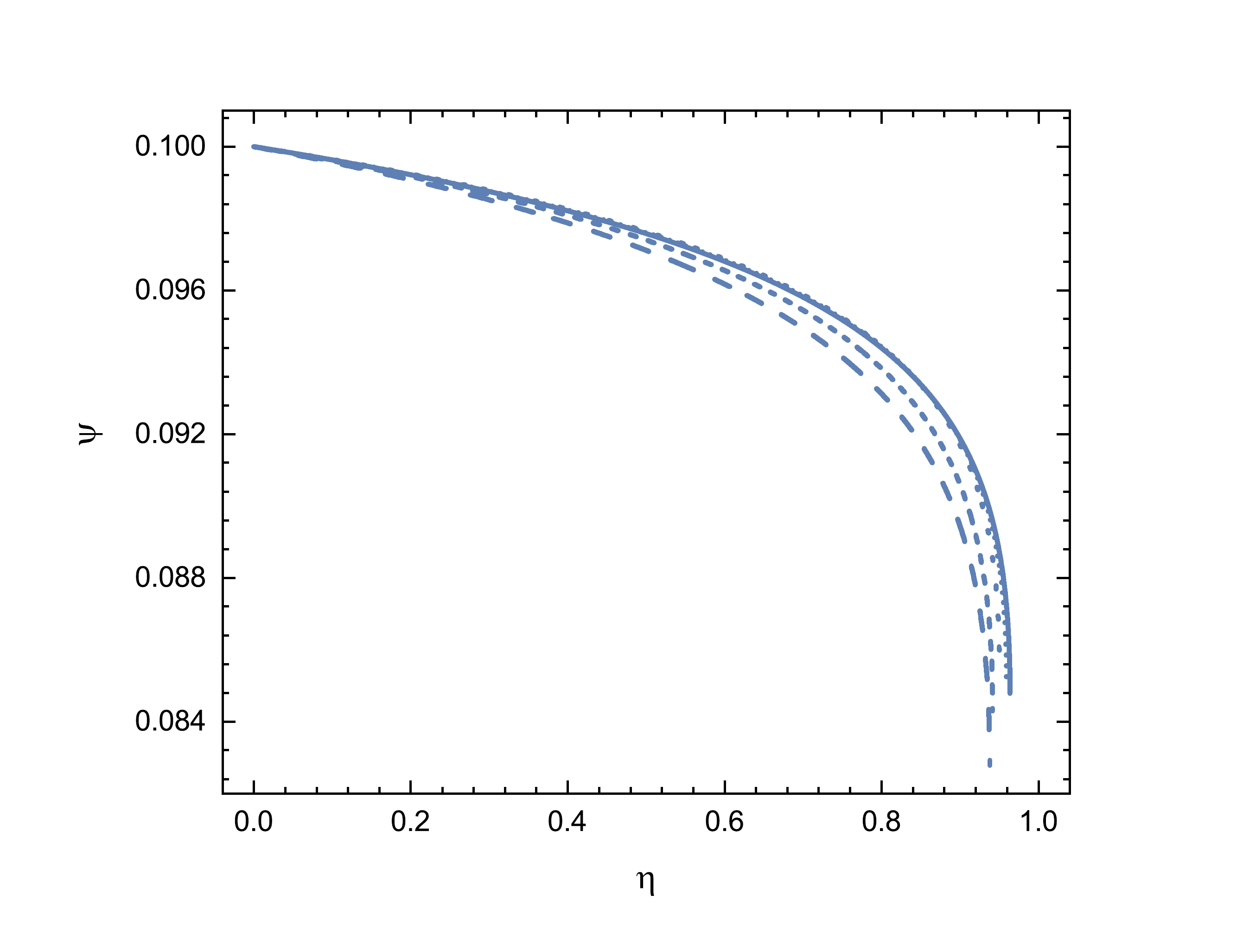}
		\caption{Variation of the rescaled scalar field $\psi$ in the presence of quantum metric fluctuations induced by the coupling between a scalar field with a Higgs type potential and the metric tensor for $\beta=0.1$ , $\psi_0=0.1$ and $\zeta_0=5\times 10^{-15}$. Left panel: $\sigma=1\times 10^{-26}$, while $\chi$ takes the values: $\chi=1\times 10^{-12}$ (solid curve), $\chi=3\times 10^{-12}$ (dotted curve), $\chi=5\times 10^{-12}$(short dashed curve), $\chi=7\times 10^{-12}$( dashed curve), $\chi=9\times 10^{-12}$ (long dashed curve). Right panel: $\sigma=9\times 10^{-26}$, while $\chi$ takes the values:: $\chi=1\times 10^{-12}$ (solid curve), $\chi=3\times 10^{-12}$ (dotted curve), $\chi=5\times 10^{-12}$ (short dashed curve), $\chi=7\times 10^{-12}$ (dashed curve), $\chi=9\times 10^{-12}$ (long dashed curve).}\label{fg11}
\end{figure*}

The variation of the rescaled scalar field $\psi$ with respect to $\eta $ is represented in Figs.~\ref{fg11}. For the adopted numerical values of the parameters the scalar field takes smaller values at the event horizon as compared to its values at infinity. Its evolution depends strongly on the initial conditions, and on the parameters of the Higgs potential.  In Table~(\ref{tt2}) we present a selected sample of the positions of the event horizon of the black hole, for fixed initial conditions at infinity, and for different values of the parameters of the Higgs potential.

\begin{table*}[htbp]
	\centering	 \begin{tabular}{|>{\centering}p{70pt}|>{\centering}p{70pt}|>{\centering}p{70pt}|>{\centering}p{70pt}|>{\centering}p{70pt}|>{\centering\arraybackslash}p{70pt}|}
		\hline
		$\chi$ & $\sigma$ & $\eta_{s}$ & $\chi$ & $\sigma$ & $\eta_{s}$\\
		\hline
		$1\times 10^{-12}$ & $1\times 10^{-26}$ & $0.96043$ & $1\times 10^{-12}$ & $1\times 10^{-26}$ & $0.96043$\\
\hline
		$3\times 10^{-12}$ & $1\times 10^{-26}$ & $0.95581$ & $1\times 10^{-12}$ & $3\times 10^{-26}$ & $0.96119$\\
\hline
		$5\times 10^{-12}$ & $1\times 10^{-26}$ & $0.94705$ & $1\times 10^{-12}$ & $5\times 10^{-26}$ & $0.96204$\\
\hline
		$7\times 10^{-12}$ & $1\times 10^{-26}$ & $0.94084$ & $1\times 10^{-12}$ & $7\times 10^{-26}$ & $0.96292$\\
\hline
		$9\times 10^{-12}$ & $1\times 10^{-26}$ & $0.93708$ & $1\times 10^{-12}$ & $9\times 10^{-26}$ & $0.96380$\\
		\hline
		\end{tabular}
	\caption{Selected numerical values of the outer apparent horizon $\eta_s$ of a black hole in the presence of quantum metric fluctuations induced by the coupling between a scalar field with a Higgs type potential and the metric tensor, for $\beta=0.1$, $\psi_0=0.1$, $\zeta_0=5\times 10^{-15}$, and different values of the parameters $\chi$ and $\sigma$.}
	\label{tt2}	
\end{table*}

\begin{figure}[htbp]
	\centering
	\includegraphics[width=8.2cm]{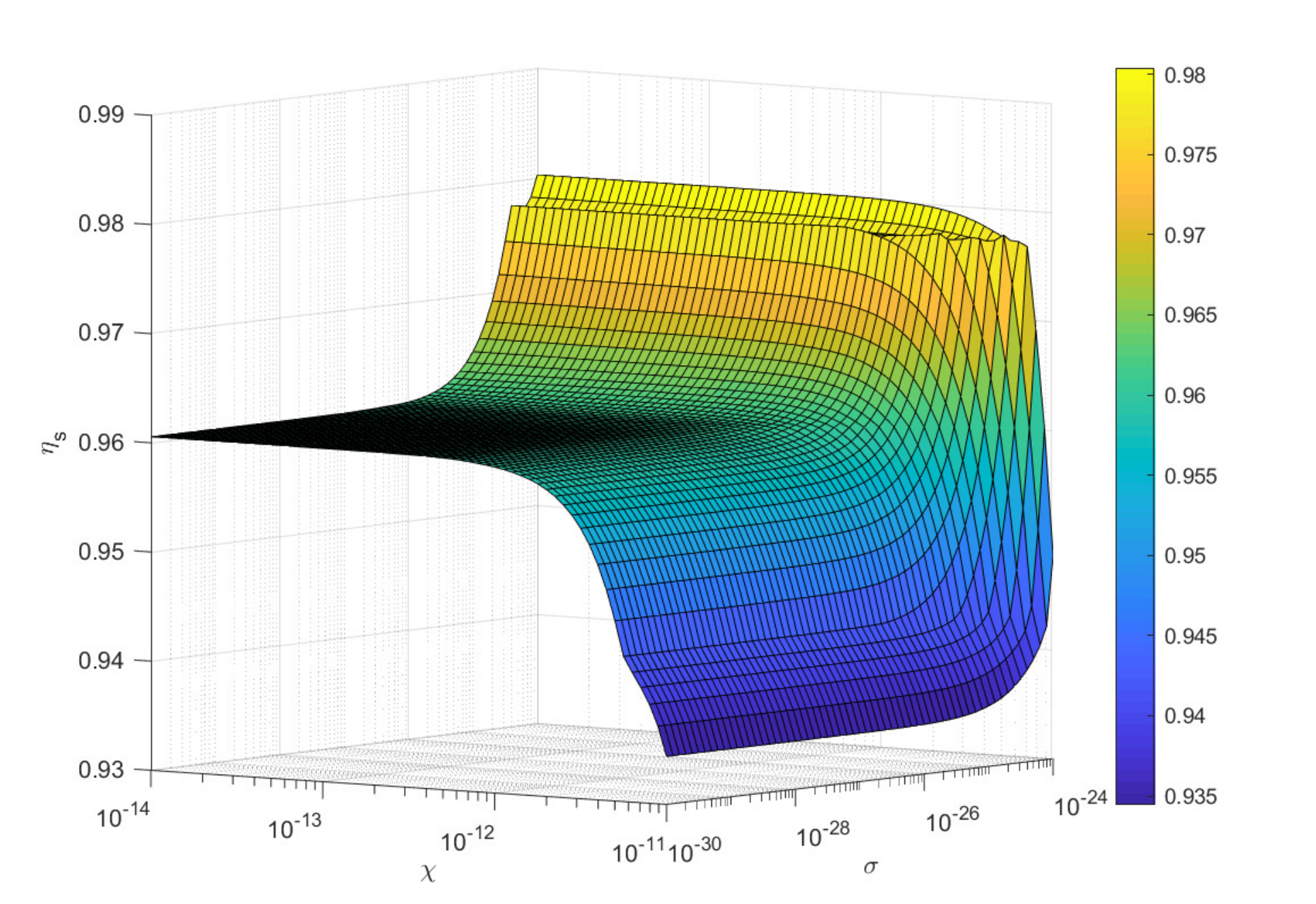}
	\caption{Variation of $\eta_s$  as a function of the parameters $\chi$ and $\sigma$ of the Higgs type potential for $\beta=0.1$, $\psi_0=0.1$ and $\zeta_0=5\times 10^{-15}$.}
	\label{ff19}
\end{figure}

The variation of the event horizon of the black hole with changing $\chi$ and $\sigma$ is represented in Fig.~\ref{ff19}.

\subsubsection{Interpolating functions}

The numerical results obtained by numerically integrating the gravitational field equations can also be fitted by some interpolating functions. In the following we will consider only the variation of the geometric and physical quantities with $\chi$ and $\sigma$, by assuming fix initial conditions. The effective mass of the black hole can be represented as
	\begin{eqnarray}
M(\eta)&=&a_4\left(\chi,\sigma\right)\eta^3+b_4\left(\chi,\sigma\right)\eta^2+\nn
&&c_4\left(\chi,\sigma\right)\eta+d_4\left(\chi,\sigma\right),
\end{eqnarray}
where the coefficients $(a_4,b_4,c_4,d_4)$ are given by
\begin{eqnarray*}
	a_4\left(\zeta _0,\psi _0\right) &=&1.695\times 10^{-2}+ 8.335\times 10^{31}\chi^3-\nn
	&& 2.325\times 10^{21}\chi^2+2.265\times 10^{10}\chi-\nn
	&& 9.61\times 10^{21} \sigma,\\
b_4\left(\zeta _0,\psi _0\right) &=&-2.004\times 10^{-2}-6.909\times 10^{31}\chi^3+\nn
	&& 1.962\times 10^{21}\chi^2-1.929\times 10^{10}\chi-\nn
	&& 7.75\times 10^{21} \sigma,\\
	c_4\left(\zeta _0,\psi _0\right) &=&8.56\times 10^{-3}+ 3.187\times 10^{31}\chi^3-\nn
	&& 9.013\times 10^{20}\chi^2+8.708\times 10^{9}\chi-\nn
	&& 2.969\times 10^{21} \sigma,\\
d_4\left(\zeta _0,\psi _0\right) &=& 1.028 +3.395\times 10^{21}\chi^2 -\nn
	&&  3.715\times 10^{9}\chi -8.284\times 10^{21}\sigma.
	\end{eqnarray*}
The Pearson correlation coefficient for the fitting is 	$R^2 = 0.9999.$
The metric tensor coefficient $e^{-\lambda}$ can be obtained as
\bea
\hspace{-0.4cm}e^{-\lambda} &=& 1-M\eta = a_5\left(\zeta _0,\psi _0\right)\eta^4+b_5\left(\zeta _0,\psi _0\right)\eta^3+\nonumber\\
\hspace{-0.4cm}&&+c_5\left(\zeta _0,\psi _0\right)\eta^2+d_5\left(\zeta _0,\psi _0\right)\eta+e_5\left(\zeta _0,\psi _0\right),\nn
\eea
where $(a_5,b_5,c_5,d_5,e_5)$ are coefficients depending on $\left(\zeta _0,\psi_0\right)$ so that $a_5=-a_4$, $b_5=-b_4$, $c_5=-c_4$, $d_5=-d_4$ and $e_5=1$, respectively.

We can fit $e^{\nu}$ with the help of the function
\begin{eqnarray}
e^{\nu(\eta)}&=&a_6\left(\zeta _0,\psi _0\right)\eta^2+b_6\left(\zeta _0,\psi _0\right)\eta+\nn
&&c_6\left(\zeta _0,\psi _0\right),
\end{eqnarray}
where the coefficients $(a_6,b_6,c_6)$ are given by
\begin{eqnarray*}
	a_6\left(\zeta _0,\psi _0\right) &=& 7.499\times 10^{-2} - 1.772\times 10^{32}\chi^3+\nn
	&& 5.593\times 10^{21}\chi^2-5.907\times 10^{10}\chi+\nn
	&& 2.327\times 10^{22} \sigma,\\
	b_6\left(\zeta _0,\psi _0\right) &=& -1.118 + 1.791\times 10^{32}\chi^3-\nn
	&& 6.197\times 10^{21}\chi^2+6.692\times 10^{10}\chi-\nn
	&& 1.684\times 10^{22} \sigma,\\
	c_6\left(\zeta _0,\psi _0\right) &=& 1.002 - 6.074\times 10^{30}\chi^3+\nn
	&& 1.945\times 10^{20}\chi^2-2.052\times 10^{9}\chi+\nn
	&& 6.935\times 10^{20} \sigma,
\end{eqnarray*}
with the correlation coefficient $R^2 = 0.9999$.

\section{Thermodynamics of black holes in modified gravity induced by quantum metric fluctuations}\label{sect5}

In the present article  article we have considered the effects of the quantum fluctuations of the metric, described by the effective fluctuation tensor, on the black hole geometry. We have investigated in detail the effects of the quantum fluctuations for the case of a static, spherically symmetric geometry, under the assumption that the scalar field and the background geometry depends only on the radial coordinate. Under these assumptions it turns out that since the black hole geometry is static, it always admits a timelike Killing vector $t^\mu$. For a static black hole possessing a Killing horizon the definition of the surface gravity $\tilde k$ is given by $t^\mu\nabla_\mu t^\nu=t^\nu \tilde{\kappa}$ \cite{BH1,BH2}. In the case of a static, spherically symmetric spacetime, the line element can be generally represented in the form
\begin{equation}
ds^2=-\tilde {\sigma}(r)f(r)c^2dt^2+\frac{dr^2}{f(r)}+r^2d\Omega^2.
\end{equation}
where the metric functions $\sigma (r)$ and $f(r)$ are functions of the radial coordinate $r$ only. By suitably normalizing the Killing vector $t^\mu=(1/\tilde \sigma_\infty,0,0,0)$, for the surface gravity of the black hole we obtain
\begin{eqnarray}
\tilde \kappa=\l(\frac{\tilde \sigma_{hor}}{\tilde \sigma_\infty}\r)\frac{c^4}{4Gm_{hor}}\l[1-\frac{2Gm'(r)}{c^2}\r]\bigg|_{hor},
\end{eqnarray}
\\
where the subscript $hor$ indicates that all physical quantities must be evaluated on the apparent horizon of the black hole. The subscript $\inf$ refers to the values at infinity of the geometrical and physical quantities.  When $\sigma=1$, and $m={\rm constant}$, we will go back to expression of the surface gravity of a Schwarzschild black hole, $\tilde {\kappa}=c^4/4Gm_{hor}$.

\subsection{The black hole temperature}

The temperature $T_{BH}$ of the black hole is defined as \cite{BH1, BH2}
\begin{equation}
T_{BH}=\frac{\hbar}{2\pi ck_B}\tilde \kappa,
\end{equation}
where by $k_B$ we have denoted Boltzmann's constant. In the dimensionless variables introduced in Eq.~(\ref{d1}), we obtain the temperature of the black hole in the form
\begin{eqnarray}
\hspace{-0.5cm}T_{BH}=\frac{T_H}{M(\eta_s)}\bigg[1+\eta^2\frac{dM(\eta)}{d\eta}\bigg]\bigg|_{\eta=\eta_s}=&T_H\theta(\eta)|_{\eta=\eta_s},
\end{eqnarray}
where
\be
\theta (\eta)=\frac{1}{M(\eta_s)}\bigg[1+\eta^2\frac{dM(\eta)}{d\eta}\bigg],
\ee
and $T_H$ denotes the Hawking temperature of the black hole, given by
\begin{equation}
T_H=\frac{\hbar c^3}{8\pi Gk_B M_0}.
\end{equation}

With the use of the fitting functions obtained in the previous Sections one can obtain the general expressions for $\theta(\eta)$ and of its first derivative in $\eta$ as follows,
\be \label{theta}
\theta(\eta)=\frac{\eta^2\l(3a_i\eta^2+2b_1\eta+c_i\r)+1}{a_i\eta^3+b_i\eta^2+c_i\eta+d_i},
\ee
\bea
\frac{d\theta(\eta)}{d\eta}&=&\frac{2\eta\l(3a_i\eta^2+2b_i\eta+c_i\r)+\eta^2\l(6a_i\eta+2b_i\r)}{a_i\eta^3+b_i\eta^2+c_i\eta+d_i}-\nn
&&\frac{\l[\eta^2\l(3a_i\eta^2+2b_i\eta+c_i\r)+1\r]\l(3a_i\eta^2+2b_i\eta+c_i\r)}{\l(a_i\eta^3+b_i\eta^2+c_i\eta+d\r)^2}.\nn
\end{eqnarray}

In the above equations  $i=1$ gives the expression of $\theta (\eta)$ and of its derivative for the zero potential case, while in the case of the Higgs potential $i$ takes the value $i=4$.

\begin{figure*}[htbp]
	\centering
	\includegraphics[width=8.9cm]{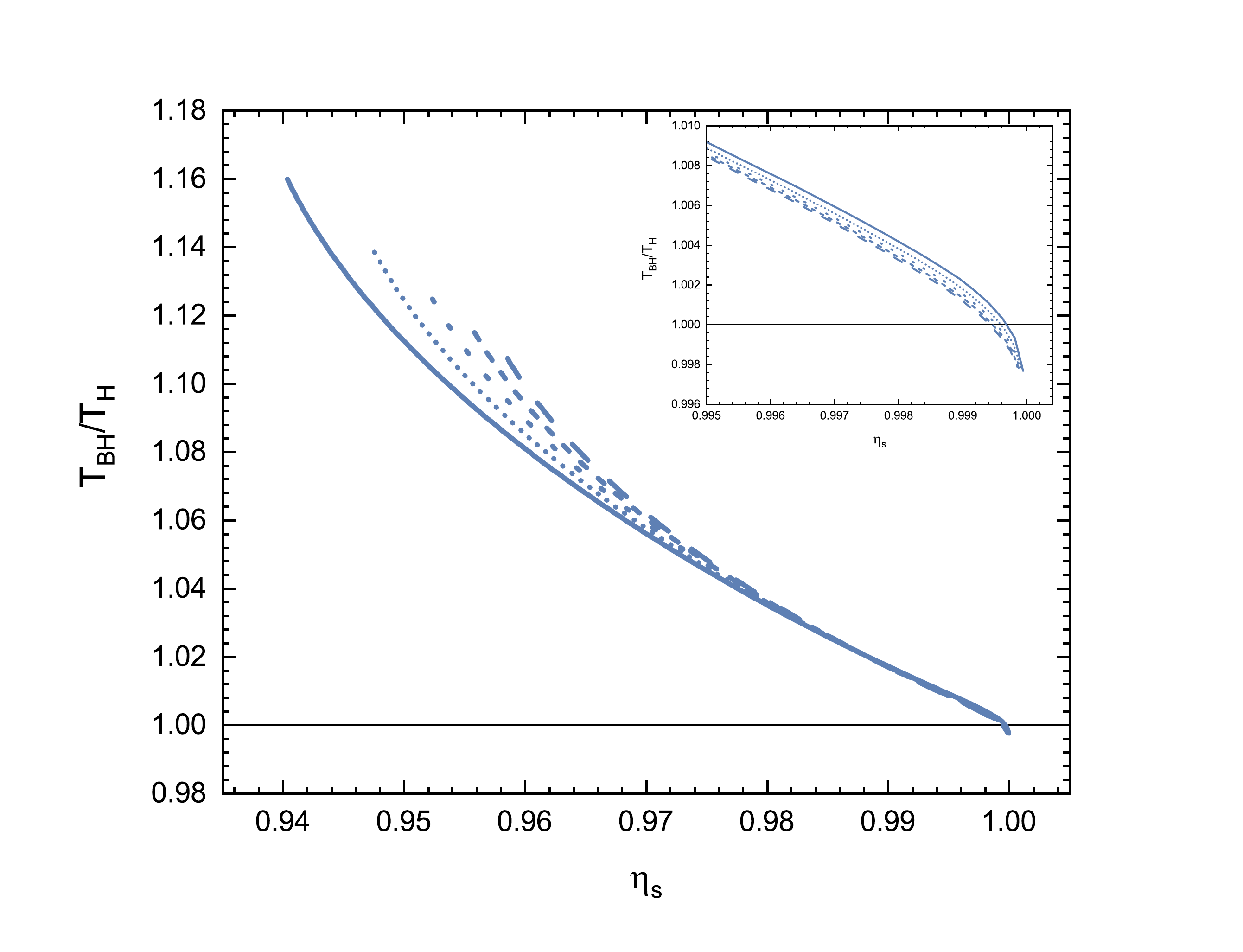}
	\includegraphics[width=8.9cm]{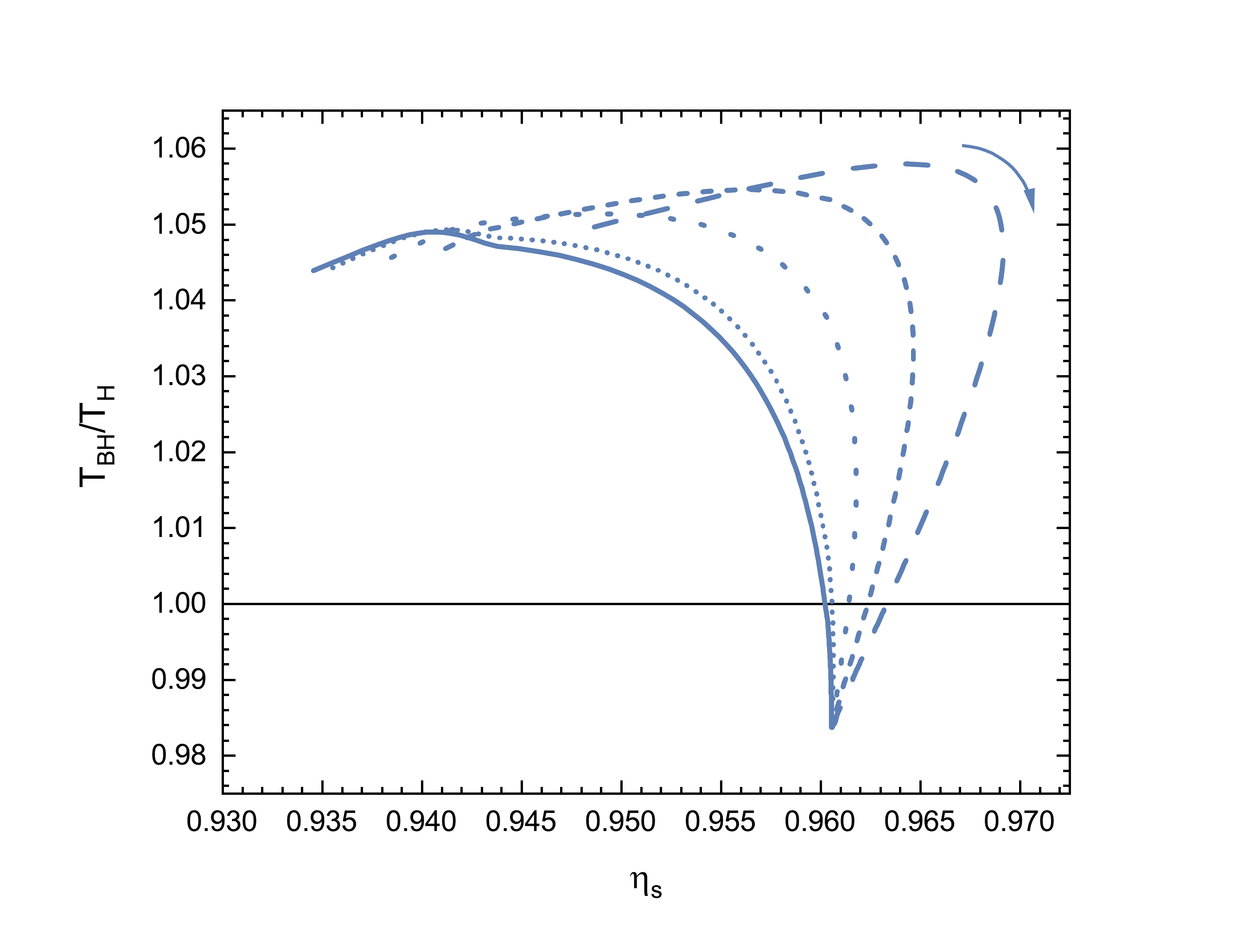}
 \caption{Variation of the dimensionless Hawking temperature $T_{BH}(\eta_s)/T_H$ as a function of the location of the event horizon $\eta_s$. Left Figure: $V(\phi)=0$,  $\beta=0.1$, and $\psi_0^2 \zeta_0 = 2\times 10^{-17}$ (solid curve), $\psi_0^2 \zeta_0 = 3\times 10^{-17}$ (dotted curve), $\psi_0^2 \zeta_0 = 4\times 10^{-17}$ (short dashed curve), $\psi_0^2 \zeta_0 = 5\times 10^{-17}$ (dashed curve), and $\psi_0^2 \zeta_0 = 6\times 10^{-17}$ (long dashed curve). Right Figure:  $V (\phi) = -\frac{\mu ^2}{2}\phi ^2+\frac{\xi }{4}\phi ^4$,  $\beta=0.1$,  $\psi_0=0.1$, $\zeta_0=5\times 10^{-15}$ and  $\sigma / \chi = 1\times 10^{-15}$ (solid curve), $\sigma / \chi = 1\times 10^{-14}$ (dotted curve), $\sigma / \chi = 4\times 10^{-14}$ (short dashed curve), $\sigma / \chi = 7\times 10^{-14}$ (dashed curve), $\sigma / \chi = 1\times 10^{-13}$ (long dashed curve). The arrow shows the direction of decreasing values.
}\label{ffnp1}
\end{figure*}

The variation of the Hawking temperature of the black hole as a function of the event horizon radius $\eta _s$ is represented in Fig.~\ref{ffnp1}.

The Hawking temperature shows a strong dependence on both initial conditions, as well as on the parameters of the scalar field potential. In the case of the vanishing scalar field potential the horizon temperature decreases with increasing $\eta _s$. The increase of the Hawking temperature for $\eta _s=0.94$, or $r_g=1.06r_{Sch}$ is of the order of 16\% as compared to the standard Schwarzschild case. In the case of the Higgs potential the dependence of the temperature on the radius of the event horizon is more complicated, The increase in the numerical values of the temperature are of the order of 6\%, with the temperature decreasing rapidly towards the Schwarzschild value with increasing $\eta _s$.

\begin{figure*}[htbp]
	\centering
	\includegraphics[width=230pt]{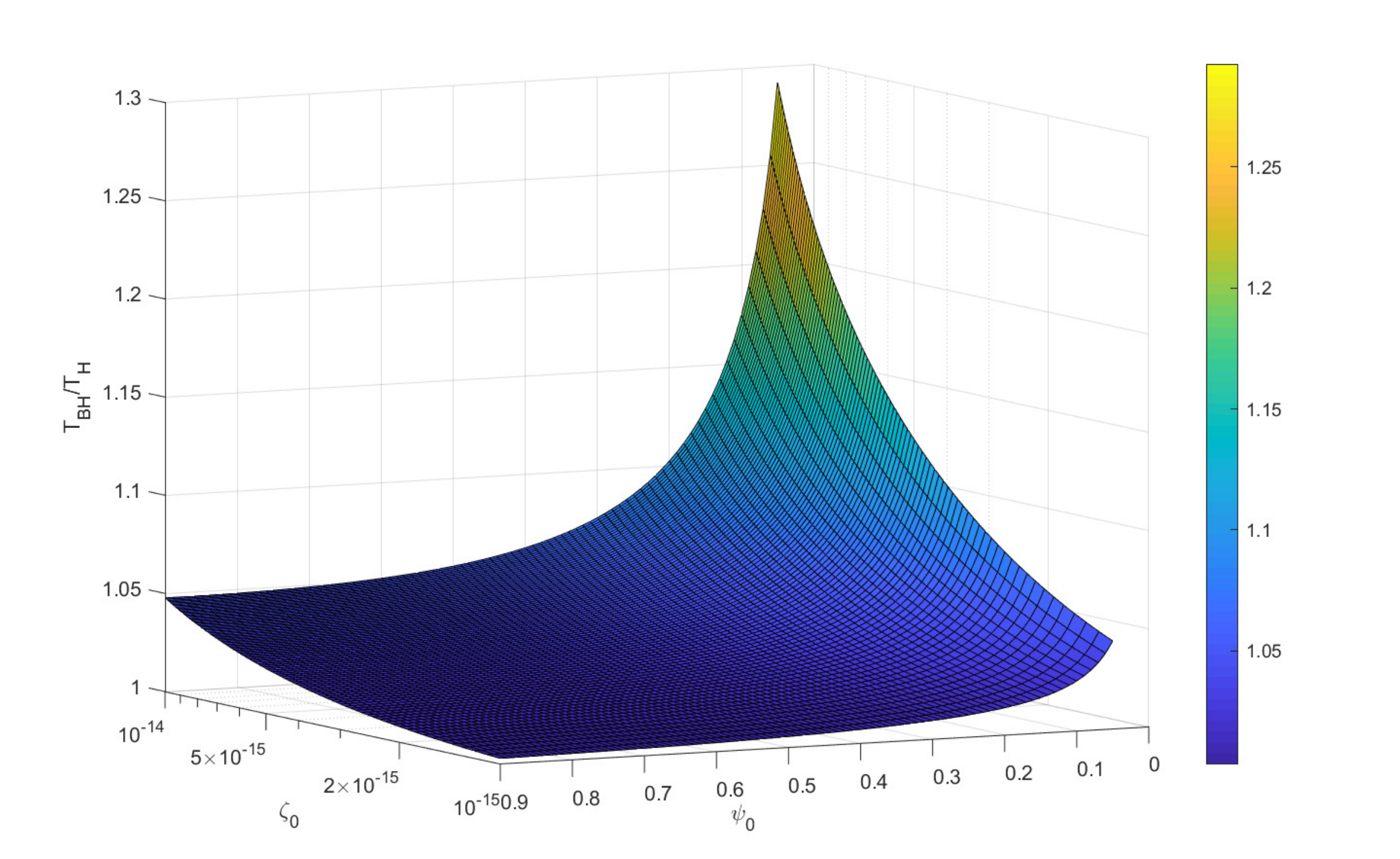}
 \includegraphics[width=230pt]{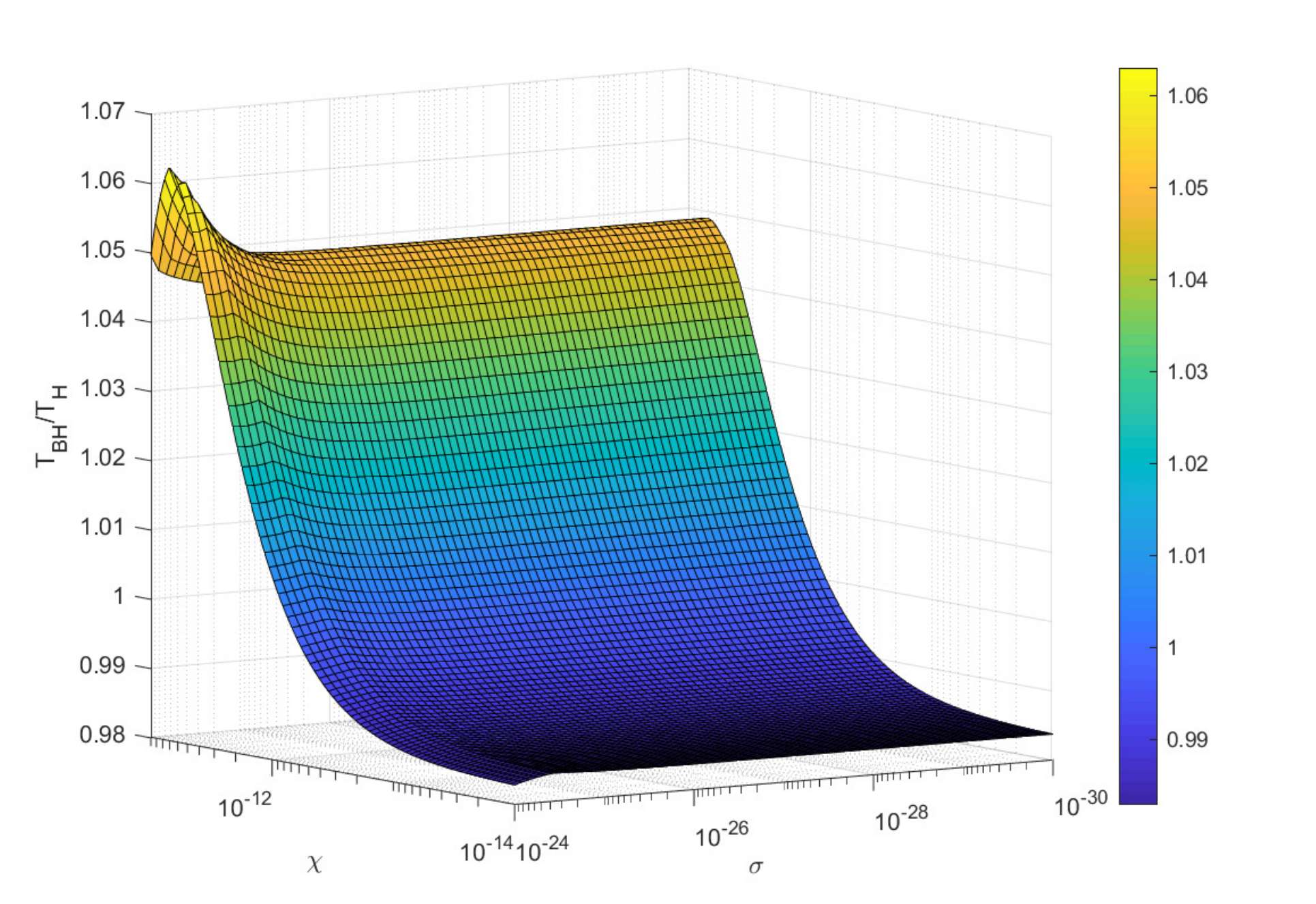}
	\caption{Variation of the black hole temperature $T_{BH}(\eta_s)/T_H$ as a function of the initial conditions of the scalar field and potential parameters. Left Figure: $V(\phi)=0$, $\beta=0.10$.
Right Figure: $V (\phi) = -\frac{\mu ^2}{2}\phi ^2+\frac{\xi }{4}\phi ^4$,  $\beta=0.10$,  $\psi_0=0.1$, $\zeta_0=5\times 10^{-15}$. }
	\label{ff31}
\end{figure*}

The variations of the dimensionless black hole temperature $T_{BH}(\eta_s)/T_H$ as a function of the initial conditions at infinity are represented, for the cases of the vanishing scalar field potential $V(\phi)=0$ and for a Higgs type potential, respectively, in Fig.~\ref{ff31}.

Some specific values of the dimensionless temperature of the black hole in the modified gravity theory induced by quantum metric fluctuations are presented in Table~\ref{tt3}.

\begin{table*}[htbp]
\centering
	 \begin{tabular}{|>{\centering}p{50pt}|>{\centering}p{60pt}|>{\centering}p{60pt}|>{\centering}p{60pt}||>{\centering}p{60pt}|>{\centering}p{60pt}|>{\centering}p{60pt}|>{\centering\arraybackslash}p{60pt}|}
		\hline
		$\quad$ & $\zeta_0=1\times 10^{-15}$ & $\zeta_0=5\times 10^{-15}$ & $\zeta_0=9\times 10^{-15}$ & $\quad$ & $\chi=1\times 10^{-14}$ & $\chi=1\times 10^{-13}$ & $\chi=1\times 10^{-12}$\\
		\hline
		$\psi_0=0.1$ & $1.0255$ & $1.0919$ & $1.1566$ & $\sigma=1\times 10^{-30}$ & $0.98369$ & $0.98583$ & $1.0049$\\
		\hline
		$\psi_0=0.5$ & $1.0057$ & $1.0278$ & $1.0497$ & $\sigma=1\times 10^{-28}$ & $0.98369$ & $0.98584$ & $1.0049$\\
		\hline
		$\psi_0=0.9$ & $1.0031$ & $1.0229$ & $1.0427$ & $\sigma=1\times 10^{-26}$ & $0.98370$ & $0.98585$ & $1.0050$\\
		\hline
	\end{tabular}
	\caption{Numerical values of the dimensionless temperature of the black hole $T_{BH}(\eta_s)/T_H$ for some specific pair of values $(\psi_0,\zeta_0)$ and $(\chi,\sigma)$. Left Table: $V(\phi)=0$, $\beta=0.10$. Right Table: $V (\phi) = -\frac{\mu ^2}{2}\phi ^2+\frac{\xi }{4}\phi ^4$,  $\beta=0.10$,  $\psi_0=0.1$, $\zeta_0=5\times 10^{-15}$.}
	\label{tt3}	
\end{table*}

\subsection{The specific heat of the black hole}

The specific heat $C_{BH}$ of the black hole is defined by \cite{BH1,BH2}
\begin{eqnarray}
C_{BH}&=&\frac{dm}{dT_{BH}}=\frac{dm}{dr}\frac{dr}{dT_{BH}}\bigg|_{r=r_{hor}}\nn
&&=\frac{M_0}{T_H}\frac{dM(\eta)}{d\eta}\frac{d\eta}{d\theta(\eta)}\bigg|_{\eta=\eta_s},
\end{eqnarray}

\begin{figure*}[htbp]
	\centering
	\includegraphics[width=8.9cm]{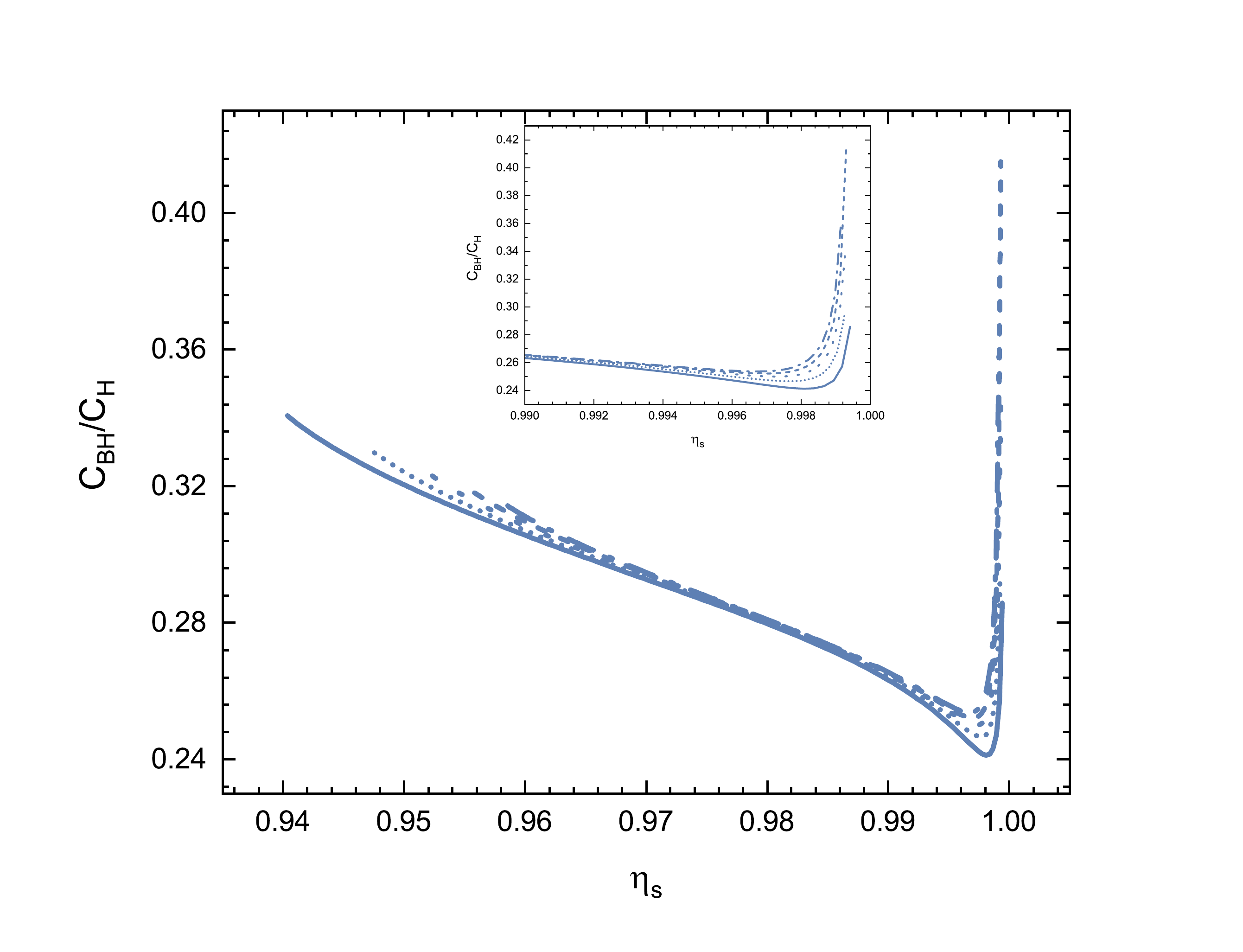}
	\includegraphics[width=8.9cm]{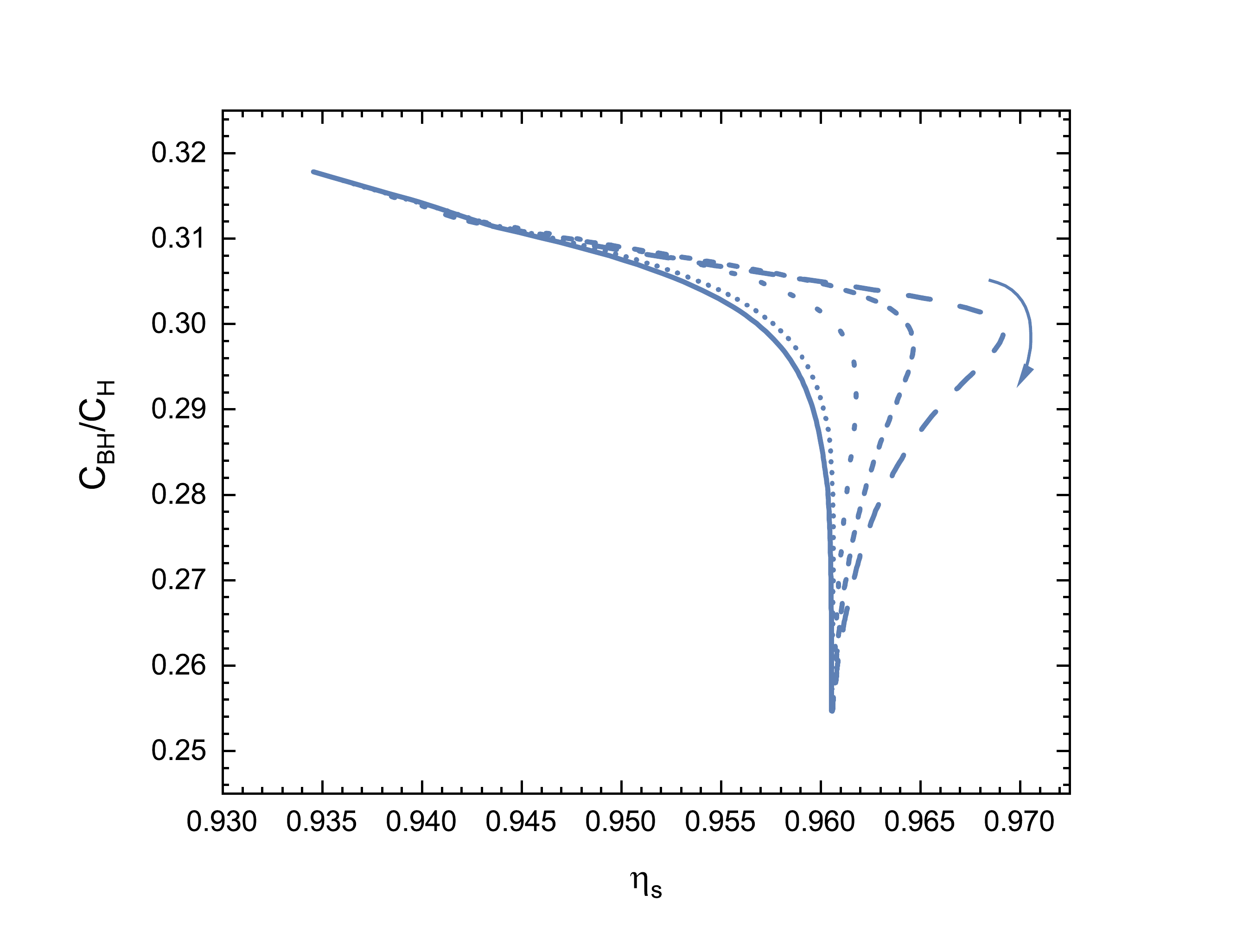}
	\caption{Variation of the dimensionless specific heat $C_{BH}(\eta_s)/C_H$ of the black hole as a function of the location of the event horizon $\eta_s$. Left Figure: $V(\phi)=0$,  $\beta=0.1$, and $\psi_0^2 \zeta_0 = 2\times 10^{-17}$ (solid curve), $\psi_0^2 \zeta_0 = 3\times 10^{-17}$ (dotted curve), $\psi_0^2 \zeta_0 = 4\times 10^{-17}$ (short dashed curve), $\psi_0^2 \zeta_0 = 5\times 10^{-17}$ (dashed curve), and $\psi_0^2 \zeta_0 = 6\times 10^{-17}$ (long dashed curve). Right Figure:  $V (\phi) = -\frac{\mu ^2}{2}\phi ^2+\frac{\xi }{4}\phi ^4$,  $\beta=0.1$,  $\psi_0=0.1$, $\zeta_0=5\times 10^{-15}$ and  $\sigma / \chi = 1\times 10^{-15}$ (solid curve), $\sigma / \chi = 1\times 10^{-14}$ (dotted curve), $\sigma / \chi = 4\times 10^{-14}$ (short dashed curve), $\sigma / \chi = 7\times 10^{-14}$ (dashed curve), $\sigma / \chi = 1\times 10^{-13}$ (long dashed curve). The arrow shows the direction of decreasing values.
	}\label{ffnp2}
\end{figure*}

In the following we define $C_{H}=M_0/T_H$. The variation of the specific heat $C_{BH}(\eta_s)/C_H$ of the black hole in the modified gravity theory induced by the quantum metric fluctuations as a function of the event horizon radius $\eta _s$ is represented in Fig.~\ref{ffnp2}.

 Similarly to the behavior of the Hawking temperature, there is a strong dependence of the specific heat of the black hole on both initial conditions, and on the parameters of the scalar field potential. In the case of the vanishing scalar field potential the specific heat decreases with increasing $\eta _s$, and reaches its maximum for $\eta _s\approx 1$.  The increase of $C_{BH}(\eta_s)/C_H$ around $\eta _s\approx 1$ is significant as compared to the values corresponding to $\eta _s\approx 0.94$. In the case of the Higgs potential the dependence of $C_{BH}(\eta_s)/C_H$ on the radius of the event horizon is rather complicated.  complicated, with the specific heat generally decreasing with increasing $\eta _s$.

\begin{figure*}[htbp]
	\centering
	\includegraphics[width=230pt]{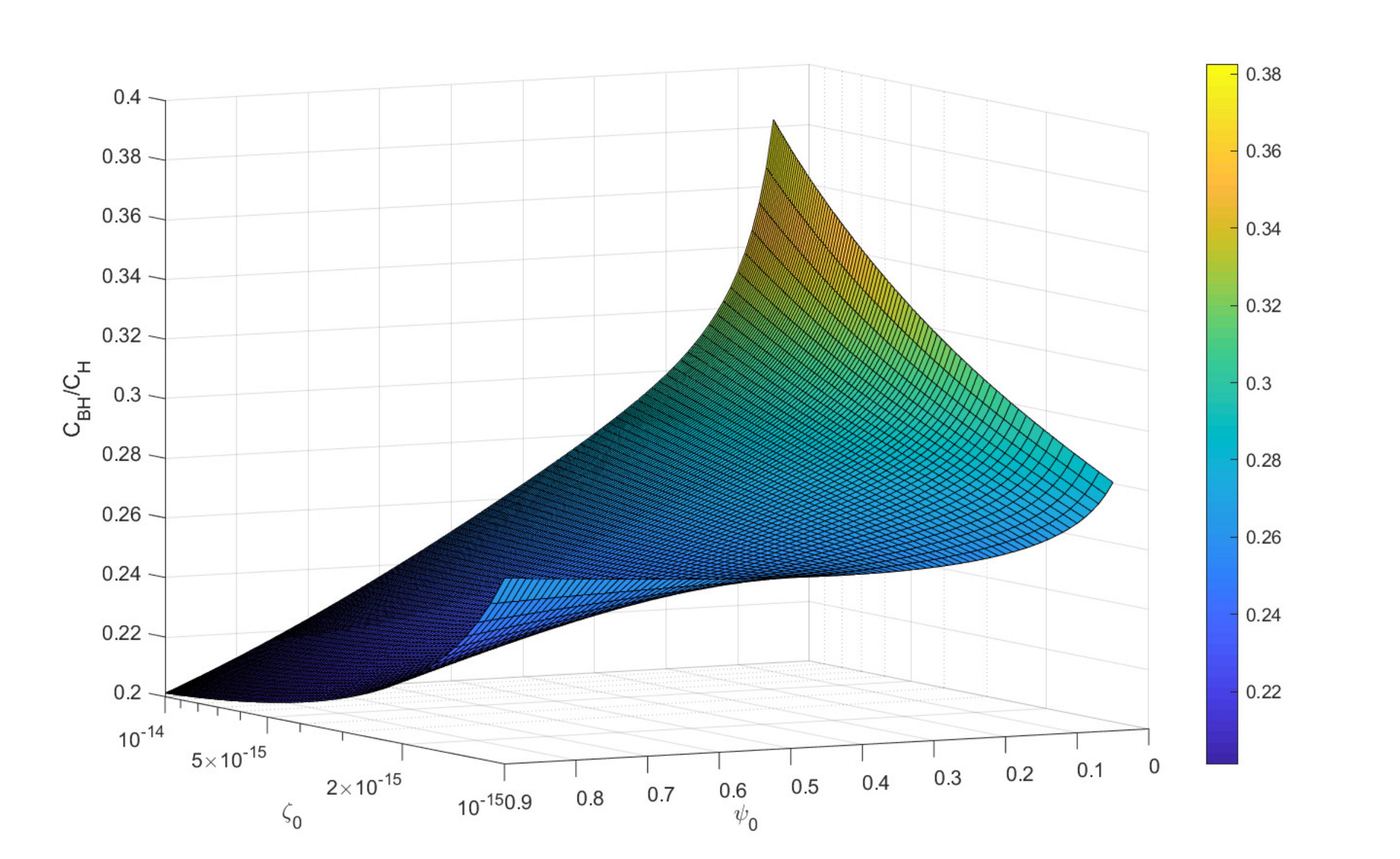}
	\includegraphics[width=230pt]{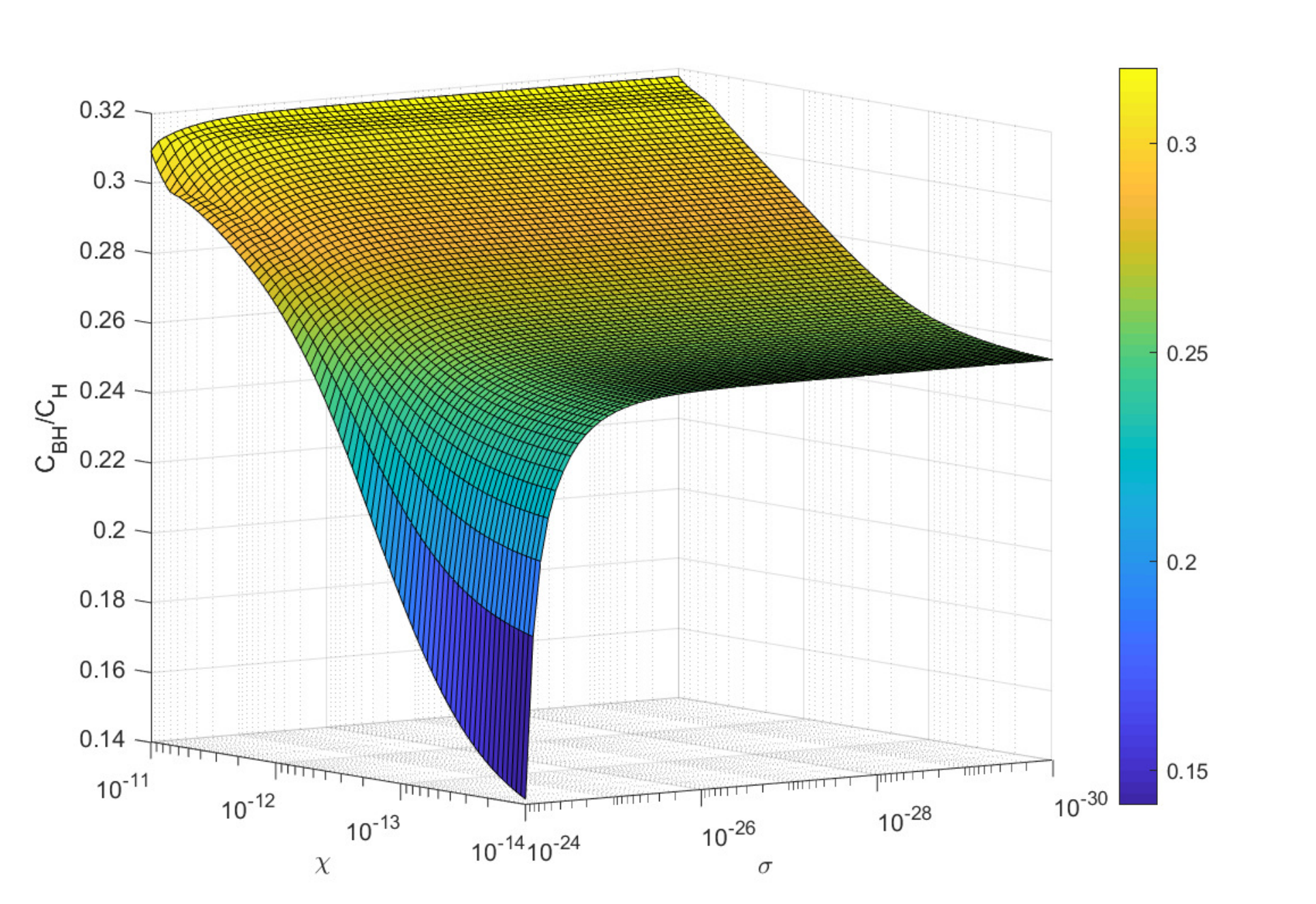}
	\caption{Variation of the specific heat $C_{BH}(\eta_s)/C_H$ as a function of the initial conditions for the scalar field and of the parameters of the scalar field potential. Left Figure: $V(\phi)=0$, $\beta=0.10$.
		Right Figure: $V (\phi) = -\frac{\mu ^2}{2}\phi ^2+\frac{\xi }{4}\phi ^4$,  $\beta=0.10$,  $\psi_0=0.1$, $\zeta_0=5\times 10^{-15}$. }
	\label{ff32}
\end{figure*}

The variations of the specific heat $C_{BH}(\eta_s)/C_H$ of the black hole in the modified gravity theory induced by the quantum metric fluctuations are represented, as a function of the initial conditions and of the potential parameters, for the cases of the vanishing scalar field potential $V(\phi)=0$ and for a Higgs type potential, respectively, in Fig.~\ref{ff32}.

Some specific numerical values of the dimensionless specific heat are presented in Table~\ref{tt4}.

\begin{table*}[htbp]
	\centering
	 \begin{tabular}{|>{\centering}p{50pt}|>{\centering}p{60pt}|>{\centering}p{60pt}|>{\centering}p{60pt}||>{\centering}p{60pt}|>{\centering}p{60pt}|>{\centering}p{60pt}|>{\centering\arraybackslash}p{60pt}|}
		\hline
		$\quad$ & $\zeta_0=1\times 10^{-15}$ & $\zeta_0=5\times 10^{-15}$ & $\zeta_0=9\times 10^{-15}$ & $\quad$ & $\chi=1\times 10^{-14}$ & $\chi=1\times 10^{-13}$ & $\chi=1\times 10^{-12}$\\
		\hline
		$\psi_0=0.1$ & $0.27025$ & $0.30871$ & $0.32809$ & $\sigma=1\times 10^{-30}$ & $0.25471$ & $0.26125$ & $0.28704$\\
		\hline
		$\psi_0=0.5$ & $0.25739$ & $0.24838$ & $0.24968$ & $\sigma=1\times 10^{-28}$ & $0.25470$ & $0.26124$ & $0.28704$\\
		\hline
		$\psi_0=0.9$ & $0.26239$ & $0.20489$ & $0.20144$ & $\sigma=1\times 10^{-26}$ & $0.25428$ & $0.26089$ & $0.28680$\\
		\hline
	\end{tabular}
	\caption{Numerical values of the specific heat of the black hole $C_{BH}(\eta_s)/C_H$ for some selected pairs $(\psi_0,\zeta_0)$ and $(\chi,\sigma)$. Left Table: $V(\phi)=0$, $\beta=0.10$. Right Table: $V (\phi) = -\frac{\mu ^2}{2}\phi ^2+\frac{\xi }{4}\phi ^4$,  $\beta=0.10$,  $\psi_0=0.1$, $\zeta_0=5\times 10^{-15}$.}
	\label{tt4}	
\end{table*}

\subsection{The black hole entropy}

The entropy $S_{BH}$ of black hole is defined by \cite{BH1,BH2}
\begin{eqnarray}
S_{BH}=\int_{r_{in}}^{r_{hor}}\frac{dm}{T_{BH}}=\int_{r_{in}}^{r_{hor}}\frac{1}{T_{BH}}\frac{dm}{dr}dr,
\end{eqnarray}
which in the dimensionless variables used in the present work can be written as
\begin{eqnarray}
S_{BH}(\eta_s)=C_H\int_{0}^{\eta_s}\frac{1}{\theta(\eta)}\frac{dM(\eta)}{d\eta}d\eta.
\end{eqnarray}

\begin{figure*}[htbp]
	\centering
	\includegraphics[width=230pt]{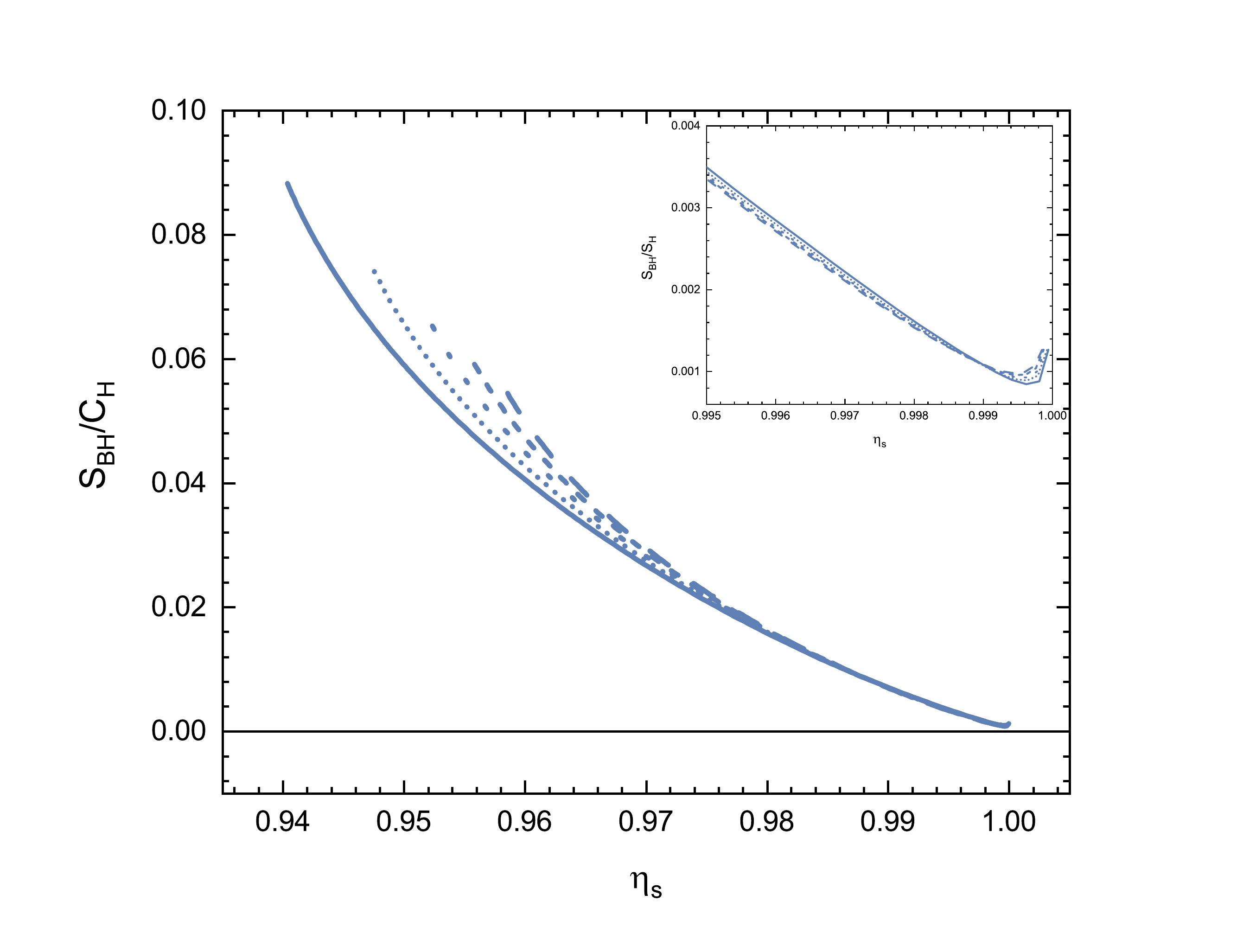}
	\includegraphics[width=230pt]{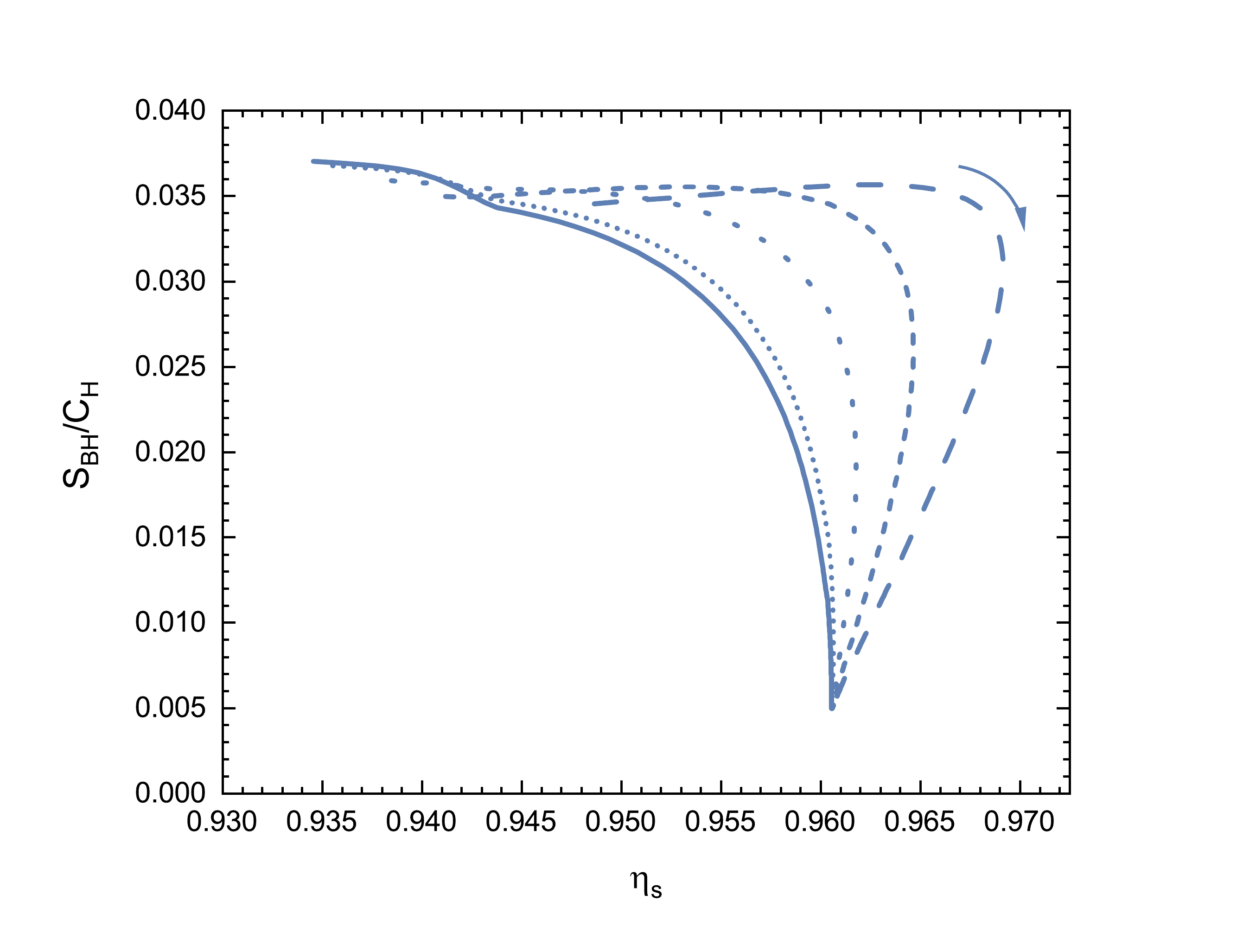}
	\caption{Variation of the entropy of the black hole $S_{BH}/C_H$ as a function of the location of the event horizon $\eta_s$. Left Figure: $V(\phi)=0$,  $\beta=0.1$, and $\psi_0^2 \zeta_0 = 2\times 10^{-17}$ (solid curve), $\psi_0^2 \zeta_0 = 3\times 10^{-17}$ (dotted curve), $\psi_0^2 \zeta_0 = 4\times 10^{-17}$ (short dashed curve), $\psi_0^2 \zeta_0 = 5\times 10^{-17}$ (dashed curve), and $\psi_0^2 \zeta_0 = 6\times 10^{-17}$ (long dashed curve). Right Figure:  $V (\phi) = -\frac{\mu ^2}{2}\phi ^2+\frac{\xi }{4}\phi ^4$,  $\beta=0.1$,  $\psi_0=0.1$, $\zeta_0=5\times 10^{-15}$ and  $\sigma / \chi = 1\times 10^{-15}$ (solid curve), $\sigma / \chi = 1\times 10^{-14}$ (dotted curve), $\sigma / \chi = 4\times 10^{-14}$ (short dashed curve), $\sigma / \chi = 7\times 10^{-14}$ (dashed curve), $\sigma / \chi = 1\times 10^{-13}$ (long dashed curve). The arrow shows the direction of decreasing values. }
	\label{ffp3}
\end{figure*}

As a function of the event horizon $\eta _s$ the variations of the dimensionless entropy $S_{BH}(\eta_s)/C_H$ are presented in Figs.~\ref{ffp3} for a vanishing and for a Higgs type scalar field potential, respectively.

The entropy of the black hole has a similar qualitative behavior as its Hawking temperature. However, the numerical values corresponding to different event horizons differ significantly from the Schwarzschild black hole values, with the entropy increasing with decreasing $\eta _s$.

\begin{figure*}[htbp]
	\centering
	\includegraphics[width=230pt]{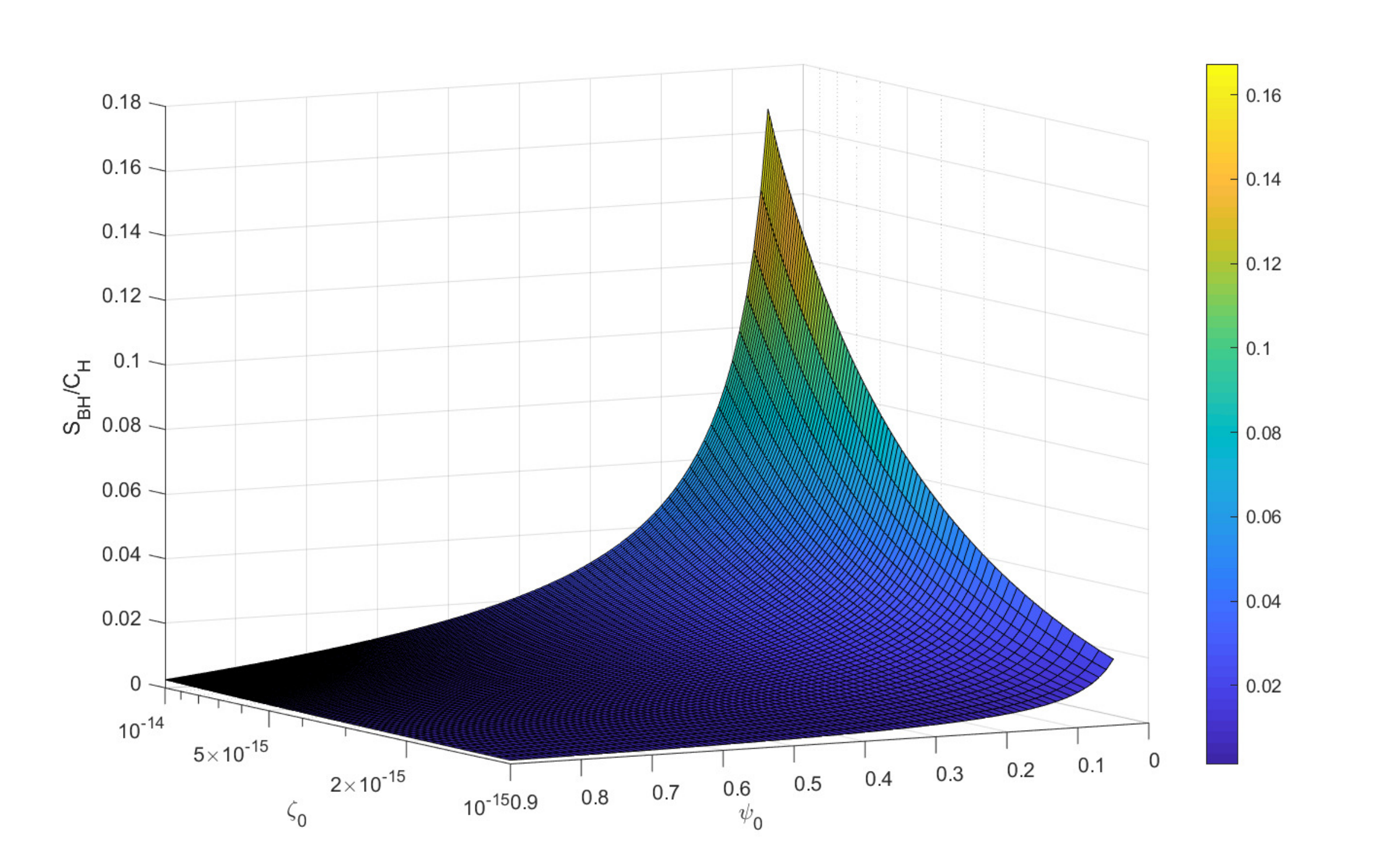}
	\includegraphics[width=230pt]{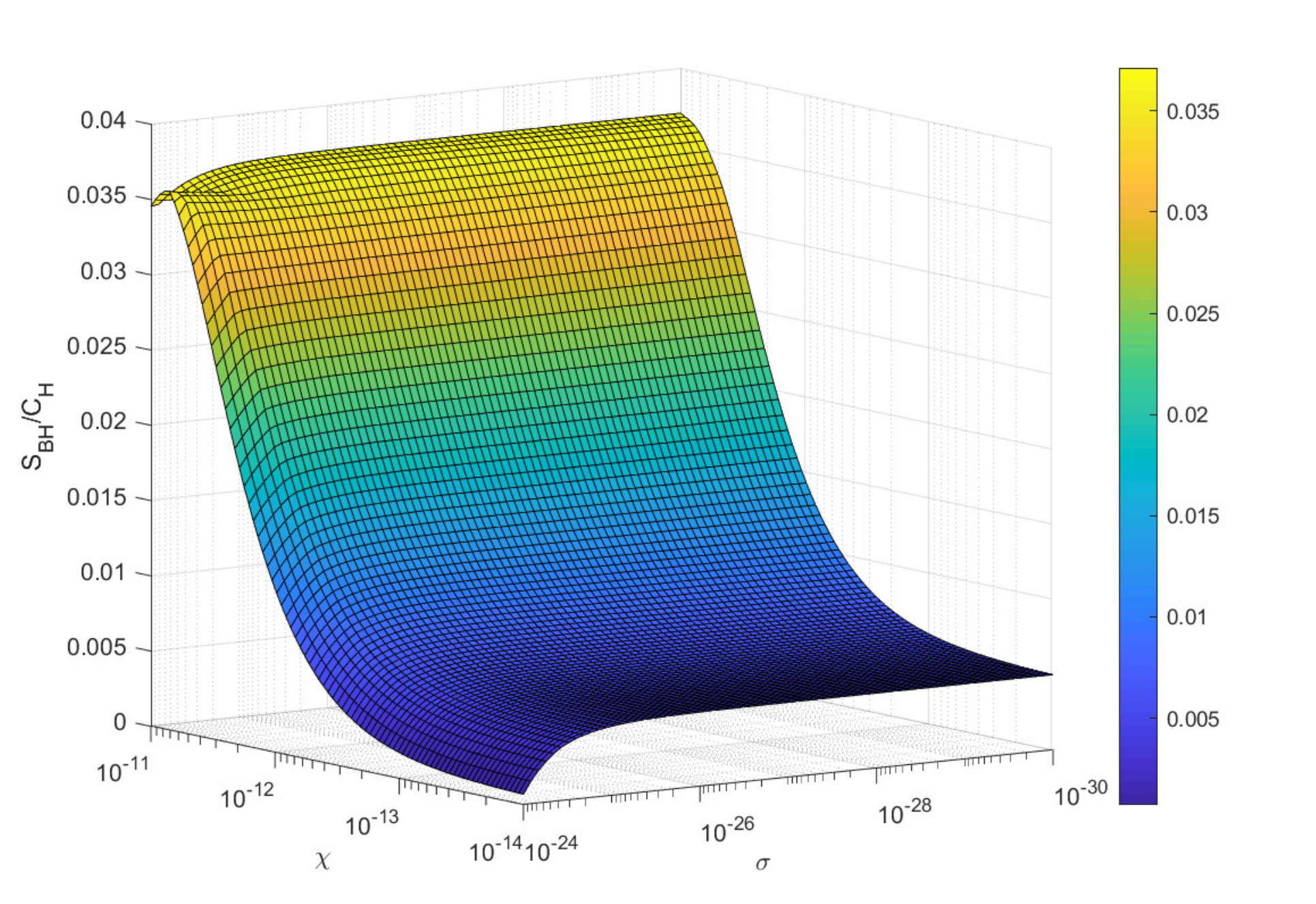}
	\caption{Variation of black hole entropy $S_{BH}(\eta_s)/C_H$ as a function of the initial conditions for the scalar field and of the scalar field potential parameters. Left Figure: $V(\phi)=0$, $\beta=0.10$.
		Right Figure: $V (\phi) = -\frac{\mu ^2}{2}\phi ^2+\frac{\xi }{4}\phi ^4$,  $\beta=0.10$,  $\psi_0=0.1$, $\zeta_0=5\times 10^{-15}$. }
	\label{ff33}
\end{figure*}

The variations of the entropy $S_{BH}(\eta_s)/C_H$ of the black hole as a function of the initial conditions at infinity for the scalar field and of the parameters of the scalar field potential  are represented, for the cases of the vanishing scalar field potential $V(\phi)=0$ and for a Higgs type potential, respectively, in Fig.~\ref{ff33}.

Some specific numerical values of the dimensionless entropy of the black hole are presented in Table~\ref{tt5}.

\begin{table*}[htbp]
	\centering
	 \begin{tabular}{|>{\centering}p{50pt}|>{\centering}p{60pt}|>{\centering}p{60pt}|>{\centering}p{60pt}||>{\centering}p{60pt}|>{\centering}p{60pt}|>{\centering}p{60pt}|>{\centering\arraybackslash}p{60pt}|}
		\hline
		$\quad$ & $\zeta_0=1\times 10^{-15}$ & $\zeta_0=5\times 10^{-15}$ & $\zeta_0=9\times 10^{-15}$ & $\quad$ & $\chi=1\times 10^{-12}$ & $\chi=1\times 10^{-13}$ & $\chi=1\times 10^{-14}$\\
		\hline
		$\psi_0=0.1$ & $1.0873\times 10^{-2}$ & $4.6043\times 10^{-2}$ & $8.0632\times 10^{-2}$ & $\sigma=1\times 10^{-30}$ & $4.9810\times 10^{-3}$ & $5.9455\times 10^{-3}$ & $1.4510\times 10^{-2}$\\
		\hline
		$\psi_0=0.5$ & $2.2684\times 10^{-3}$ & $8.1100\times 10^{-3}$ & $1.3838\times 10^{-2}$ & $\sigma=1\times 10^{-28}$ & $4.9806\times 10^{-3}$ & $5.9452\times 10^{-3}$ & $1.4510\times 10^{-2}$\\
		\hline
		$\psi_0=0.9$ & $1.3173\times 10^{-3}$ & $1.8739\times 10^{-3}$ & $2.3603\times 10^{-3}$ & $\sigma=1\times 10^{-26}$ & $4.9436\times 10^{-3}$ & $5.9094\times 10^{-3}$ & $1.4492\times 10^{-2}$\\
		\hline
	\end{tabular}
	\caption{Numerical values of the entropy of the black hole $S_{BH}(\eta_s)/C_H$ for some specific pairs $(\psi_0,\zeta_0)$ and $(\chi,\sigma)$. Left Table: $V(\phi)=0$, $\beta=0.10$. Right Table: $V (\phi) = -\frac{\mu ^2}{2}\phi ^2+\frac{\xi }{4}\phi ^4$,  $\beta=0.10$,  $\psi_0=0.1$, $\zeta_0=5\times 10^{-15}$.}
	\label{tt5}	
\end{table*}

\subsection{Black hole luminosity and evaporation time}

\begin{figure*}[htbp]
	\centering
	\includegraphics[width=230pt]{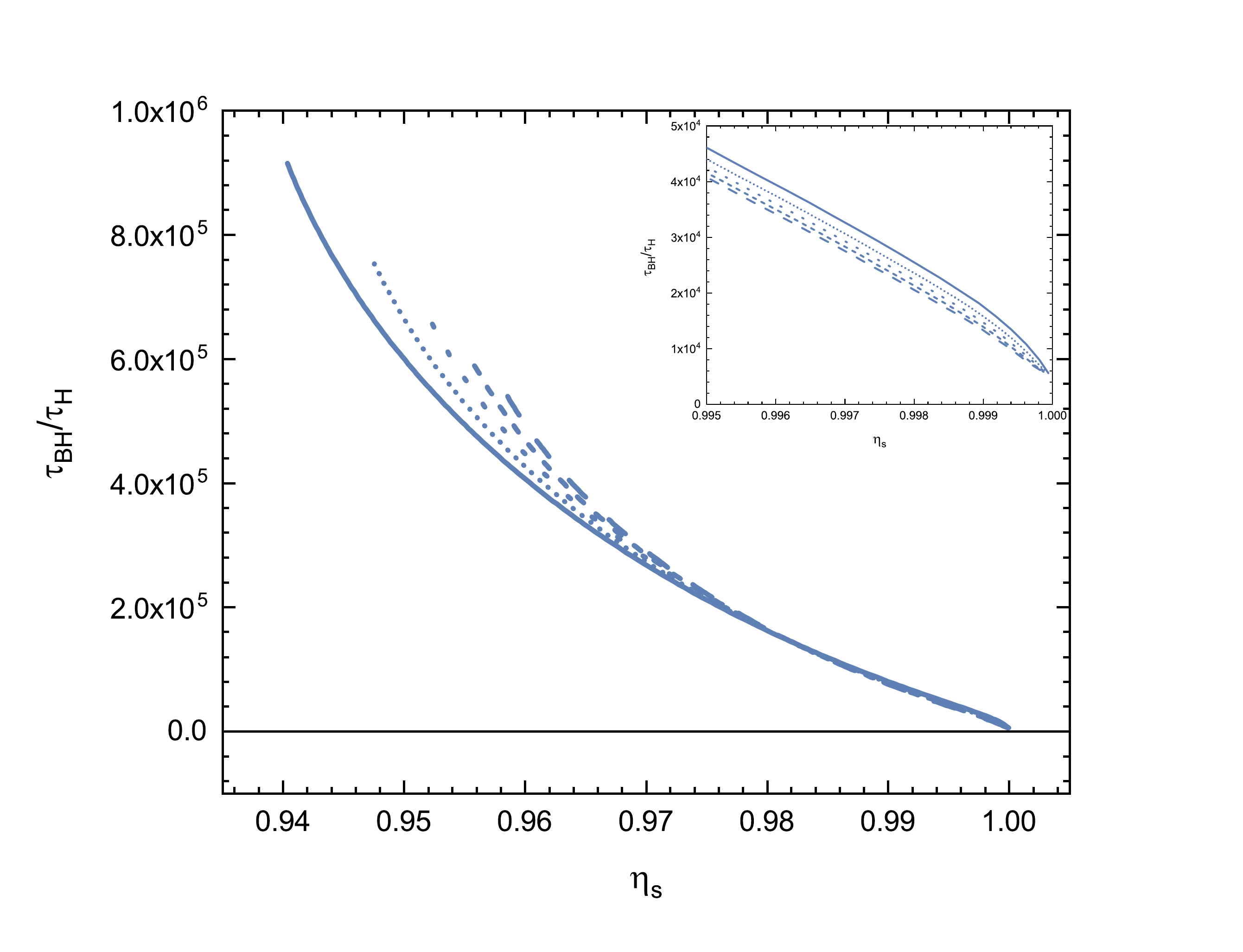}
	\includegraphics[width=230pt]{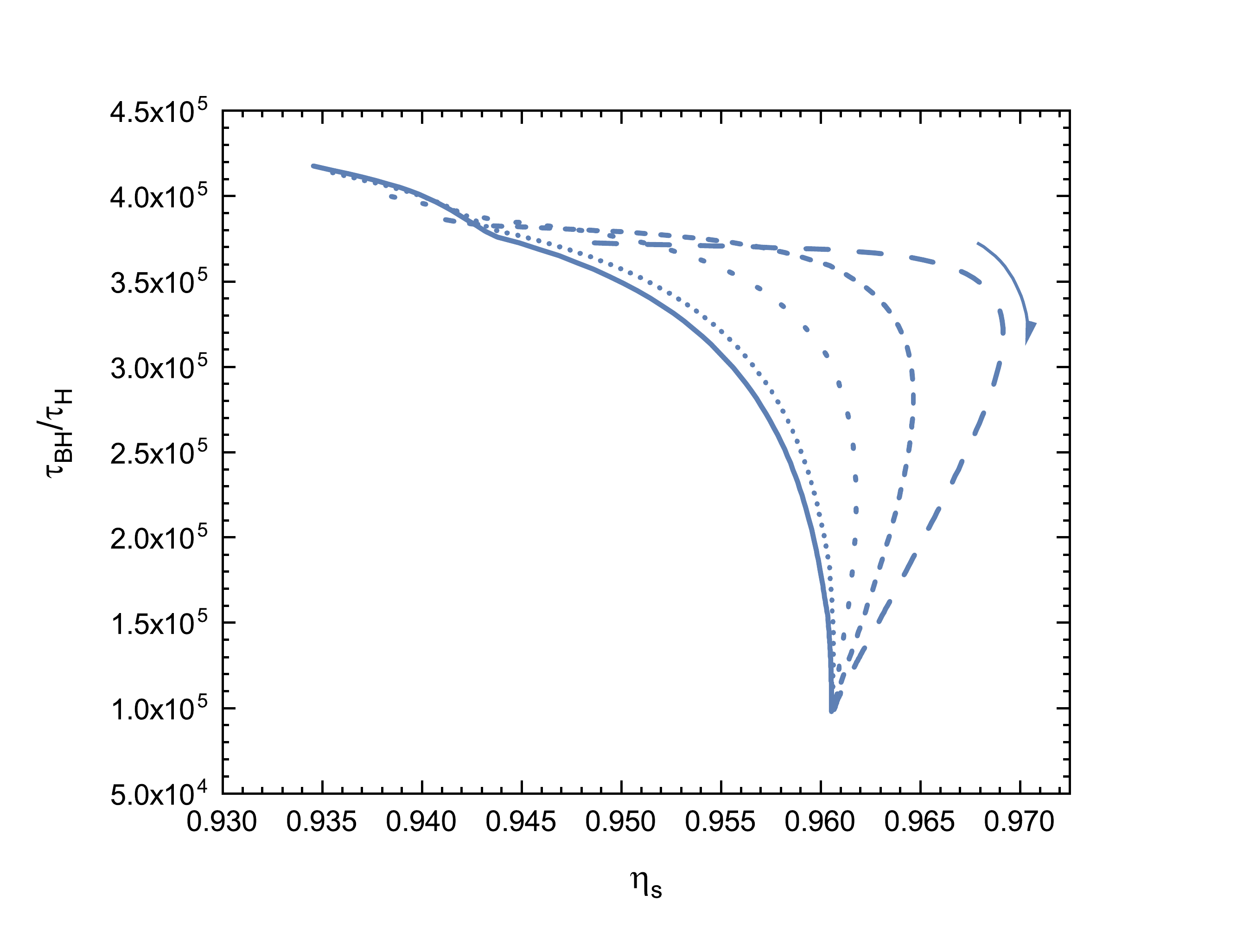}
	\caption{Variation of the dimensionless evaporation time $\tau_{BH}(\eta_s)/\tau_H$ as a function of the location of the event horizon $\eta_s$. Left Figure: $V(\phi)=0$,  $\beta=0.1$, and $\psi_0^2 \zeta_0 = 2\times 10^{-17}$ (solid curve), $\psi_0^2 \zeta_0 = 3\times 10^{-17}$ (dotted curve), $\psi_0^2 \zeta_0 = 4\times 10^{-17}$ (short dashed curve), $\psi_0^2 \zeta_0 = 5\times 10^{-17}$ (dashed curve), and $\psi_0^2 \zeta_0 = 6\times 10^{-17}$ (long dashed curve). Right Figure:  $V (\phi) = -\frac{\mu ^2}{2}\phi ^2+\frac{\xi }{4}\phi ^4$,  $\beta=0.1$,  $\psi_0=0.1$, $\zeta_0=5\times 10^{-15}$ and  $\sigma / \chi = 1\times 10^{-15}$ (solid curve), $\sigma / \chi = 1\times 10^{-14}$ (dotted curve), $\sigma / \chi = 4\times 10^{-14}$ (short dashed curve), $\sigma / \chi = 7\times 10^{-14}$ (dashed curve), $\sigma / \chi = 1\times 10^{-13}$ (long dashed curve). The arrow shows the direction of decreasing values. }
	\label{ffp4}
\end{figure*}

The Black hole luminosity due to the Hawking evaporation is defined as \cite{BH1,BH2}
\begin{eqnarray}
L_{BH}=\frac{dm}{dt}=-\sigma A_{BH}T^4_{BH},
\end{eqnarray}
where $\sigma$ is a model dependent parameter, and $A_{BH}=4\pi r_{hor}^2$ is the area of the event horizon. The black hole evaporation time $\tau$ can be obtained then as
\begin{eqnarray}
\tau=\int_{t_{in}}^{t_{fin}}dt=-\frac{1}{4\pi\sigma}\int_{t_{in}}^{t_{fin}}\frac{dm}{r^2_{hor}T^4_{BH}},
\end{eqnarray}
or, in the dimensionless variable $\eta$, in the equivalent form
\begin{eqnarray}
\tau(\eta_s)=-\tau_H\int_{\eta_s}^{0}\frac{1}{\eta^2\theta^4(\eta)}\frac{dM(\eta)}{d\eta}d\eta,
\end{eqnarray}
 where we have denoted $\tau_H=c^4/8\pi G^2\sigma M_0T^4_{BH}$. As a function of the position of the event horizon the evaporation time of the black holes is represented, for the vanishing and for the Higgs type potential of the scalar field, in Figs.~\ref{ffp4}. For both cases of the vanishing and Higgs scalar field potential the evaporation time decreases with increasing $\eta _s$. In the case of the Higgs potential, the evaporation time decreases from $\tau _{BH}\approx 4.2\times 10^{5}\tau _H$ for $\eta_s=\approx 0.93$ to $\tau _{BH}\approx 1.0\times 10^5$ for $\eta _s\approx 0.96$. By taking into account that the Hawking evaporation time of a standard general relativistic black hole is of the order of magnitude of $\tau_H\approx 4.8\times 10^{-27}\times \left(M_0/{\rm g}\right)^3$, it follows that the evaporation time of black holes in the modified gravity induced by the quantum metric fluctuations could be
four or five orders of magnitude higher. But even in this case the evaporation time for astrophysical size black holes via Hawking radiation in the presence of quantum fluctuations of the metric remains very high. In the case of a black hole having three solar masses, the evaporation time is of the order of $10^{58}-10^{59}$ years.

\begin{figure*}[htbp]
	\centering
	\includegraphics[width=230pt]{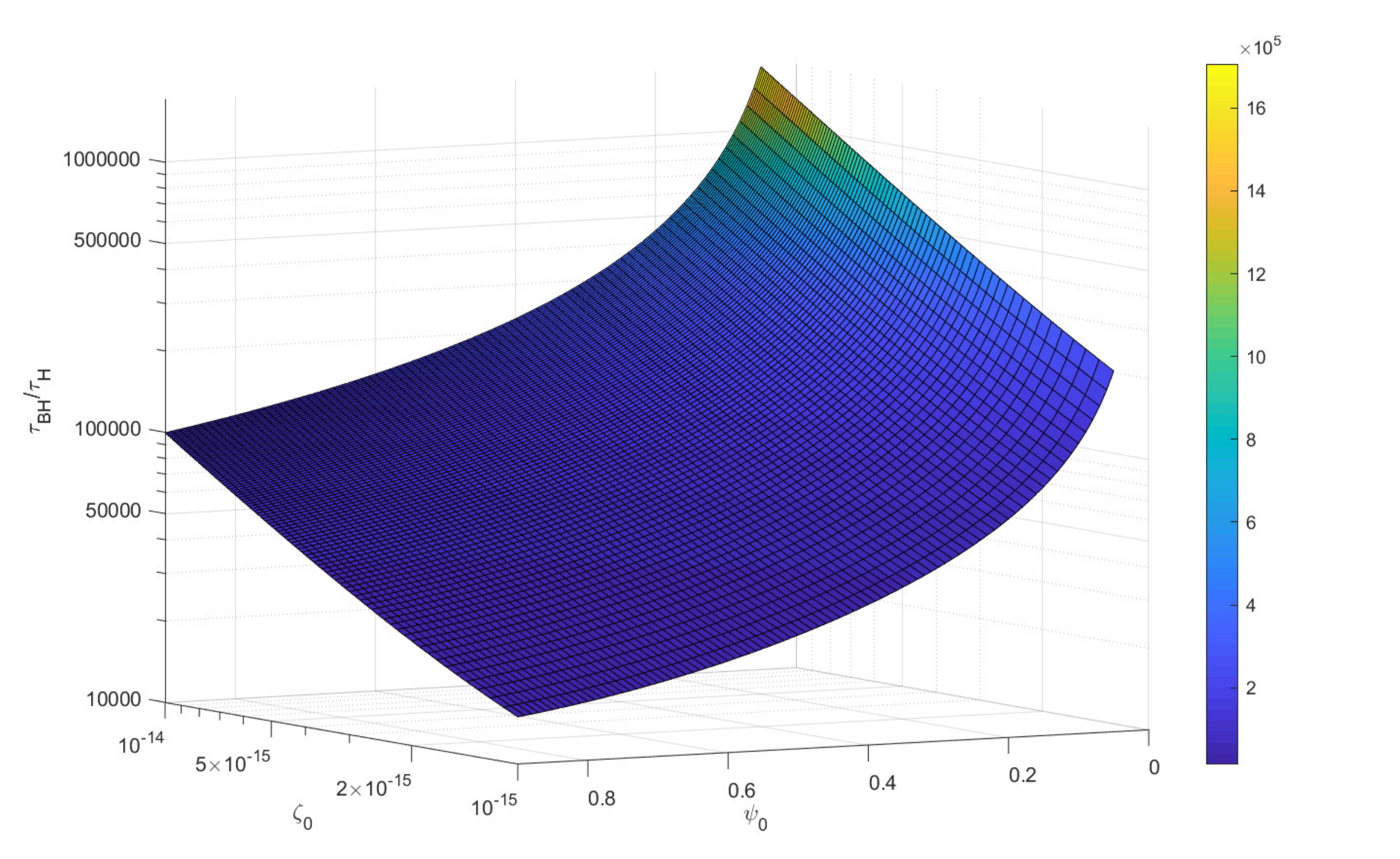}
	\includegraphics[width=230pt]{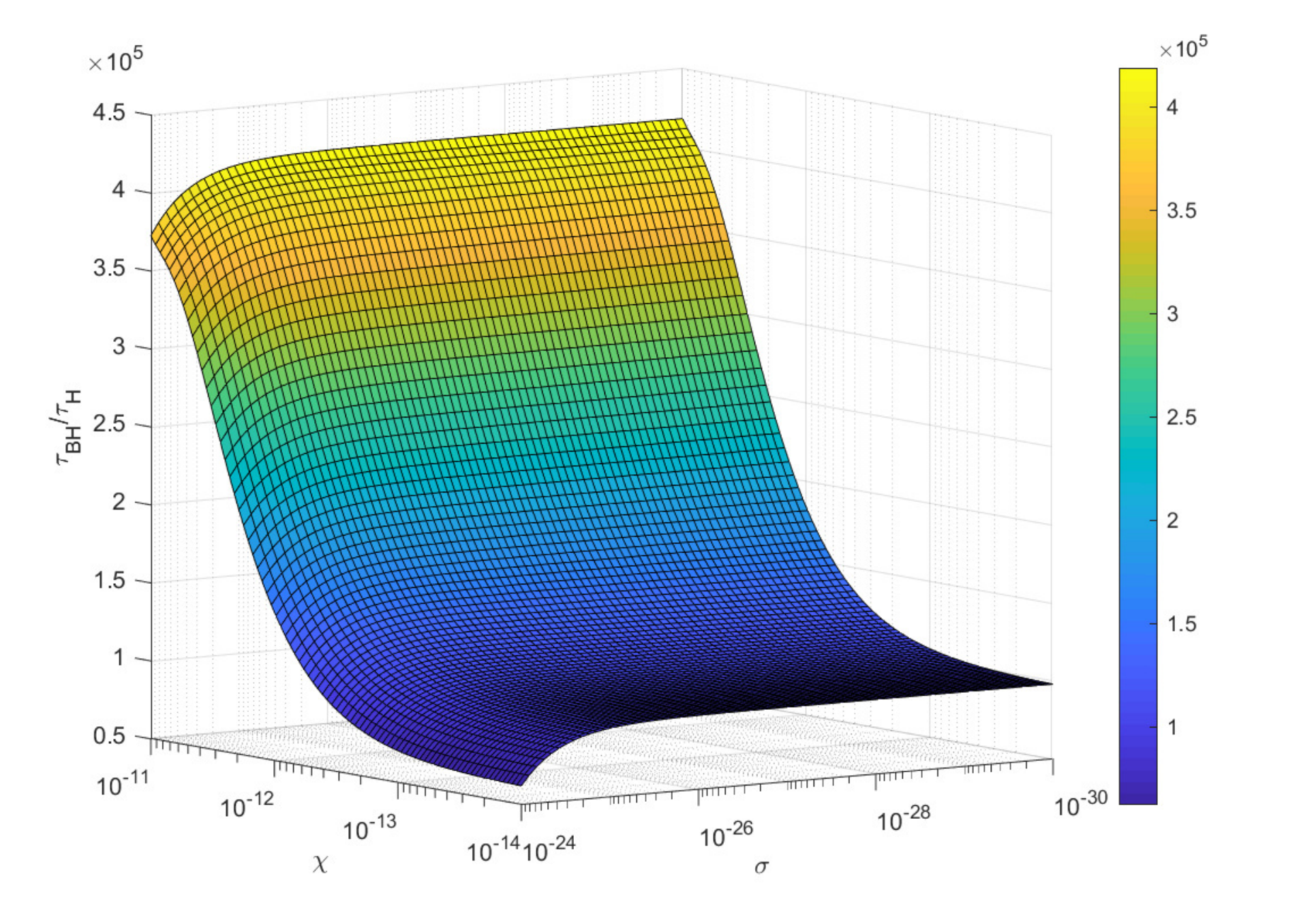}
	\caption{Variations of the dimensionless evaporation time $\tau_{BH}(\eta_s)/\tau_H$ as a function of the initial values at infinity of the scalar field and of the parameters of the scalar field potential. Left Figure: $V(\phi)=0$, $\beta=0.10$.
		Right Figure: $V (\phi) = -\frac{\mu ^2}{2}\phi ^2+\frac{\xi }{4}\phi ^4$,  $\beta=0.10$,  $\psi_0=0.1$, $\zeta_0=5\times 10^{-15}$. }
	\label{ff34}
\end{figure*}

The variations of the evaporation time $\tau_{BH}(\eta_s)/\tau_H$ as a function of the initial conditions of the scalar field at infinity and of the parameters of the potential of the scalar field are represented, for the cases of the vanishing scalar field potential $V(\phi)=0$ and for a Higgs type potential, respectively, in Fig.~\ref{ff34}.

Some specific values of the evaporation time of the black holes in the modified gravity induced by the quantum metric fluctuations  are presented in Table~\ref{tt6}.

\begin{table*}[htbp]
	\centering \begin{tabular}{|>{\centering}p{50pt}|>{\centering}p{60pt}|>{\centering}p{60pt}|>{\centering}p{60pt}||>{\centering}p{60pt}|>{\centering}p{60pt}|>{\centering}p{60pt}|>{\centering\arraybackslash}p{60pt}|}
		\hline
		$\quad$ & $\zeta_0=1\times 10^{-15}$ & $\zeta_0=5\times 10^{-15}$ & $\zeta_0=9\times 10^{-15}$ & $\quad$ & $\chi=1\times 10^{-14}$ & $\chi=1\times 10^{-13}$ & $\chi=1\times 10^{-12}$\\
		\hline
		$\psi_0=0.1$ & $1.1884\times 10^{5}$ & $4.5608\times 10^{5}$ & $7.8994\times 10^{5}$ & $\sigma=1\times 10^{-30}$ & $9.7938\times 10^{4}$ & $1.0658\times 10^{5}$ & $1.8334\times 10^{5}$\\
		\hline
		$\psi_0=0.5$ & $2.6415\times 10^{4}$ & $9.4256\times 10^{4}$ & $1.6184\times 10^{5}$ & $\sigma=1\times 10^{-28}$ & $9.7934\times 10^{4}$ & $1.0657\times 10^{5}$ & $1.8333\times 10^{5}$\\
		\hline
		$\psi_0=0.9$ & $1.5002\times 10^{4}$ & $5.2685\times 10^{4}$ & $9.0206\times 10^{4}$ & $\sigma=1\times 10^{-26}$ & $9.7570\times 10^{4}$ & $1.0621\times 10^{5}$ & $1.8294\times 10^{5}$\\
		\hline
	\end{tabular}
	\caption{Numerical values of the dimensionless evaporation time of the black hole $\tau_{BH}(\eta_s)/\tau_H$ for some specific pairs $(\psi_0,\zeta_0)$ and $(\chi,\sigma)$. Left Table: $V(\phi)=0$, $\beta=0.10$. Right Table: $V (\phi) = -\frac{\mu ^2}{2}\phi ^2+\frac{\xi }{4}\phi ^4$,  $\beta=0.10$,  $\psi_0=0.1$, $\zeta_0=5\times 10^{-15}$.}
	\label{tt6}	
\end{table*}

\section{Discussions and final remarks}\label{sect6}

In the present work we have analyzed black hole type solutions in a specific semiclassical approach to Einstein gravity  theory, in which the quantum effects in the metric are introduced via a fluctuating, second order tensor component $\delta \hat{g}_{\mu \nu}$. By substituting the stochastic operator $\delta \hat{g}_{\mu \nu}$ by its average value $\left<\delta \hat{g}_{\mu \nu}\right>=K_{\mu\nu}$ we obtain a modified gravity theory, with geometry-matter coupling , and non-conservation of the matter energy-momentum tensor. In the present theoretical approach the fluctuation tensor remains arbitrary, and its form cannot be fixed from geometrical or physical considerations. In this paper we have considered the simplest possible form of $K_{\mu \nu}$, by assuming that it is given by the direct coupling between a scalar field and the metric tensor, so that $K_{\mu \nu}=\alpha \phi (x)g_{\mu \nu}$. We have written down the action, and obtained the full form of the field equations for this model.

As a physical/astrophysical application of the modified gravity theory induced by the quantum metric fluctuations we have considered one of the simplest cases, by adopting a vacuum static and spherically symmetric geometry. However, even in this simple theoretical model the gravitational field equations of the theory turn out to be extremely complicated from a mathematical point of view, consisting of a system of highly nonlinear differential equations. Hence, in order to obtain solutions of the field equations we need to use numerical methods. In order to facilitate the numerical integration procedures  we have reformulated the static spherically symmetric gravitational field equations of the model in a dimensionless form. We have also introduced as independent variable $\eta=1/r$ the inverse of the radial coordinate $r$. This transformation significantly simplifies the numerical integration procedure,  which also implies the fixing the numerical values of the scalar field $\phi$, of its derivative $\phi'$, and of the effective mass $M$ at infinity. The presence of a singular behavior in the Einstein gravitational field equations, or, more exactly, in the two metric tensor components $e^{-\lambda}$ and $e^{\nu}$, indicates the existence of an event horizon encompassing the massive central object, and, consequently, of a black hole type astrophysical structure. The mass of the black hole is obtained from the gravitational field equations, and it corresponds to the effective mass of the model. The mass of the black hole is thus obtained as the total contribution from the mass-energy of the scalar field, coupled to the metric tensor, plus the ordinary mass of the black hole.

In the presence of quantum fluctuations coupled to the metric tensor via a scalar field the solutions of the gravitational field equations of the modified Einstein gravity  theory essentially depend on the scalar field potential. In our analysis we have considered two choices of $V\left(\phi\right)$, corresponding to two choices, namely, the vanishing potential $V(\phi)=0$, and the Higgs-type potential, respectively. In both cases we have numerically integrated the field equations for different initial conditions, and different potential parameters. Our results indicate the presence of a singularity at a finite value of the inverse of the radial coordinate $\eta _s$, and therefore the formation of an event horizon, showing the existence of a  black hole. The location of the event horizon depends on the initial conditions at infinity, given by the values of the scalar field and of its derivative at infinity.  Hence the presence of the effective quantum fluctuations of the metric generates a complex relation between the scalar field, characterizing the fluctuations,  and the black hole properties. In particular, for some specific initial conditions, in the zero potential case the event  horizon can be located at (inverse) distances of the order of $\eta _z=0.8$ from the black hole, corresponding to a gravitational radius of the order of $r_g=1.25r_{Sch}$, where $r_{Sch}=2GM/c^2$ is
of the standard Schwarzschild radius. This result indicates that in the presence of quantum metric fluctuations,  more compact black holes can be formed as compared to the standard general relativistic case. For the Higgs-type potential the location of the event horizon is strongly dependent not only on the initial conditions at infinity, but also on the numerical values of the dimensionless parameters $\chi $ and $\sigma $ of the potential, showing a multi-parametric dependence of the black hole properties on the scalar field, and its potential.

In the two cases we have extensively studied we have found that the numerical results can be fitted well by some simple analytical functions. In the zero potential case the metric tensor component $e^{\nu(r)}$ can be represented by a function of the form $e^{\nu (r)}=B_1/r+C_1/r^2+D_1$, with $B_1,C_1,D_1$ constants that depend on the initial conditions at infinity of the scalar field and of its derivative. The metric tensor component can be obtained as $e^{-\lambda (r)}=1-B_2/r-C_2/r^2-D_2/r^3-E_2/r^4$, with $A_2,B_2,C_2,D_2$ constants depending on the initial conditions at infinity. Similar results are
obtained for the case of the Higgs-type scalar field potential. The analytical representations are very useful in the study of the thermodynamic properties of the black holes in the presence of quantum fluctuation of the metric, and also for the investigations  of the dynamical properties  and motion of the massive  particles around them. In particular, the analytical representations of the metric may be used for the study of the electromagnetic properties of the accretion disks existing around black holes \cite{disk1,disk2,disk3,disk4}. Hence by using the emissivity properties of the accretion disks one could, at least in principle, discriminate between the properties of the black holes in the modified gravity theory with quantum metric fluctuations, and of their general relativistic counterparts. Moreover,  one could use disk properties for obtaining some observational constraints on the model parameters.

The thermodynamic properties of the obtained numerical black hole solutions have been also investigated in detail. The Hawking temperature is one of the basic and essential physical property of black holes, and it relates their classical and quantum properties. The horizon temperature of the black holes in modified gravity in the presence of quantum fluctuation show a strong dependence on the initial conditions at infinity, and the properties of the scalar field potential. This represents an important difference between the Hawking temperature of the standard general relativistic black holes, and those in modified gravity with geometry-matter coupling.  Similar dependencies on the initial conditions and scalar field potential do appear for the specific heat, entropy and evaporation time of the black holes in  the presence of quantum fluctuation of the metric.  In particular, the black hole  evaporation times may be very different in in the presence of the quantum fluctuations of the metric as compared to standard general relativity. Of course the results presented in this study on the thermodynamics of black holes, obtained for only two classes of scalar field potentials, the zero potential, and the Higgs-type potential, respectively, and for a limited set of initial conditions at infinity, may be considered of preliminary nature only. A detailed and systematic study involving a larger class of potentials and initial conditions would be necessary to fully clarify the thermodynamic properties of this type of black holes. But even the present limited investigation already points towards  the complexity of the gravitational phenomena in the presence of quantum fluctuations, and of the interesting physics these theories lead to. Another result related to our numerical investigations is that they did not indicate the presence of any globally regular solutions of the field equations of the modified gravity theory induced by the quantum fluctuations of the metric.

A fundamental result in the theory of black holes is the no-hair theorem \cite{Bek, Bek1, Ad, Bek2}. This important theorem  states that around asymptotically flat black holes no external nontrivial  scalar fields with non-negative field potential $V (\phi)$ can exists. The numerical results obtained in the present investigation indicate that in its standard formulation the no-hair theorem  cannot be extended to modified gravity theories in the presence of quantum fluctuation theory and with geometry-matter coupling. All the numerical black hole solutions we have obtained are asymptotically flat, with the metric becoming Minkowskian at infinity. However, scalar fields with positive potentials do exist around them. On the other hand the answer to the question if such black hole - scalar field configurations do result from a particular choice of the initial conditions at infinity, of the model parameters and of the scalar field potentials, or if they are inherent properties of the basic gravitational theory can be obtained only through further investigations.

 The black holes obtained as solutions of the modified gravity  theory induced by the presence of quantum metric fluctuations does present a much richer theoretical structure as compared with the standard general relativistic black holes. Their properties are also associated with a very rich external dynamics, determined  by the presence of the complex coupling between the metric and the scalar field,  which leads to a set of strongly nonlinear field equations, characterized by a high level of mathematical complexity. The new black hole solutions have some specific astrophysical signatures, which, once detected observationally, could open new perspectives in gravitational physics and astrophysics. The possible astrophysical/observational implications of the quantum metric fluctuation  effects on black holes properties will be considered in a future publication.

\section*{Acknowledgments}

We would like to thank the two anonymous referees for comments and suggestions that helped us to improve our manuscript. T. H. would like to thank the Yat Sen School of the Sun Yat Sen University in Guangzhou, P. R. China, for the kind hospitality offered during the preparation of this work.

\end{document}